\documentclass[useAMS,usenatbib,usegraphicx]{mn2e}
\usepackage{pslatex,color,bm,url,amsmath,amsfonts,amssymb,verbatim}
\def\idm#1{{\mbox{\scriptsize #1}}}
\def\vec#1{{\pmb #1}}
\def\kepler{{\em Kepler}}
\def\code#1{{\sc #1}}
\def\Y{\langle Y \rangle}

\def\vec#1{{\pmb #1}}  
\DeclareMathAlphabet{\mathsfit}{\encodingdefault}{\sfdefault}{m}{sl}
\SetMathAlphabet{\mathsfit}{bold}{\encodingdefault}{\sfdefault}{bx}{sl}

\newcommand{\au}{\mbox{au}} 

\newcommand{\Rsun}{\mbox{R}_{\odot}}
\newcommand{\mE}{\mbox{M}_{\oplus}}

\newcommand{\RE}{\mbox{R}_{\oplus}}

\newcommand{\Pb}{P_{\idm{1}}}
\newcommand{\Pc}{P_{\idm{2}}}
\newcommand{\Pd}{P_{\idm{3}}}

\newcount\exno

\definecolor{myred}{rgb}{0.7, 0.0, 0.13}
\definecolor{myblue}{rgb}{0.1,0.0,0.6} 
\definecolor{mybrown}{rgb}{0.9,0.4,0.3}
\newcommand\ednote[1]{{\global\advance\exno by 1\
    \color{myred}$\spadesuit$({\bfseries\the\exno}).
\color{myblue}\bfseries\em #1}}
\newcommand\corre[1]{{ #1}} 

\newcommand\hide[1]{}

\def\kepler{{\sc Kepler}}
\def\code#1{{\tt #1}}

\def\idm#1{{\mbox{\scriptsize #1}}}
\newcommand\Chi{{\chi^2_\nu}}

\newcommand\gccm{{\,g$\cdot${}cm$^{-3}$}}
\def\kepler{{\sc Kepler}}
\title[The Kepler-30 planetary system]%
{The architecture and formation of the Kepler-30 planetary system}
\author[F.~Panichi, K.~Go\'zdziewski, C.~Migaszewski \& E.~Szuszkiewicz]{
F. Panichi$^{1}$\thanks{e-mail: federico.panichi@stud.usz.edu.pl},
K. Go\'zdziewski$^{2}$\thanks{e-mail: krzysztof.gozdziewski@umk.pl},
C. Migaszewski$^{2}$\thanks{e-mail: migaszewski@umk.pl} \&
E. Szuszkiewicz$^{1}$\thanks{e-mail: szusz@univ.szczecin.pl}\\
$^{1}$Institute of Physics and CASA*, Faculty of Mathematics and Physics,
University of Szczecin,  Wielkopolska 15, 70-451 Szczecin, Poland\\
$^{2}$Centre for Astronomy, Faculty of Physics, Astronomy and Informatics,
Nicolaus Copernicus University, Grudziadzka 5, 87-100 Toru\'n, Poland \\
}
\begin{document}
%
\date{Accepted .... Received ...; in original form ...}
\pagerange{\pageref{firstpage}--\pageref{lastpage}} \pubyear{2017}
\maketitle
\label{firstpage}
\begin{abstract}
We study the orbital architecture, physical characteristics of planets,  formation and long-term evolution of the Kepler-30 planetary system, detected and announced in 2012 by the \kepler{} team.  We show  that \corre{the Kepler-30 system} belongs to a particular class of very compact and quasi-resonant, yet long-term stable planetary systems. 
We re-analyse the light curves of the host star spanning Q1-Q17 quarters of the \kepler{} mission. A huge variability of the Transit Timing Variations (TTV) exceeding 2~days is induced by a massive Jovian planet located between two Neptune-like companions. The innermost pair is near to the 2:1 mean motion resonance (MMR), and the outermost pair is close to \corre{higher order MMRs, such as 17:7 and 7:3}. 
Our re-analysis of photometric data allows us to constrain, better than before, the orbital elements, planets' radii and masses, which are $9.2 \pm 0.1$, $536 \pm 5$, and $23.7 \pm 1.3$ Earth masses for Kepler-30b, Kepler-30c and Kepler-30d, respectively. The masses of the inner planets are determined within $\sim 1\%$ uncertainty. We infer the internal structures of the Kepler-30 planets and their bulk densities in a wide range from $ (0.19 \pm 0.01)$\,\gccm{} for Kepler-30d, $(0.96\pm0.15)$\,\gccm{} for Kepler-30b,  to $(1.71 \pm 0.13)$\,\gccm{} for the Jovian planet Kepler-30c.  
We attempt to explain the origin of this unique planetary system and a deviation of the orbits from exact MMRs through the planetary migration scenario. We anticipate that the Jupiter-like planet plays an important role in determining the present dynamical state of this system.
\end{abstract}
%
\section{Introduction}
%
Among transiting exoplanets detected by the \kepler{} mission, only a small fraction of \corre{multi-planet systems is characterized with the orbital elements and planets' masses determined with the uncertainties smaller than a few percent} \citep[e.g.,][]{JontofHutter2016}. The \corre{physical and dynamical proprieties for those few multi-planet systems} were mostly obtained with the Transit Timing Variation (TTV) method \citep{Agol2005,Holman2005}. When the TTV amplitude has a significant signal-to-noise (S/N) ratio, it is possible to uniquely constrain the masses and orbital elements. This is the case, for instance, of the Kepler-9  \citep{Holman2010} and of the \corre{Kepler-88} systems \citep{Nesvorny2013} for which a large full-amplitude of the TTVs reaching 24~hours ensures precise estimates of their dynamical and physical parameters. The Radial Velocity (RV) and  photometric follow-up observations of such systems make it possible to further refine their orbital parameters and masses \citep[e.g.,][]{Borsato2014,Barros2014}.

In this paper, we report on a new and up-to-date characterization of the Kepler-30 system composed of three planets. These planets have been validated by \cite{Fabrycky2012} on the basis of the very initial Q1-Q6 \kepler{} data quarters ($\simeq 600$~days) \corre{as well as by \cite{Tingley2011} on the basis of the first Q0-Q2 quarters.} The early studies on this system were focused on the orbital-spin alignment and star-spots detection \citep{SanchisOjeda2012} or on differential rotation of the parent star \citep{Lanza2014}.  Here, we take full advantage of the available  light curves of Kepler-30, spanning Q1-Q17 quarters ($\simeq 1600$~days) to constrain better the masses and orbital elements of the planets.  
 
We find the Kepler-30 system particularly interesting and unique in the \kepler{} sample for several reasons. One of them is the huge TTV full-amplitude of the innermost planet of $\simeq 2$~days (48 hours), which is \corre{two times larger} than the TTV full-amplitude of $\simeq$1~day for the innermost planet in the Kepler-88 system \citep{Nesvorny2013} dubbed as ``the king of the transit variations'' in their work\footnote{\corre{The enormous TTV amplitude in the Kepler-30 system present over the whole Q1-Q17 \kepler{} data quarters inspired us to call the Kepler-30 system as ``the Cesar of TTVs'' in the original manuscript title.}}. 
Curiously enough, the innermost planets' (Kepler-30b) TTV amplitude is so large that (apparently) it has been overlooked in the \cite{Rowe2015} TTVs catalogue -- as we suppose, due to the too narrow time-window for detecting transits.
\corre{In a more recent work by \cite{Holczer2016}, which we refer to in our preliminary analysis of the dynamical features of the system, the Kepler-30b has the largest TTV amplitude reported.}
\corre{The TTV full-amplitude of 2~days for Kepler-30b} has been also identified in the discovery paper by \cite{Fabrycky2012}, as well as by Howard Relles, an independent TTV researcher\footnote{\url{http://www.exoplanet-science.com/}}. 
\corre{\cite{Hadden2017} have studied the TTVs signals of the Kepler-30 system  determined from long cadence \kepler{} photometry in \cite{Rowe2015} with analytical approximations and direct $N$-body simulations. However, they considered only the two innermost planets in their analysis. Our new light-curve and TTV analysis updates and extends the results of previous studies.}

\corre{Another reason for a particular interest in the Kepler-30 system is its} orbital architecture. Unlike typical multiple-planet \kepler{} systems \citep{Weiss2017,Millholland2017}, the Kepler-30 planets have individually very different masses, i.e., a Super-Earth or mini-Neptune  (planet~b, or planet~1), a large Jovian planet (planet~c, or planet~2) and a Neptune-like planet (planet~d, or planet~3). 
\corre{The planets orbit around a Sun-mass star, roughly within the Mercury orbit in the Solar System, with the orbital periods from the shortest one of $\sim 29$~days, to the middle of $\sim 60$~days, and to the longest one of $\sim 143$~days, respectively.}
Moreover, the inner planets are close to two-body 2:1 \corre{and the outer ones to higher order} MMRs \corre{such as 7:3 and 17:7}.
\corre{The proximity to the 2:1 MMR between a relatively massive Jovian planet and a much smaller, mini-Neptune (or super-Earth) is the common feature present in both Kepler-30 and Kepler-88 systems.}
\corre{As we show further in this paper, the Kepler-30 system appears dynamically active, yet is non-resonant, and surprisingly, very stable.}
\corre{Even if the system appears as non-resonant, we attempt to explain its origin through the planetary migration scenario.}

Given significant discrepancy between the TTV measurements in \cite{Rowe2015} and \cite{Holczer2016}, in this work we re-analyse the available light curves of the Kepler-30 in the Mikulski Archive for Space Telescopes (MAST\footnote{\url{http://archive.stsci.edu/kepler}}) \citep[]{Padovani1998,Conti2011}. We attempt to verify the previous TTV measurements and to determine as accurately as possible mid-transit timings for all transits. Through dynamical modeling of the TTV signal \citep{Agol2005,Deck2015}, we aim to obtain the dynamical masses, and constrain the geometry of orbits, as well as investigate the dynamical architecture imposed by the likely initial conditions. 

This paper is structured as follows. In Section \ref{sec2:Newfit}, we re-analyse the Kepler-30 light curves to verify and update the TTV measurements in \cite{Holczer2016}. We also estimate the planets' radii, which agree with those given in \cite{SanchisOjeda2012}.  In particular, we find the Kepler-30b radius equal to $3.75~\RE$, i.e., much larger than $1.91~\RE$ reported in \cite{Rowe2015}. Our value of $3.75~\RE$ \corre{is in a good agreement with} the estimate in \cite{SanchisOjeda2012}, i.e., $3.9 \pm 0.2~\RE$. This radius implies that Kepler-30b is a Neptune-like planet rather than a super-Earth.

Optimization of the TTV model recalled in Sec.~\ref{sec2:Newfit} results in precision mass estimates of Kepler-30c and Kepler-30b which are, respectively, $20\%$ and $25\%$ smaller when compared with data in \cite{SanchisOjeda2012},  and in the determination of the eccentricity for Kepler-30b whose value is almost twice as large ($e_1=0.077$). \corre{The calculated masses for Kepler-30b and Kepler-30c agree well with the recent data in \cite{Hadden2017} though they omitted the outermost planet in their TTV models.}

The updated parameters\corre{, in the present work are used to investigate} the dynamical state of the system, which we discuss in Section \ref{sec3:maps}. One-- and two--dimensional dynamical maps unveil an intriguing web of two-body and three-body mean motion resonances, some of low-order, however the system is safely separated from them. In Sect.~\ref{sec4:radiusMass}, the internal structure of the planets and densities are deduced, thanks to the updated determinations of masses and radii. We find that Kepler-30b has a bulk density of $(0.96\pm0.15)$\,\gccm{}, Kepler-30c may be  a gaseous giant planet $70\%$ more massive than Jupiter, and Kepler-30d is one of the lowest density Neptune-size planets discovered up to now. Section \ref{sec5:migration} is devoted to the planetary migration simulations aiming to explain the formation of the observed orbital architecture. We present our conclusions \corre{in Section \ref{conclusion}.} 
\corre{Our TTV measurements, and the results of our independent light-curve analysis, as well as graphical material illustrating the MCMC sampling of four variants of the orbital TTV model are available on-line in Appendix.}
%
\section{Q1-Q17 quarters TTV data revisited}
\label{sec2:Newfit}
%
\subsection{The $N$-body model of the TTVs}

In order to interpret the TTV measurements, we follow a simplified dynamical model described in \cite{Gozdziewski2016} for the determination of the Keplerian osculating elements on the basis of the TTV data. In particular, we assume that the planetary system is coplanar with inclination $I_i = 90^{\circ}$ and nodal longitudes $\Omega_i = 0^{\circ}$, where $i = 1, 2, 3$ for subsequent planets. We use the Poincar\'e elements $x_i = e_i\cos\varpi_i$, $y_i = e_i\sin\varpi_i$ to avoid the geometric singularity of the periastron longitude for nearly-circular orbits.

To determine and select the best-fitting models, we use a hybrid optimization  described in \cite{Gozdziewski2016}. It is based on two steps. At first, with the help of  genetic and evolutionary algorithms \citep[GEA from hereafter,][]{Storn1977,Izzo2010}, we search for the maximum of the likelihood function
\begin{equation}
 \log {\cal L} =  
-\frac{1}{2} \sum_{i,t}
\frac{{\mbox{(O-C)}}_{i,t}^2}{\sigma_{i,t}^2}
- \frac{1}{2}\sum_{i,t} \log {\sigma_{i,t}^2} 
- \frac{1}{2} N_{obs.} \log{2\pi},
\label{eq:Lfun}
\end{equation}
where $(\mbox{O-C})_{i,t}$ is the deviation of the observed $t$-th transit moment of an $i$-th planet from its $N$-body ephemeris, and $N_{obs.}$ is the total number of TTV observations. In general, we modify the raw TTV uncertainties $\sigma_{i,t}^2 \rightarrow \sigma_{i,t}^2+\sigma_f^2$ with the so called error floor, $\sigma_f$. It makes it possible to account for, possibly, underestimated uncertainties of the measurements. Assuming that the uncertainties are Gaussian (normal), the ${\cal L}$-function is defined such that $\Chi \simeq 1$, i.e., close to a value expected for statistically valid model in such a case \citep[e.g.,][]{Baluev2008}. 

After this quasi-global optimization step, which helps to detect possibly non-unique solutions, we choose the best-fitting models in the sense of ${\cal L}$ (or $\Chi$). Next, to determine realistic parameter uncertainties, and to analyse the local topology of the parameter space, we apply the Bayesian approach \citep[e.g.,][]{Sivia2006}.  Most of the priors are selected as uniform in this work, and we only restrict the parameters within reasonable bounds. The posterior probability distribution is sampled with the Markov Chain Monte Carlo (MCMC) affine sampler \citep{Goodman2010},  developed in the \code{emcee} package by \cite{ForemanMackey2013}. We note that the first step (GEA) is very useful to reduce CPU overhead, since we initiate the MCMC sampling nearby the proper solutions.

We note here that  $(x,y) \equiv (x_i,y_i), i=1,2,3$ exhibit pairwise much weaker linear correlations for the specific Kepler-30 TTV measurements, unlike for the Kepler-60 \citep{Gozdziewski2016}  and Kepler-29 \citep{Migaszewski2017} systems, respectively. Therefore we decided to set both uniform and non-uniform (Gaussian) priors on $(x,y)$ or on the eccentricities indirectly. We set the priors with the zero mean and a small variance $\sigma_{x,y} \simeq 0.05$, which could be expected for compact \kepler{} systems evolved due to migration. (A detailed discussion is presented in Sect.~\ref{sec5:migration}). Yet, as we report further, we find that the $(x,y)$ Gaussian priors have negligible influence on the resulting best-fitting models.

\subsection{A preliminary modeling of the TTVs}

Using the dynamical Model~I (no $\sigma_{x,y}$ priors and no error floor $\sigma_f$), we first analysed the Kepler-30 TTVs data-set in the H16 catalogue. It consists of $77$ measurements spanning the Q1-Q17 quarters. We used, whenever possible, the short cadence timing data, derived on the most dense photometric sampling, and possibly precise determination of the TTVs and their uncertainties. The timing measurements display a very clear sinusoidal modulation for the two innermost planets, yet their full TTV amplitudes of $\simeq 48$ and $\simeq 1.5$~hours, respectively, differ by a factor more than 30; for the outermost one, with the longest orbital period of $\simeq 143$~days, the TTV signal is much less regular and spans $\simeq 2$~hours (see Fig.~\ref{fig:figure1}).

The best-fitting solutions to the H16 data-set  yield $\Chi \simeq 2.6$, which might indicate that the uncertainties are underestimated, indeed. In such a case the error floor would be $\simeq 3$~minutes. We also found a few measurements outlying from the best-fitting model prediction. The error floor model does not improve that solution, either.

Another possibility of statistically ``imperfect'' $\Chi$ and  measurements outlying from the model might be the presence of an additional, yet non-detected planet in the system. We first focused on that possibility, and we tried to improve the best-fitting model by extending it with such an unseen, fourth planetary companion.  We tested orbital configurations including the fourth hypothetical planet within wide ranges of the semi-major axis, $a_4 \in(0.1, 0.9)~\au$, eccentricity $e_4 \in[0,0.15)$, and mass $m_4\in(0.0001,100)~\mE$.  Although the four-planet model fits the TTV data-set slightly better, as it yields $\Chi \simeq 1.9$, such a value might be still considered statistically as too large. 

Moreover, all the best-fitting four-planet solutions appear dynamically unstable with the orbital crossing-time of just a few tens of years. We also confirmed the dynamical instability \corre{with the Mean Exponential Growth Factor of Nearby Orbits} \cite[MEGNO,][]{Cincotta2000,Gozdziewski2001}, which is a variant of the maximum Lyapunov Characteristic Exponent \cite[mLCE,][]{Benettin1980}.

%
\subsection{Q1-Q17 quarters light curves re-analysed}
%
%
Since our attempts to extend the planetary model to four planets failed, we
decided to re-compute the TTVs from  the most recent Q1-Q17 light curves of Kepler-30. A few outlying points and the relatively large $\Chi \sim 2.6$ for the best-fitting three-planet model are the primary concern.

We re-analyse the Pre-search Data Conditioning Simple Aperture Photometry (PDCSAP) light curves in the {MAST} archive. The PDCSAP light curves are automatically de-trended from instrumental noise and systematic artifacts by \kepler{} PDC algorithm \citep{Smith2012,Stumpe2012,Kinemuchi2012}. We then normalize the individual PDCSAP light curves, and we split each of them in fragments centered around a particular transit, with a width of $\pm 0.5$~days. Each single fragment is then re-normalized through a third order polynomial to de-trend the out-of-transit parts of the light curve from the effect of stellar activity and instrumental noise. 

As a second step, we folded the light curves for each planet and, to reduce the CPU overhead, we averaged the photometric data in 100 bins. \corre{To correctly fold the light curve,  we subtract the mid-transit time calculated with the \code{exonailer}\footnote{\url{https://github.com/nespinoza/exonailer}} package \citep{Espinoza2016} from \corre{each single, previously selected fragment}. In this way, we attempt to correctly determine each transit shape and duration}. 

We follow  \cite{Gajdos2017} to derive the TTVs and transit parameters and we applied the \code{exonailer} code for computations. The limb-darkening and light curve models are derived with the help of  the \code{batman} package \citep[]{Kreidberg2015}.

The average flux is equal to the median in each bin while the uncertainties are estimated through the median absolute deviation (MAD). In particular, for Gaussian errors, the standard deviation of each binned flux is equal to 1.486 times the MAD \citep{Kipping2011,Masuda2017}. 

Using the \code{exonailer} package, we fit the transit model to the folded and binned light curves to infer the planet-to-star radius ratio ($p$), the inclination ($I$) and the limb-darkening coefficients ($q_1,q_2$) while keeping the photometric noise $\sigma_w$ fixed, as determined with the off-transit fragments of the light curve. We investigate a possible influence of non-Gaussian photometric errors \citep{Rousseeuw1993}, yet we did not find any substantial or ``alarming'' deviations.

\corre{Fixing the limb-darkening coefficients to some theoretical value, inferred from the stellar spectral type, might underestimate the planet-to-star ratio size \citep{Espinoza2015}. Therefore we fitted the limb-darkening coefficients using the folded light-curve for each planet individually, to check for any discrepancies. The determined limb-darkening coefficients are the same for all the three folded light-curves within 1$\sigma$ uncertainties.  In this way, we could directly compare our results with the \citep{SanchisOjeda2012} paper. } 

Finally, we fit each mid-transit time for the three planets with the transit parameters ($p,q_1,q_2, I,\sigma_w)$ kept constant at their best-fitting values from the previous step. 

We note that all best-fitting parameters of the transit model and their uncertainties are derived with the MCMC analysis of each single fragment of the light curve. The same \code{emcee} package is used for the MCMC sampling, as for the dynamical TTV model. We applied uniform priors to determine the mid-transit times, and sampled the posterior with 256 \code{emcee} walkers for up to 28,000 iterations (see the next Section for more details).

\corre{The results of the light curve analysis as well as priors used in the MCMC sampling, mid-transit times, their uncertainties and measured TTVs are reported in Appendix}. (We present a detailed discussion of the new determined radii of the planets in Sect.~\ref{sec4:radiusMass}).

\subsection{Orbital three-planet model}
Having the updated TTV data, we performed GEA experiments, to search for the best-fitting solutions. We report here on the results for the following dynamical models derived from Eq.~\ref{eq:Lfun}: Model~I without the error floor \corre{and without Gaussian priors set for $(x,y)$}, Model~II with the error floor and without Gaussian priors for $(x,y)$, Model~III with both the error floor and Gaussian priors $\sigma_{x,y}$ set for $(x,y)$, and Model~IV without the error floor but with Gaussian priors \corre{for $(x,y)$  with $\sigma_{x,y}$}. 

We started from canonical Model~I. A few repeated GEA runs converged to one particular best-fitting solution, and we considered it as reference one. We then performed the MCMC sampling around this best-fitting model, to examine parameter correlations and to derive their uncertainties. The best-fitting solutions are similarly derived for three remaining variants of the dynamical model, i.e.,  regarding the presence of the Gaussian $(x,y)$ priors and the error floor. The best-fitting parameters are determined as the median values of MCMC samples, after removing first $\sim 10\%$ of ``\corre{burn-in}'' samples in each, particular MCMC run. 
\begin{figure*}
\centerline{
\vbox{
\hbox{
\includegraphics[width=0.33\textwidth]{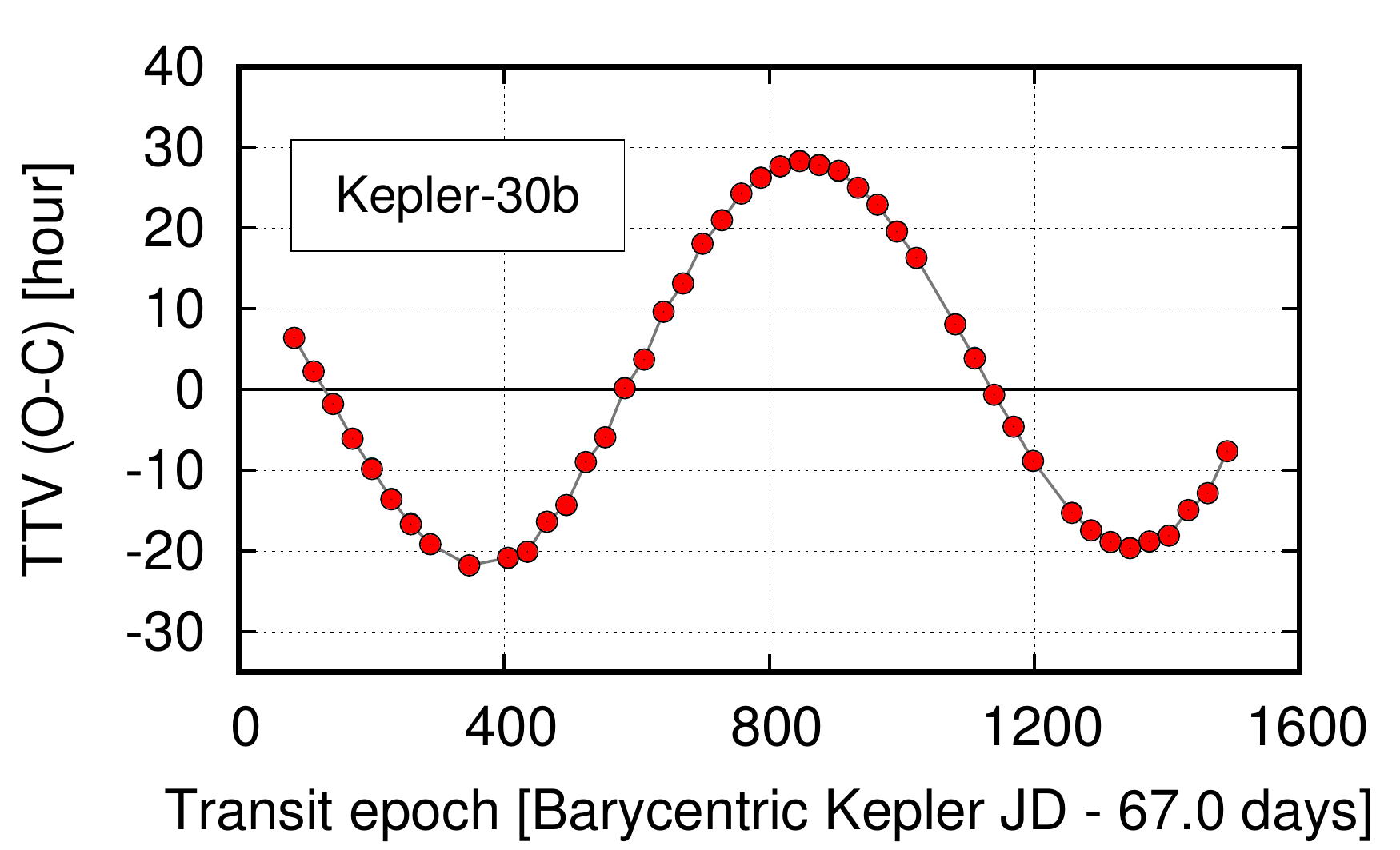}
\includegraphics[width=0.33\textwidth]{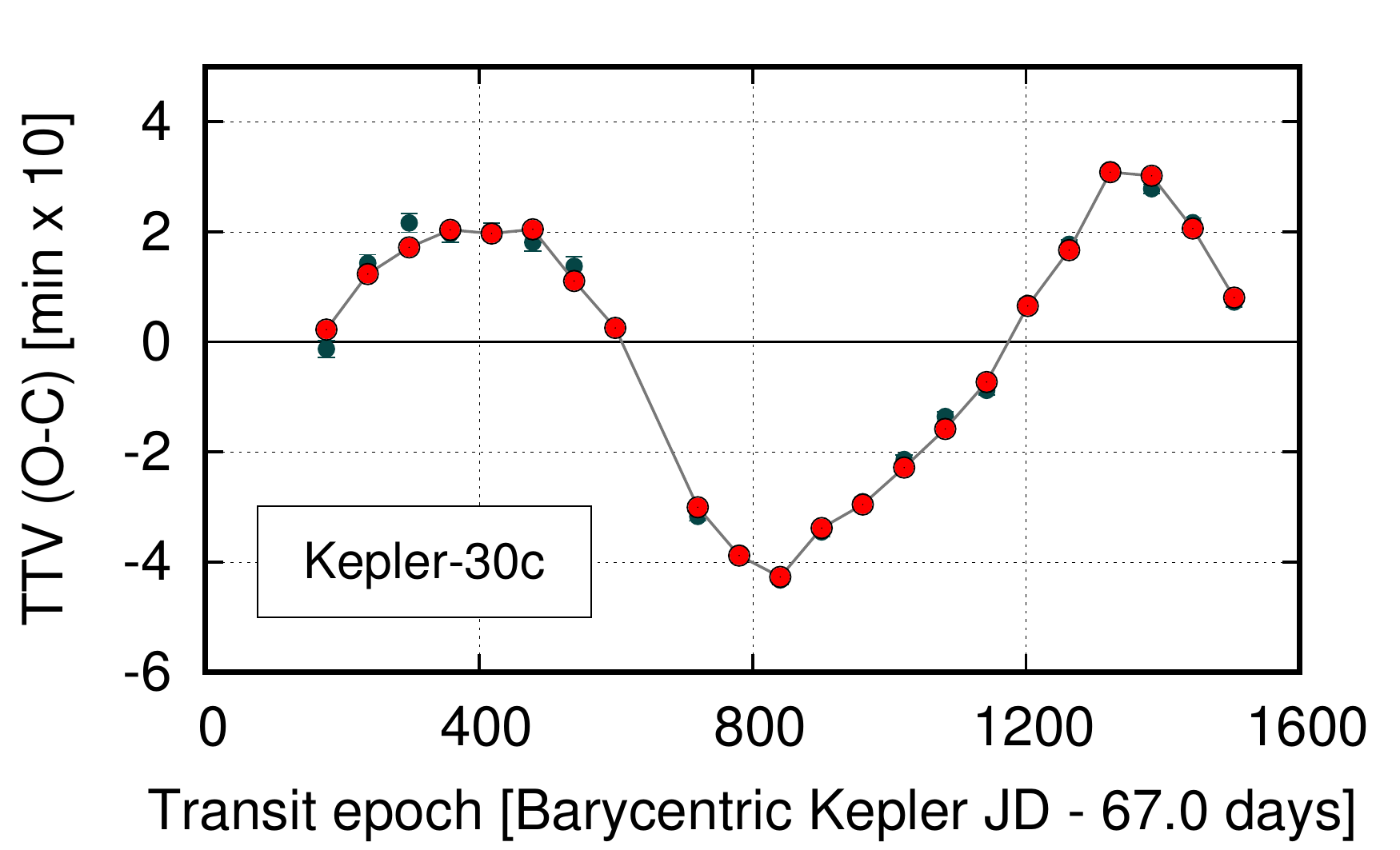}
\includegraphics[width=0.33\textwidth]{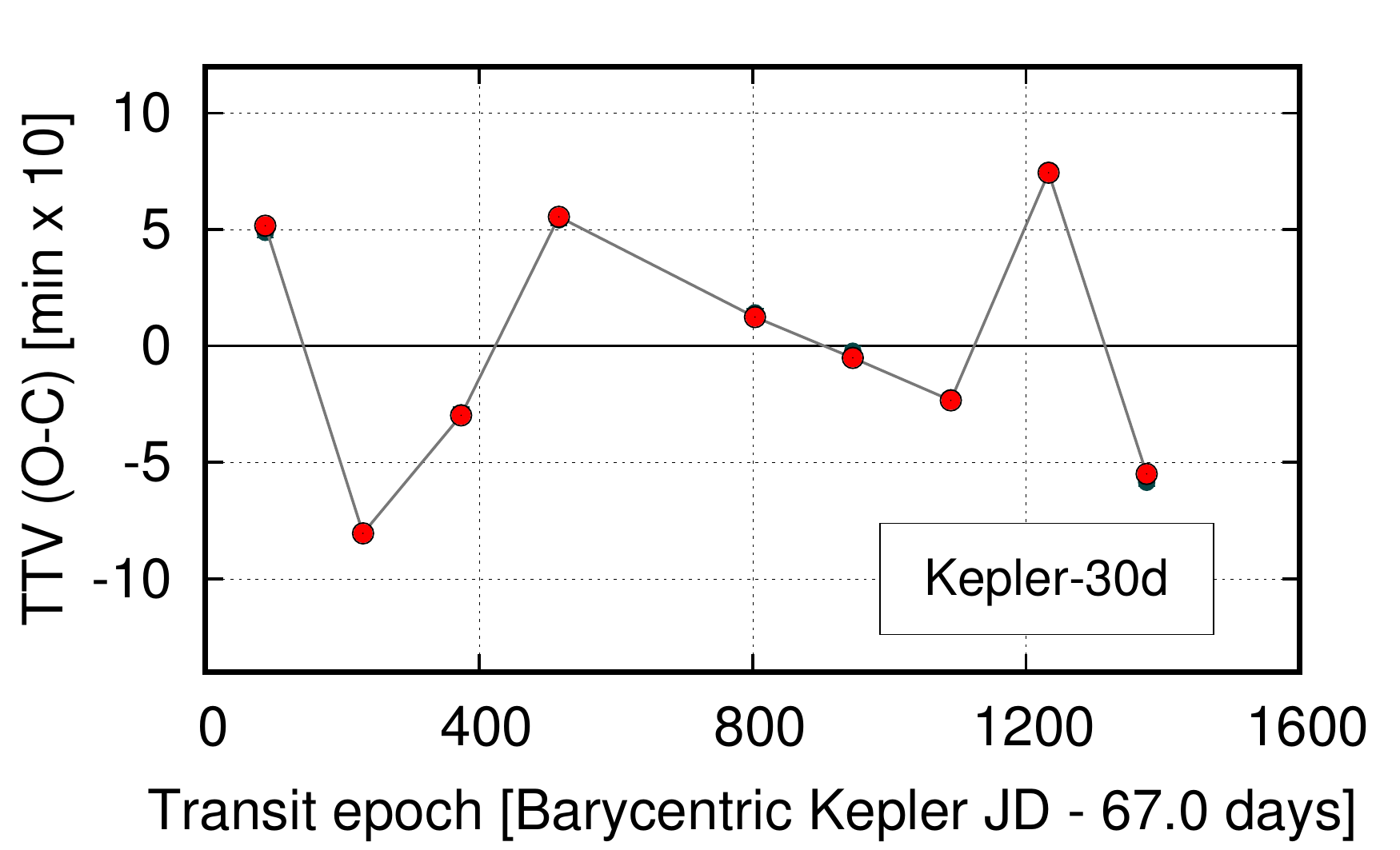}
}
\hbox{
\includegraphics[width=0.33\textwidth]{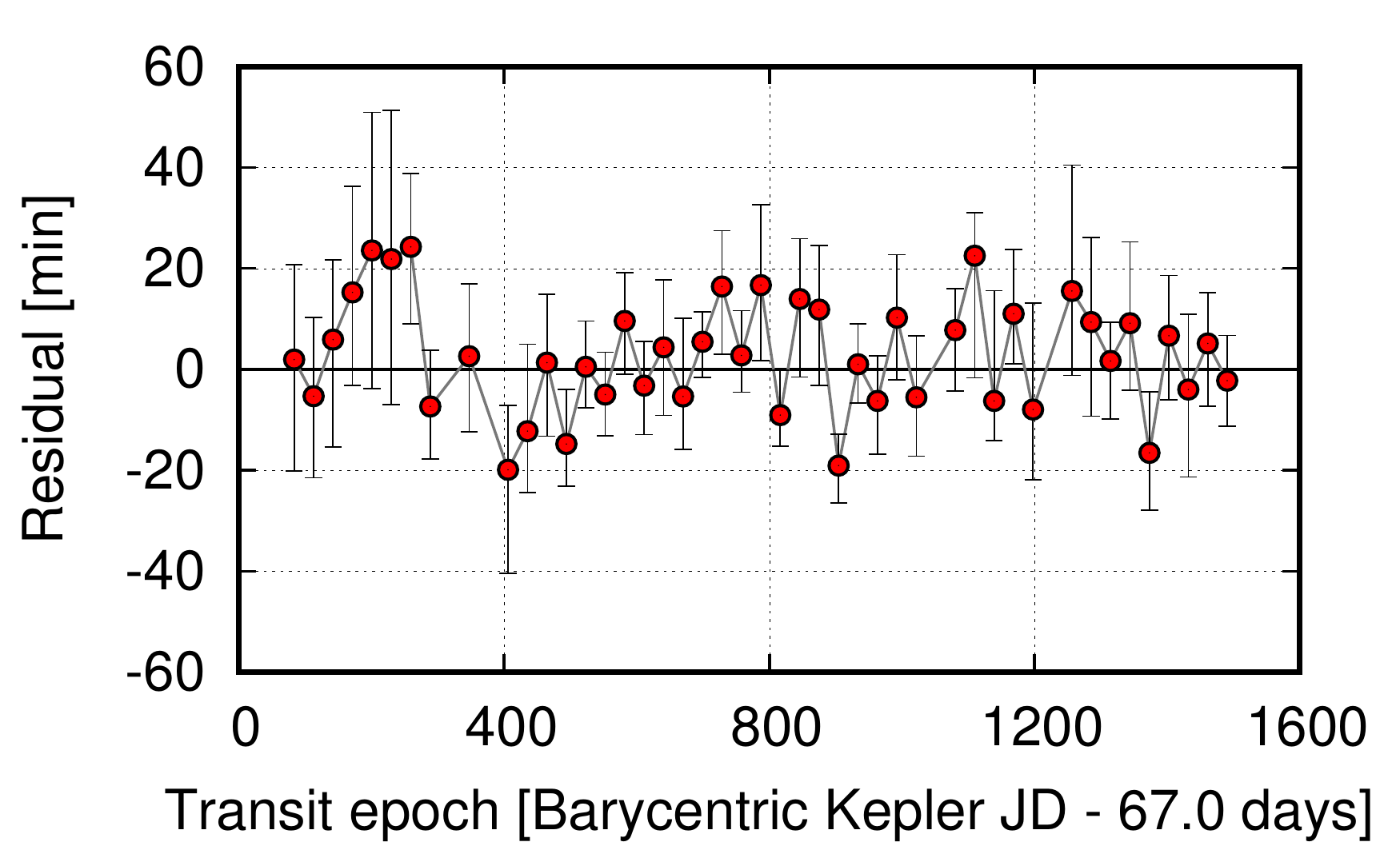}
\includegraphics[width=0.33\textwidth]{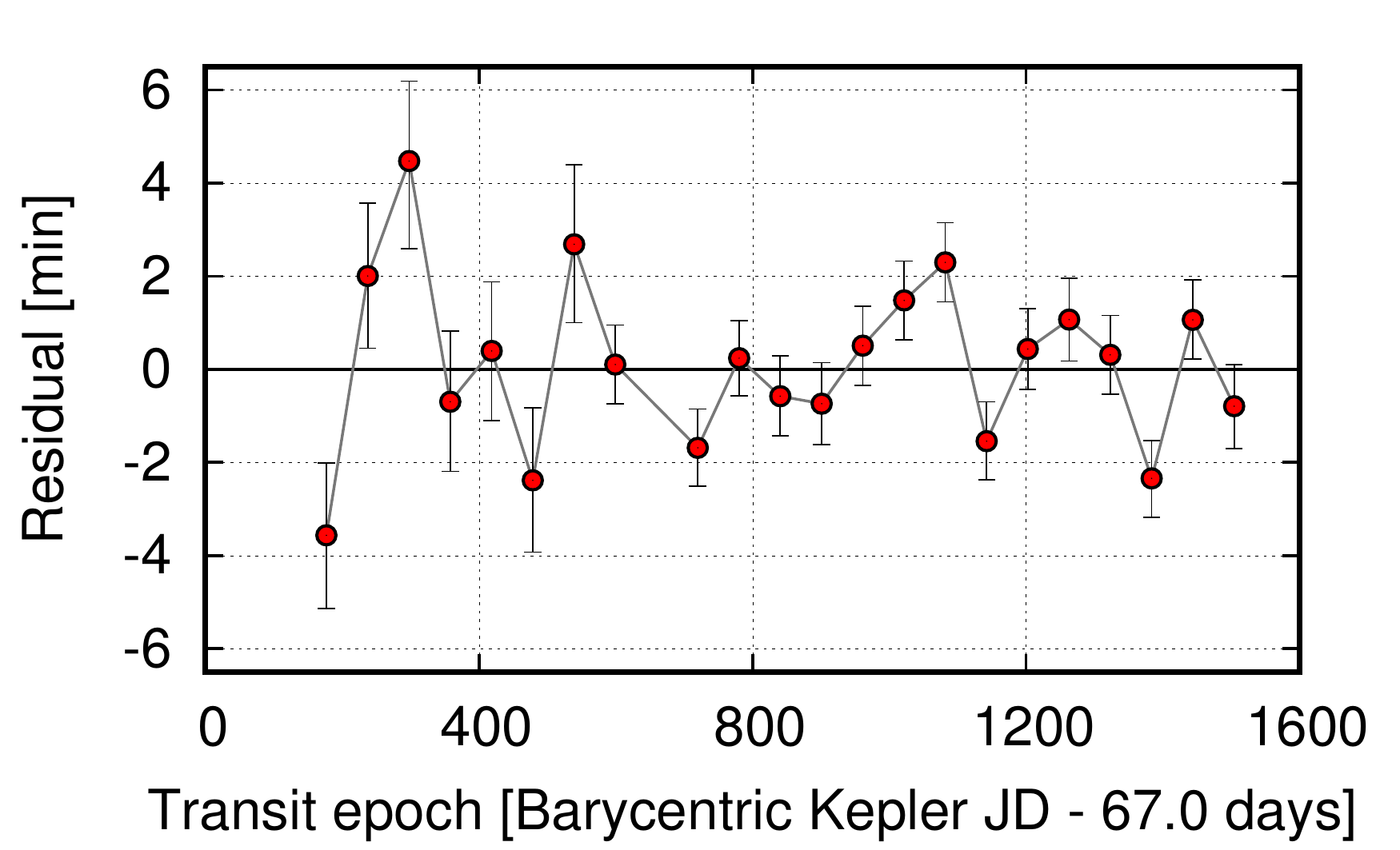}
\includegraphics[width=0.33\textwidth]{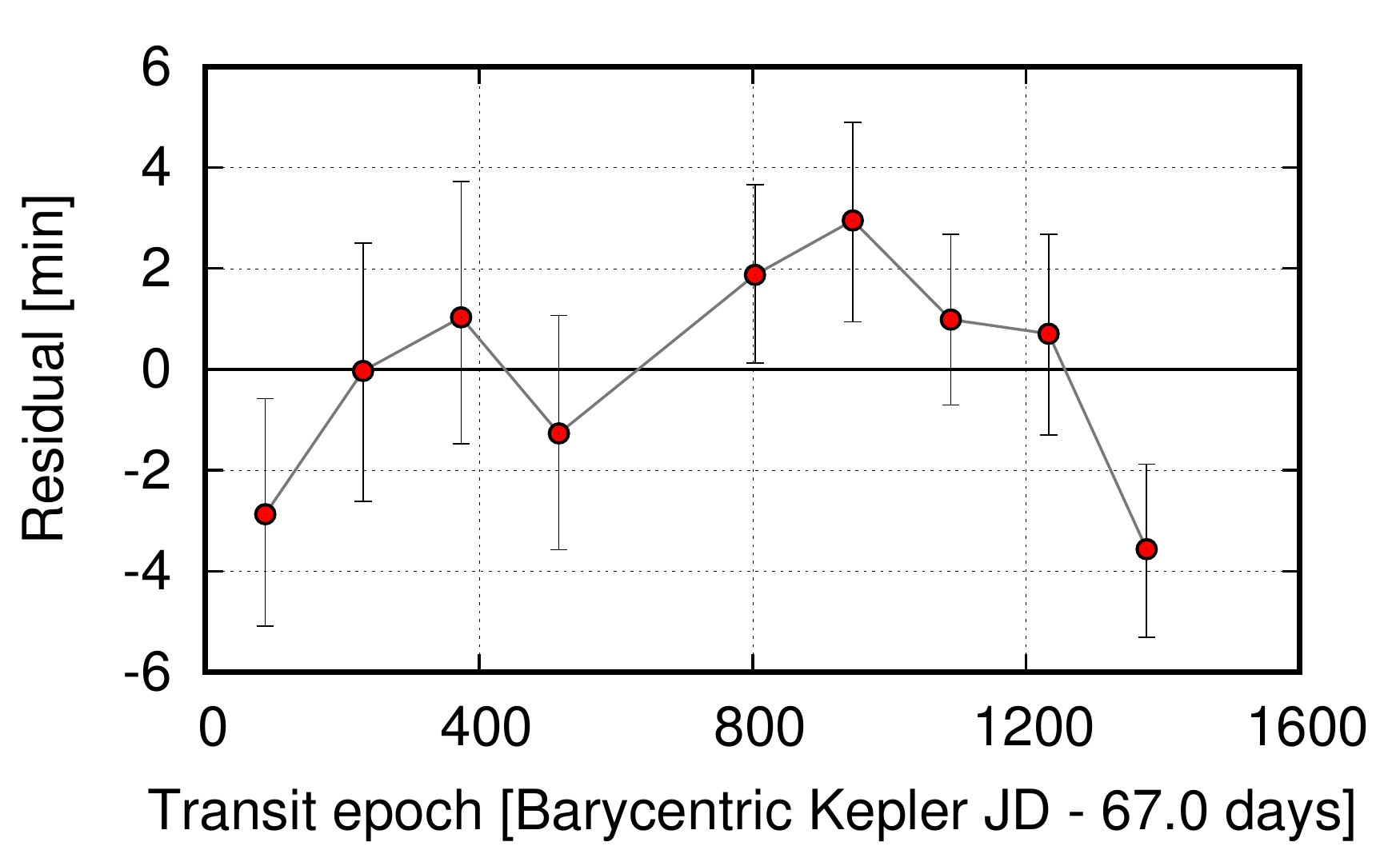}
}
}
}
\caption{
The best-fitting Model~I for Kepler-30 and its residuals, see Tab.~\ref{tab:MCMCb}. {\em Top-row}: synthetic TTV signals (red points connected with grey lines to guide eye) over-plotted on the TTV measurements (dark grey points with error-bars). See Tab.~\ref{tab:tab3} and Tab.~\ref{tab:tab4} for the TTV measurements. From the left to the right: data for Kepler-30b, Kepler-30c,  and Kepler-30d, respectively. The Keplerian osculating elements are also displayed in Tab.~\ref{tab:MCMCb}. {\em Bottom-row}: residuals of the best-fitting solution w.r.t. the observations. 
}
\label{fig:figure1}
\end{figure*}
%
%
To determine the uncertainties, we performed extensive MCMC sampling for various choices of the chain lengths (iterations), i.e., 128,000 up to 256,000 and for $140$ to $560$ \code{emcee} ``walkers'' (initial sets of parameters selected in a small hyper-ball around the best-fitting model derived by the GEA experiments). The chain lengths may be considered excessively large, once the acceptance fraction is $\simeq 0.3$. However we wanted to directly verify that autocorrelation is not significant and the posterior distribution does not change if the chain length increases. The results for 560 walkers and 192,000 samples are reported graphically in Appendix \corre{(on-line material)}. Besides canonical, fitted orbital parameters $P_{1,2}, x_{1,2}, y_{1,2}, T_{1,2}$ and masses $m_{1,2}$, Table~\ref{tab:MCMCb} displays osculating Keplerian elements inferred from the fitted parameters, as well as their uncertainties. \corre{The star mass and radius taken from \citep{Rowe2015} agree within $1\sigma$ uncertainties with more recent estimates in \cite{Johnson2017} as well as in \cite{SanchisOjeda2012}. }

We note that all these best-fitting solutions overlap both  with the initial GEA derived Model~I, as well as they overlap with themselves within the formal uncertainties. For a reference, the best-fitting solution for canonical  Model~I (all parameter priors set uniform, no error-floor $\sigma_f$) and the residuals are illustrated in Fig.~\ref{fig:figure1}. The residuals span a few minutes for the two outer planets, while of the order of 10~minutes for the innermost one. 
%
%
\begin{table*}
\caption{
The best-fitting parameters and their uncertainties derived for four variants of the TTV dynamical model through the MCMC sampling. The uncertainties are estimated as the 16th and 84th percentile of the samples around the median values. The inferred osculating orbital elements and their uncertainties are also displayed. The epoch is Barycentric Kepler Julian Day epoch (BJKD) $-$ 67~days. We adopted the mass of the host star as 0.986~$M_{\odot}$ \citep{Rowe2015}. See the text for more details.
}
\label{tab:MCMCb}
\begin{tabular}{l r r r}
\hline\hline
Planet & Kepler-30\,b & Kepler-30\,c & Kepler-30\,d  \\
\hline\hline
\multicolumn{4}{c}{{\bf Model~I}: without error floor $(\sigma_f)$ and without Gaussian priors for $(x,y)$  } \\ 
\hline
 $P\,$[d] & 29.2187$\pm$0.0009 & 60.32503$\pm$0.00010 &142.642$\pm$0.006 \\
 $e \cos\varpi$ & -0.05860$\pm$0.00024 & -0.0085$\pm$0.0005 & 0.0258$\pm$0.0011 \\
 $e \sin\varpi$ & -0.04988$\pm$0.00028 & 0.0077$\pm$0.0005 & 0.0087$\pm$0.0012 \\
$T\,$[d] & 83.709$\pm$0.004 & 116.5490$\pm$0.0004 & 87.2610$\pm$0.0011\\
$m_p\,[\mE]$  &9.2$\pm$0.1 & 536$\pm$5 &23.7$\pm$1.3 \\
\hline
$a\,[\au]$  & 0.18479$\pm$0.000004 & 0.29977$\pm$0.000001 & 0.53178$\pm$0.00001\\
$e$  & 0.0770$\pm$0.0003 & 0.0115$\pm$0.0005 & 0.0272$\pm$0.0024 \\
$\omega\,$[deg] & -139.6$\pm$0.2& 137.8$\pm$2.4 & 18.6$\pm$2.3 \\
${\cal M}\,$[deg] & -2.6$\pm$0.2 & -86.9$\pm$2.4  & 236.1$\pm$2.3 \\
\hline\hline
\multicolumn{4}{c}{{\bf Model~II}: with error floor $(\sigma_f)$ and without Gaussian priors for $(x,y)$  } \\ 
\hline
 $P\,$[d] & 29.2194$\pm$0.0010 & 60.32497$\pm$0.00014 &142.645$\pm$0.006 \\
 $e \cos\varpi$ & -0.05858$\pm$-0.00029 & -0.0084$\pm$0.0006 & 0.0258$\pm$0.0014 \\
 $e \sin\varpi$ & -0.0496$\pm$0.0003 & 0.0085$\pm$0.0007 & 0.0103$\pm$0.0016 \\
$T\,$[d] & 83.710$\pm$0.004 & 116.5490$\pm$0.0008 & 87.2614$\pm$0.0015\\
$m_p\,[\mE]$  &9.2$\pm$0.2 & 536$\pm$5 &23.3$\pm$3.0 \\
$\sigma_f\,$[d] & \multicolumn{3}{c}{0.0015$\pm$0.0004} \\
\hline
$a\,[\au]$  & 0.184790$\pm$0.000003 & 0.299774$\pm$0.000002 & 0.53179$\pm$0.00002\\
$e$  & 0.0768$\pm$0.0003 & 0.0120$\pm$0.0006 & 0.0278$\pm$0.0030 \\
$\omega\,$[deg] & -139.7$\pm$0.2 & 134.6$\pm$3.0 & 21.8$\pm$2.9 \\
${\cal M}\,$[deg] & -2.4$\pm$0.2 & -83.7$\pm$3.0  & 232.9$\pm$2.8 \\
\hline\hline
\multicolumn{4}{c}{{\bf Model~III}: with error floor $(\sigma_f)$ and with Gaussian priors for $(x,y)$  } \\ 
\hline
$P\,$[d] & 29.2194$\pm$0.0010 & 60.32497$\pm$0.00014 & 142.645$\pm$0.006 \\
$e \cos\varpi$ & -0.05858$\pm$0.00029 & -0.0084$\pm$0.0006 & 0.0258$\pm$0.0014 \\
$e \sin\varpi$ & -0.0496$\pm$0.0003 & 0.0085$\pm$0.0007 & 0.0103$\pm$0.0016 \\
$T\,$[d] & 83.710$\pm$0.004 & 116.5490$\pm$0.0008 & 87.2614$\pm$0.0015\\
$m_p\,[\mE]$  &9.2$\pm$0.2 & 536$\pm$5 &23.3$\pm$3.0 \\
$\sigma_f\,$[d] & \multicolumn{3}{c}{0.0015$\pm$0.0004} \\
\hline
$a\,[\au]$  & 0.184790$\pm$0.000004 & 0.299774$\pm$0.000002 & 0.53179$\pm$0.00002\\
$e$  & 0.0768$\pm$0.0003 & 0.0120$\pm$0.0006 & 0.0278$\pm$0.0030 \\
$\omega\,$[deg] & -139.7$\pm$0.2& 134.6$\pm$3.0 & 21.8$\pm$2.9 \\
${\cal M}\,$[deg] & -2.4$\pm$0.2 & -83.7$\pm$3.0 & 232.9$\pm$2.8 \\
\hline\hline
\multicolumn{4}{c}{{\bf Model~IV}: without error floor $(\sigma_f)$ and with Gaussian priors for $(x,y)$  } \\ 
\hline
$P\,$[d] & 29.2187$\pm$0.0009 & 60.32503$\pm$0.00010 &142.642$\pm$0.006 \\
$e \cos\varpi$ & -0.05860$\pm$0.00024 & -0.0085$\pm$0.0005 & 0.0258$\pm$0.0011 \\
$e \sin\varpi$ & -0.04988$\pm$0.00028 & 0.0077$\pm$0.0005 & 0.0087$\pm$0.0012 \\
$T\,$[d] & 83.709$\pm$0.004 & 116.5490$\pm$0.0004 & 87.2610$\pm$0.0011\\
$m_p\,[\mE]$  &9.2$\pm$0.1 & 536$\pm$5 &23.6$\pm$1.3 \\
\hline
$a\,[\au]$  & 0.184787$\pm$0.000004 & 0.299774$\pm$0.000001 & 0.53178$\pm$0.00001\\
$e$  & 0.0770$\pm$0.0003 & 0.0115$\pm$0.0005 & 0.0271$\pm$0.0024 \\
$\omega\,$[deg] & -139.6$\pm$0.2& 137.8$\pm$2.4 & 18.6$\pm$2.3 \\
${\cal M}\,$[deg] & -2.4$\pm$0.2 & -83.7$\pm$2.9 & 232.9$\pm$2.8 \\
\hline\hline
\end{tabular}
\end{table*}
We find that the $(x,y)$ parameters (eccentricities) are  constrained within of $0.01$ range in Models~I and II. The largest osculating eccentricity $e_1 = 0.077$ appears for Kepler-30b. The posterior histograms in Appendix (on-line material), see Figs.~\ref{fig:figureA1}--\ref{fig:figureA5}, \ref{fig:figureB1}--\ref{fig:figureB5}, \ref{fig:figureC1}--\ref{fig:figureC5}, and \ref{fig:figureD1}--\ref{fig:figureD5}, show that all eccentricities are meaningfully shifted from circular orbits, which may be consistent with the results for the \kepler{} sample of multiple systems \cite[i.e.,][]{Mills2016,MacDonald2016,JontofHutter2016}. \corre{However, the Kepler-30 system exhibits an interesting architecture of three planets with very different planetary masses, including a large jovian planet which might induce strong perturbations on orbits of the two much smaller companions}. Therefore, to avoid statistical biases on the eccentricities, we tested Models~I and II, in which we did not impose Gaussian priors, which are usually required to constrain the eccentricities in physical ranges and to not ``overfit'' the TTV data \citep{Migaszewski2017}. Indeed, as may be seen in Figs.~\ref{fig:figureA4}, \ref{fig:figureB4}, \ref{fig:figureC4} and \ref{fig:figureD4}, although $(x_i,x_j)$ and $(y_i,y_j)$, $i,j=1,2,3$ exhibit linear correlations, they are not so vivid as for the Kepler-29 and Kepler-60 systems.  We also verified this feature with imposing Gaussian priors on all $(x,y)$ with the mean of zero and the standard deviation of $\sigma_{x,y}=0.05$ (Models~III and~IV), and we obtained similar results. 

The lack of strong linear correlations between the Poincar\'e parameters is one more, somehow unique feature of the Kepler-30 system. It may be explained by very strong gravitational interactions induced by the Jovian planet and relatively ``clean'' TTV signals. We also note that we further interpret the derived eccentricities through the migration mechanism and its likely outcomes are consistent with the observations (Sect.~\ref{sec5:migration}).

Regarding the error floor $\sigma_f$ for all TTV measurements, we find  $\sigma_f \simeq 2$~minutes for Models~II and~III. However, it does not influence the best-fitting parameters beyond self-consistent uncertainties. In the Kepler-30 case, it may be questionable to use one and the same $\sigma_f$ for all planets, since their transits exhibit different shapes, and the timing precision is not uniform. Therefore, the final parameters are derived from the dynamical model without the $\sigma_f$ correction, which also reduces the number of the degrees of freedom. We recall this canonical model as Model~I (see Tab.~\ref{tab:MCMCb}). 

\subsection{A comparison of the H16 and new TTVs}
We find that the re-analysed TTV data-set fits statistically better the three-planet model than the H16 data-set. Our best-fitting model of Kepler-30 yields relatively small $\Chi \simeq 1.6$. The uncertainties of measurements in the new TTV data-set (see the on-line material) are larger by a factor of $2$ with respect to data in H16.

\begin{figure*}
\centerline{
\vbox{
\hbox{
\includegraphics[width=0.33\textwidth]{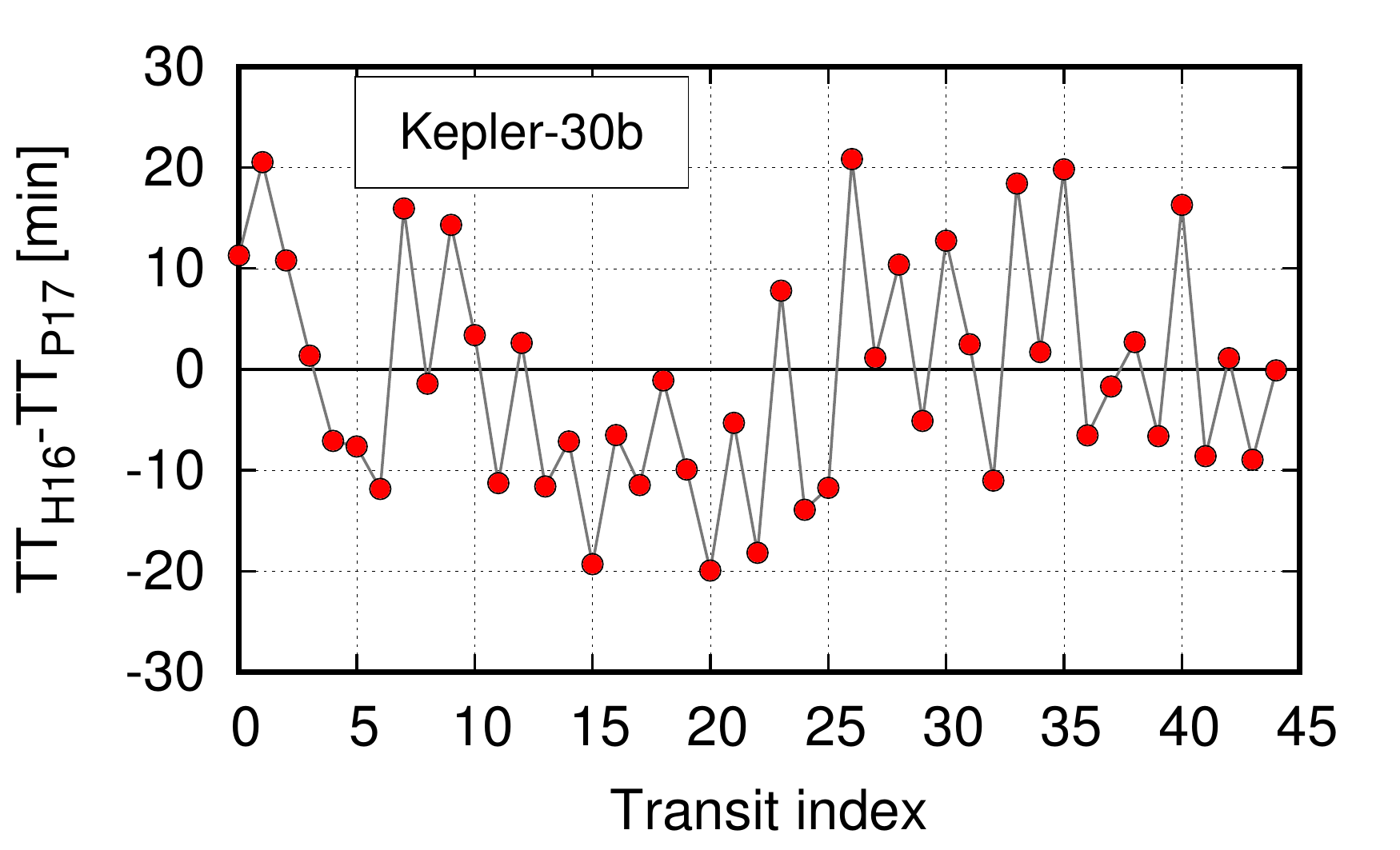}
\includegraphics[width=0.33\textwidth]{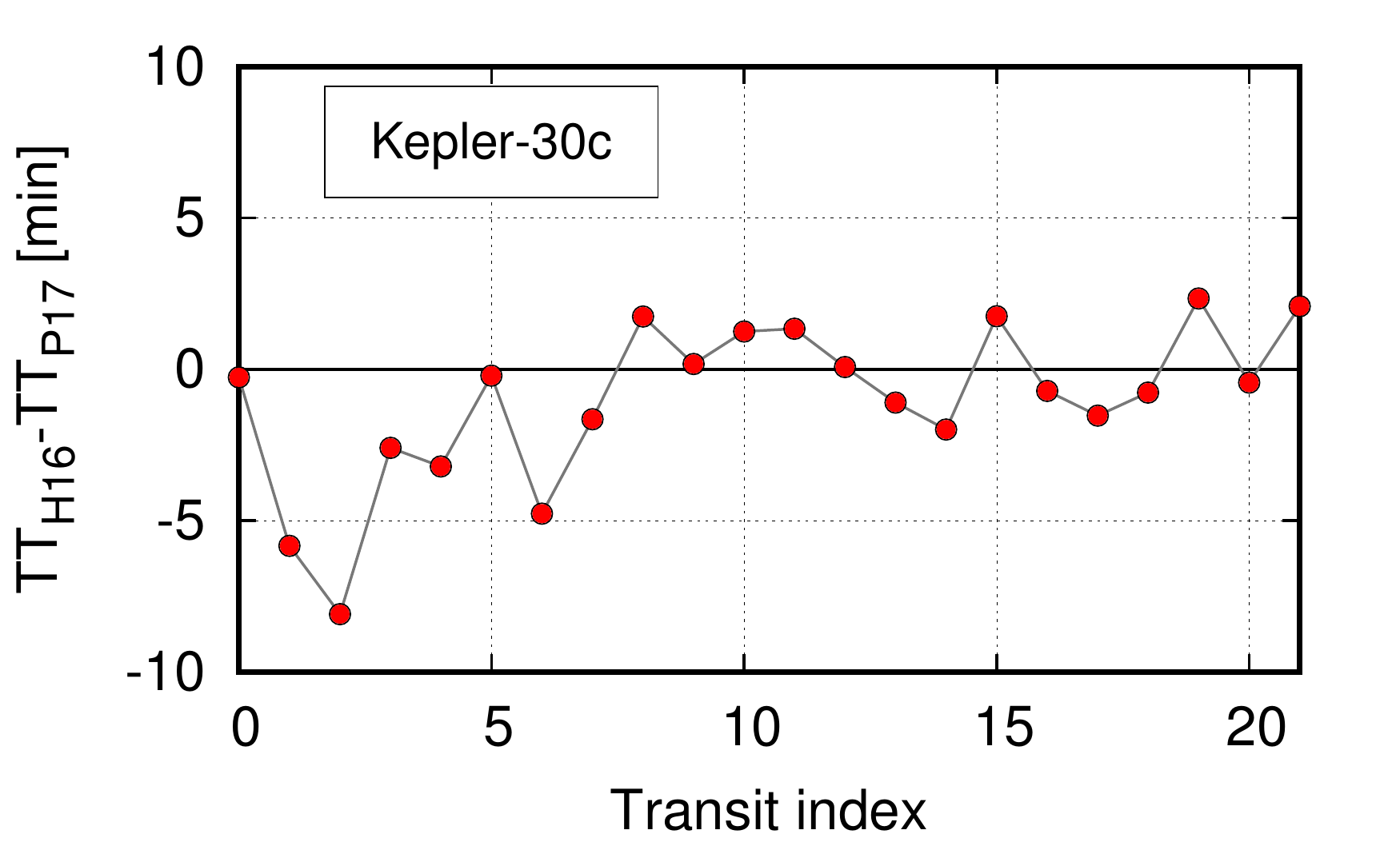}
\includegraphics[width=0.33\textwidth]{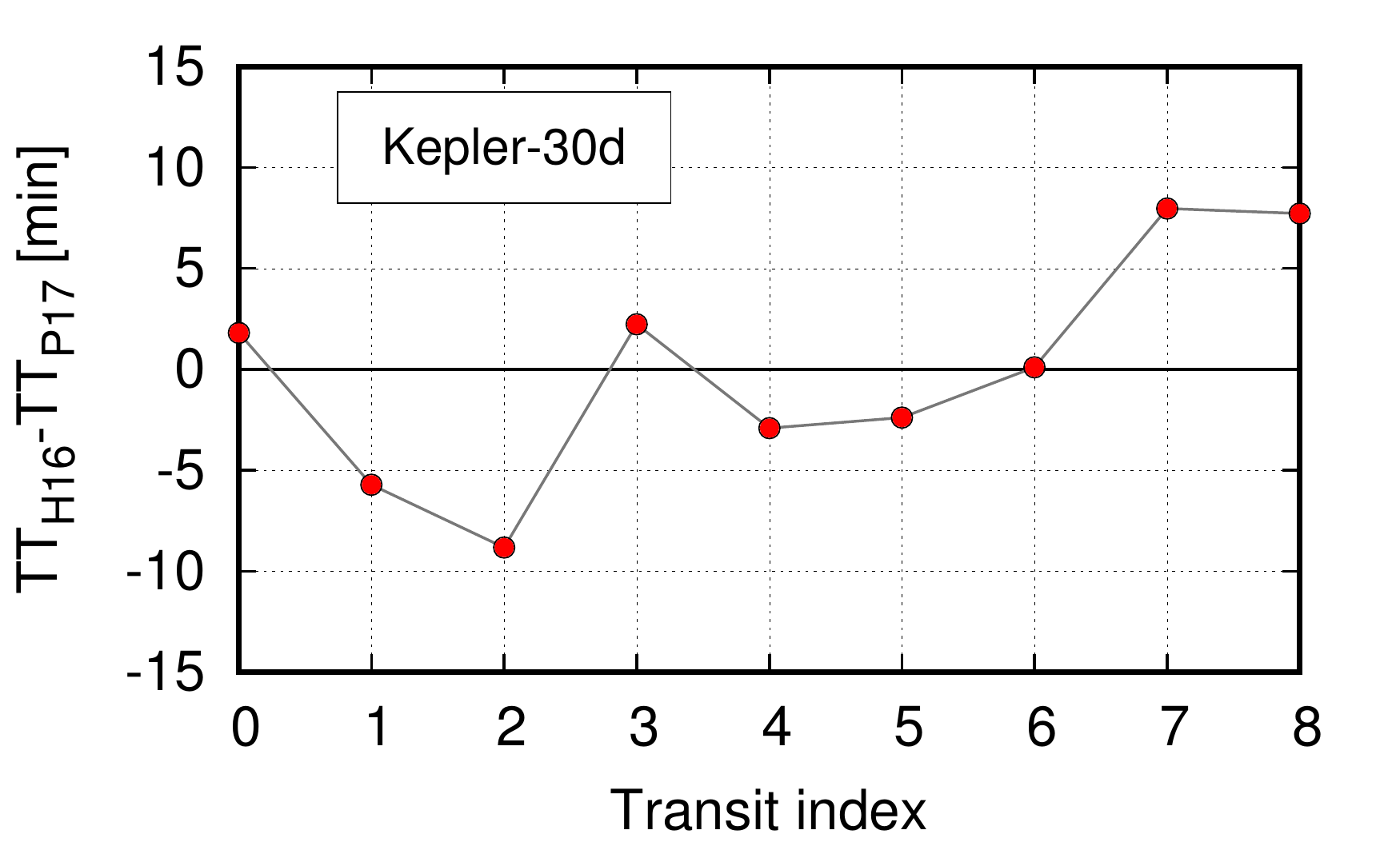}
}
}
}
\caption{
Differences of the mid-transit times between the Holczer et al. catalogue H16 ({\rm TT}$_{\rm H16}$) and this work ({\rm TT}$_{\rm P17}$) for the three-planet model of Kepler-30. In both cases each mid-transit time includes the relative transit timing variation.
}
\label{fig:figure2}
\end{figure*}

We also find that \corre{at the location of} the 33rd transit for Kepler-30b listed in H16 \corre{the PDCSAP light curve does not contain any flux at all. Therefore we could not calculate nor include the related TTV datum in our set of TTV measurements.}
\corre{A comparison of the mid-transit times in H16 and in our work is presented in Figure~\ref{fig:figure2}. For Kepler-30b, the maximum difference between the two TTVs sets is of the order of $\sim 20$ minutes. For Kepler-30c and Kepler-30d, the maximum difference is of the order of $\sim 8$ minutes. We note that  while the maximum difference might be quite large, the average difference is less than one minute for all three planets.}
%
\section{Dynamical setup of the Kepler-30 system}
\label{sec3:maps}
%
In this section we study the dynamical evolution and the stability of the best-fitting configurations as well as the structure of the phase space. We use the direct $N$-body integrations and two fast indicators, called MEGNO (mentioned earlier) and a new one, called the Reversibility Error Method (REM). We recall here that MEGNO \citep{Cincotta2000,Gozdziewski2001} saturates to a well defined value of $\Y \simeq 2$ for the Lyapunov-exponent stable orbits, and diverges linearly for chaotic orbits. Here, we computed MEGNO with the symplectic, fourth-order integrator scheme SABA$_4$ \citep{Laskar2001} and the symplectic tangent map \citep{Mikkola1999,Gozdziewski2008}.  The REM fast indicator \citep{Faranda2012,Panichi2016} \corre{is the maximum Lyapunov exponent-like, CPU efficient fast indicator optimised to compute high-resolution dynamical maps for low-eccentric \kepler{} systems}. The REM is based on the loose of time-reversibility propriety of chaotic orbits in conservative Hamiltonian systems. We applied a REM variant that makes use of an accurate Keplerian-drift \citep{Wisdom2016} and symplectic correctors \citep{Wisdom2006}. The integrator time-step of $1.0$~day is used in all dynamical maps.

\begin{figure}
    \centering
\vbox{
   \hbox{\includegraphics[width=0.47\textwidth]{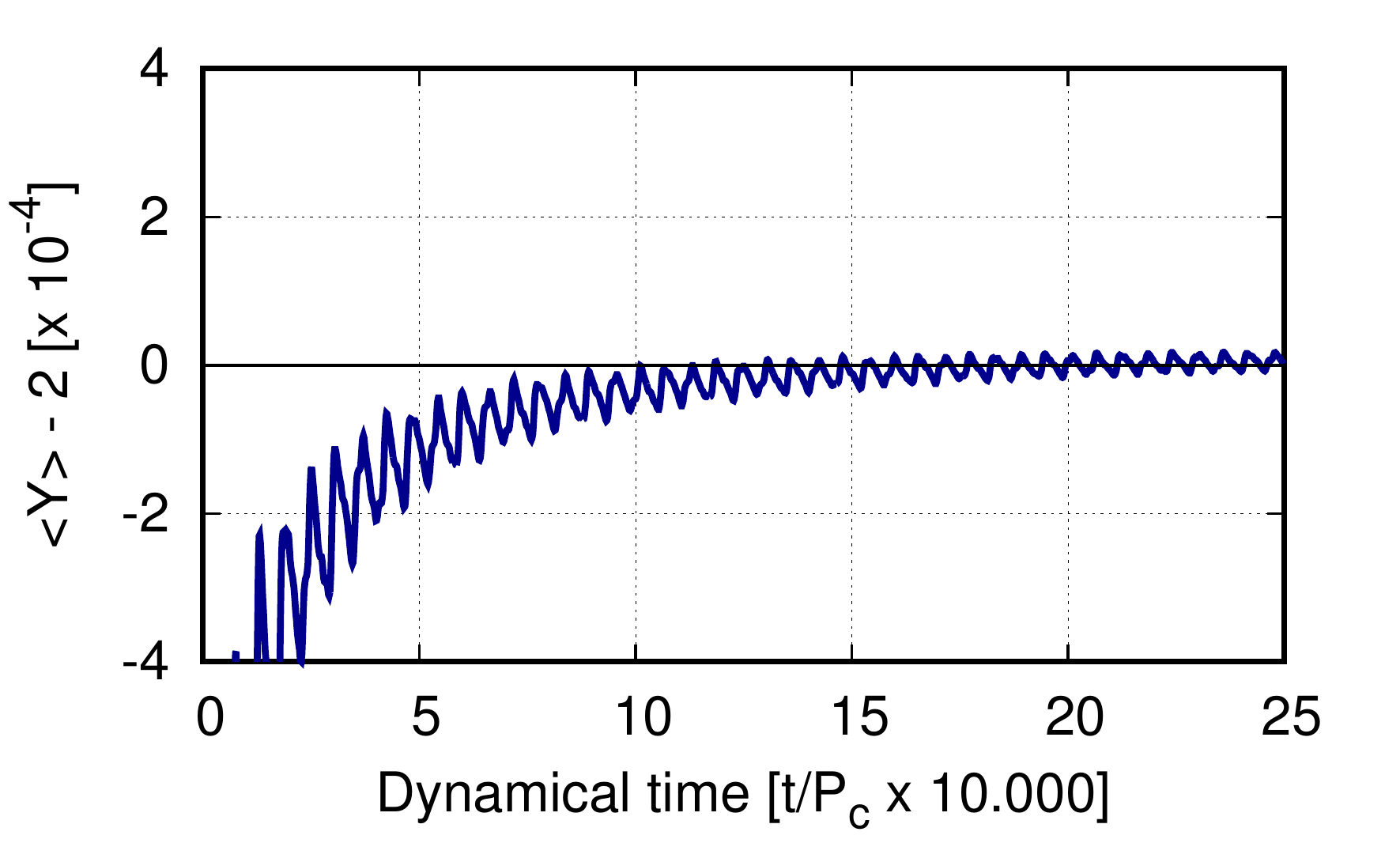}}
}    
\caption{
Temporal MEGNO evolution for the Kepler-30 system (Model~I in Tab.~\ref{tab:MCMCb}). 
The fourth order symplectic scheme SABA$_4$ was used in this run.
The integration time is equal to $100$~kyrs ($2.5\times10^5$ outermost orbital periods), the time-step is equal to $0.5$~days. 
}
\label{fig:figure3}
\end{figure}

In Figure~\ref{fig:figure3} we present an example of the temporal MEGNO evolution for the initial Keplerian osculating elements in Tab.~\ref{tab:MCMCb} (Model~I). Despite low-amplitude initial oscillations of $\Y$, the indicator smoothly converges $\Y \simeq 2$ after a few kyrs, indicating a stable, quasi-periodic configuration. We note that  the integration interval of $\sim 2\times 10^5$ outermost periods is long enough to account for gravitational interactions in low- and higher-order MMRs, and is also sufficient to detect the fine structures of the phase space.
Osculating elements derived from solutions selected at the reference epoch within $\pm~1\sigma$ of the canonical best-fitting Model~I in Table~\ref{tab:MCMCb} exhibit $\Delta\omega_{\rm 1,2}\simeq 80^{\circ}$ and $\Delta\omega_{\rm 2,3}\simeq 125^{\circ}$  with a small scatter. Regarding that, the Kepler-30 configuration is exceptional among other well studied \kepler{} systems having aligned and librating apses \citep{JontofHutter2016,Migaszewski2017}. For the Kepler-30 system, the direct $N$-body integrations reveal circulating $\Delta\omega$ angles.  For the inner pair of planets, $\Delta\omega$ circulates in the prograde direction with a period  $P\left(\Delta\omega_{\rm 1,2}\right)\simeq32$~yr,  while for the outer pair it circulates in the retrograde direction with a period  $P\left(\Delta\omega_{\rm 2,3}\right)\simeq500$ yrs. We also compute $\Delta\omega_{\rm 1,3}$ between the apsides  of the innermost and outermost planets, which circulates similarly to the innermost pair, with a period of $P\left(\Delta\omega_{\rm 1,3}\right)\simeq33$ yrs. 

We determined the osculating period ratios of adjacent pairs of planets showing that the innermost pair  (Kepler-30b and Kepler-30c) is located close to the 2:1~MMR border. The osculating orbital period ratio is $\simeq 2.06$. For the outer pair it is  $2.36$.  The outer pair may be also found nearby to higher order two-body mean motion resonances, such as 7:3~MMR and 17:7~MMR. All two-body MMR critical angles circulate for the best-fitting configuration. 

\begin{figure}
\centerline{
\vbox{
\hbox{\includegraphics[width=0.47\textwidth]{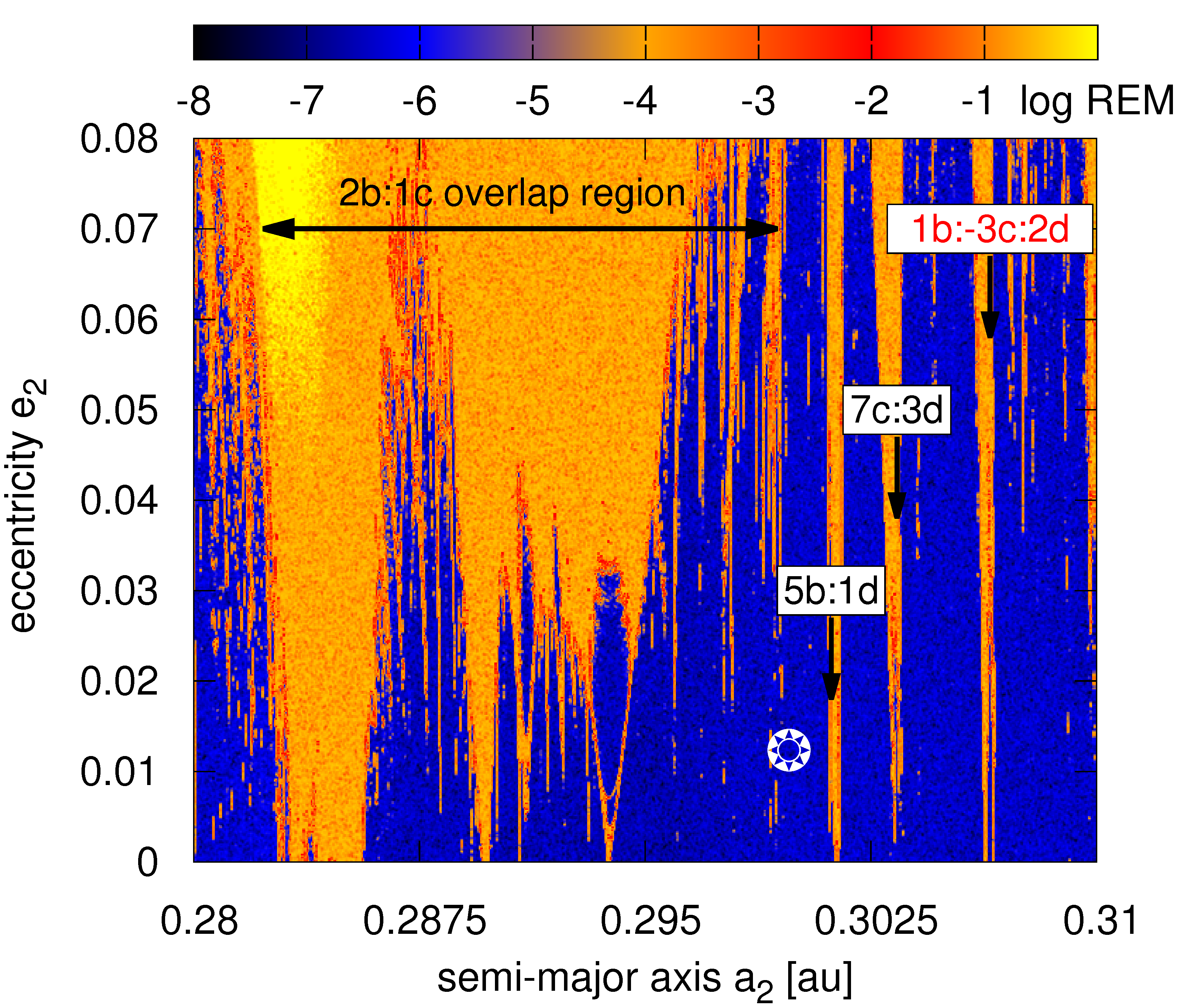}}
}
}
\caption{
Two--dimensional REM dynamical map for the best-fitting Kepler-30 Model~I in the $(a_{\rm 2},e_{\rm 2})$-plane. A white star symbol denotes the best-fitting solution in Table~\ref{tab:MCMCb}. The most relevant MMRs are labeled with black arrows and osculating period ratios. The total integration time is $72$~kyrs, the SABA$_4$ integrator time-step is equal to $1.0$~day. The grid has the resolution of $640\times480$~data points.
}
\label{fig:figure4}
\end{figure}

The structure of the phase space in the neighborhood of the best-fitting model may be visualized in a selected (semi-major axis,eccentricity)--plane. Such two--dimensional representation of the phase space unveils a complex structure of stable and unstable regions, emerging due to overlapping of two-body and three-body MMRs. Fig.~\ref{fig:figure4} shows the REM dynamical map in the $(a_{\rm 2},e_{\rm 2})$-plane. The most prominent two-body MMRs are labeled. The best-fitting system, marked with the star symbol, is located between them, and clearly lies outside the 2:1~MMR. It confirms that it is dynamically non-resonant. Yet the best-fitting system lies safely in a  stable zone, significantly separated from all low-order MMRs labeled in the $(a_{\rm 2},e_{\rm 2})$-scan. 

\begin{figure}
\centering
\vbox{
\hbox{\includegraphics[width=0.47\textwidth]{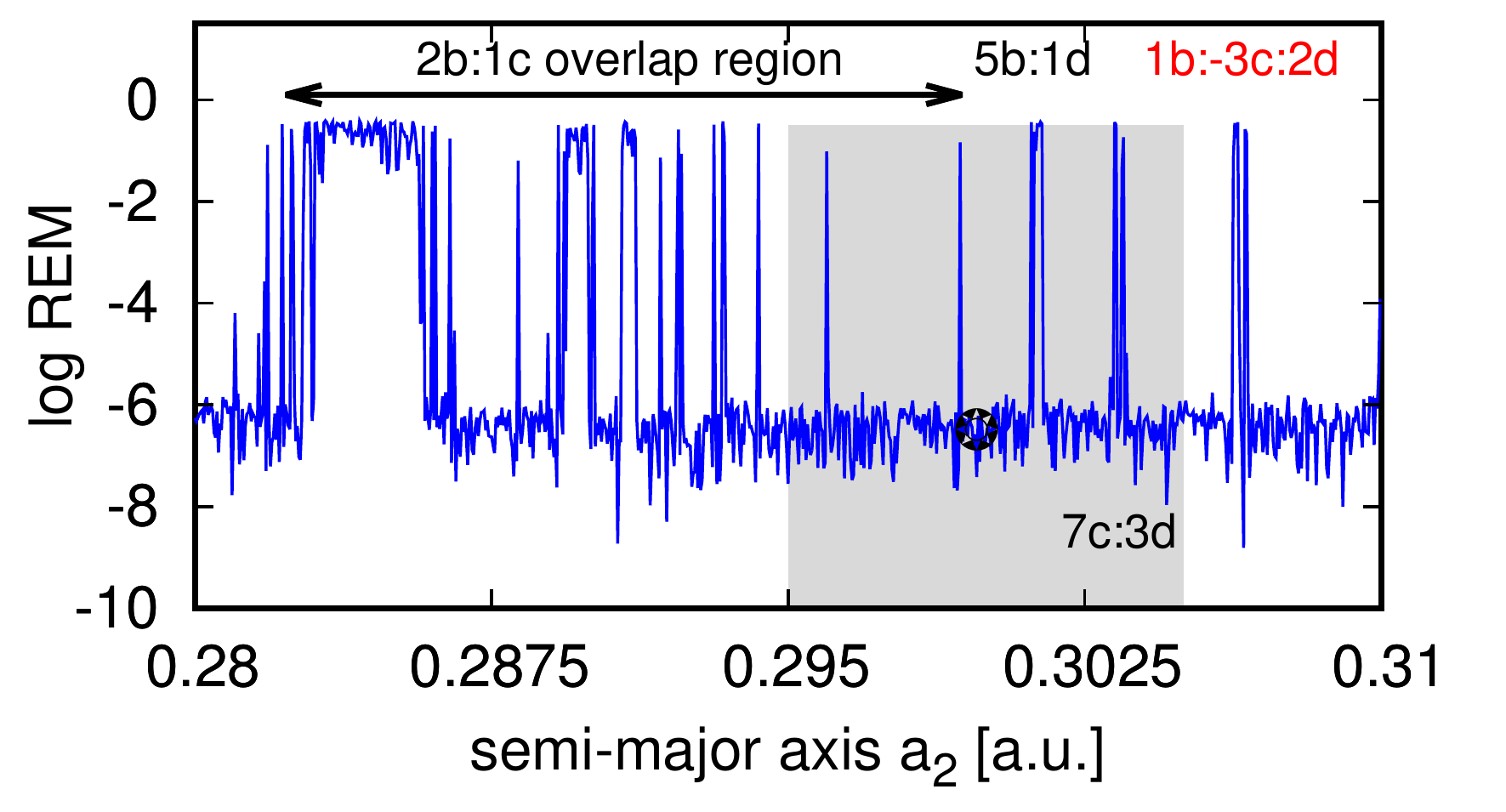}}
\hbox{\includegraphics[width=0.47\textwidth]{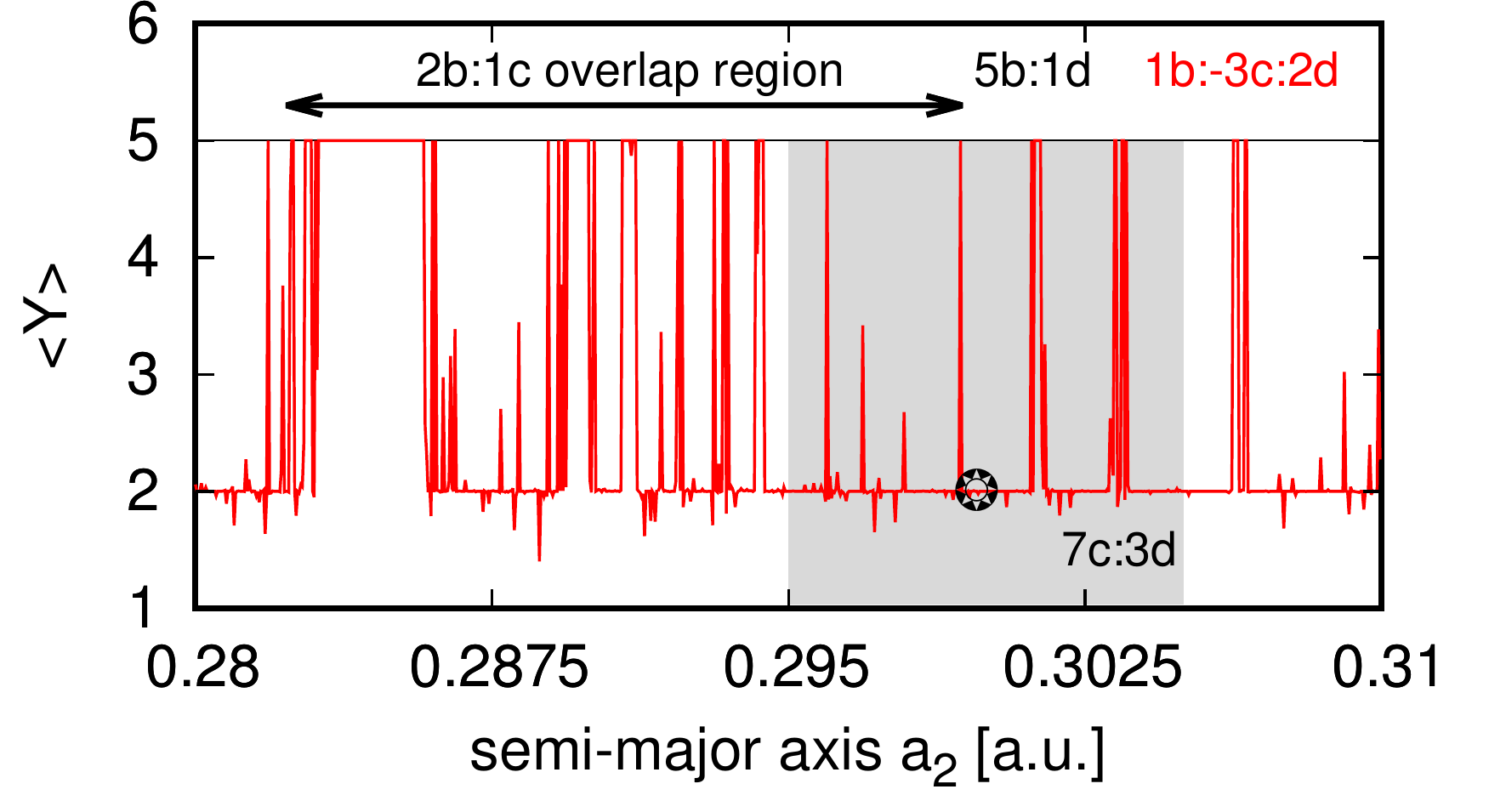}}
}
\caption{
One--dimensional dynamical maps for Kepler-30 along the $(a_{\rm 2})$ semi-major axis. The black star denotes the best-fitting solution, and nearby two-body MMRs are labeled. \emph{ Top-panel:} A flat REM floor  of $\simeq 10^{-6}$ represents  stable orbits, while a few orders of magnitude larger REM values $\simeq 1$ are for unstable orbits.  \emph{Bottom-panel:} MEGNO scan with the maximum of  $5$ for unstable orbits (thin black horizontal line). The MEGNO scan is obtained with the SABA$_4$ scheme while the REM one is obtained with the second-order modified leapfrog \citep{Panichi2017}. For both the algorithms, the time-step is equal to $1.0$~day, the forward integration time for REM is equal to $36$~kyrs ($\sim10^5$ outermost orbits) and the scan resolution is equal to $800$~points. A grey rectangle marks a neighborhood of the best-fitting solution (Model~I in Tab.~\ref{tab:MCMCb}).
}
\label{fig:figure5}
\end{figure}
Figure~\ref{fig:figure5} shows the results obtained with the two indicators, a relatively wide unstable 2:1~MMR region overlapping with higher MMRs, as well as the higher order resonances are clearly represented. The nearby two-body MMR inferred from Fig.~\ref{fig:figure5} is the \corre{2:1~MMR between Kepler-30b and Kepler-30c}.  Other types of resonances present in this scan are the three-body MMRs \citep[e.g.,][]{Nesvorny1998}. We identified these resonances through determining the proper mean motions (fundamental frequencies) of the system with the modified frequency analysis \citep{Laskar1992,Nesvorny1996} and searching for the linear combination of these frequencies. We identified the position of the Laplace resonance (1:-3:2), a few low-order three-body MMRs, like (1:-5:7), (2:-5:2) and high-order ones, such as (13:-25:6), as well as the two-body MMRs near to the best-fitting solution. Some of two- and three-body MMRs are identified in Fig.~\ref{fig:figure5}.

\begin{figure}
\centering
 \vbox{
    \hbox{\includegraphics[width=0.47\textwidth]{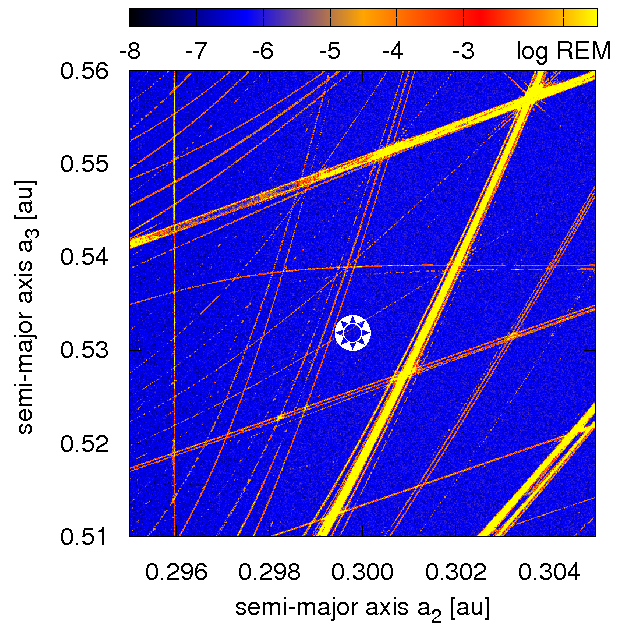}}
    \hbox{\includegraphics[width=0.47\textwidth]{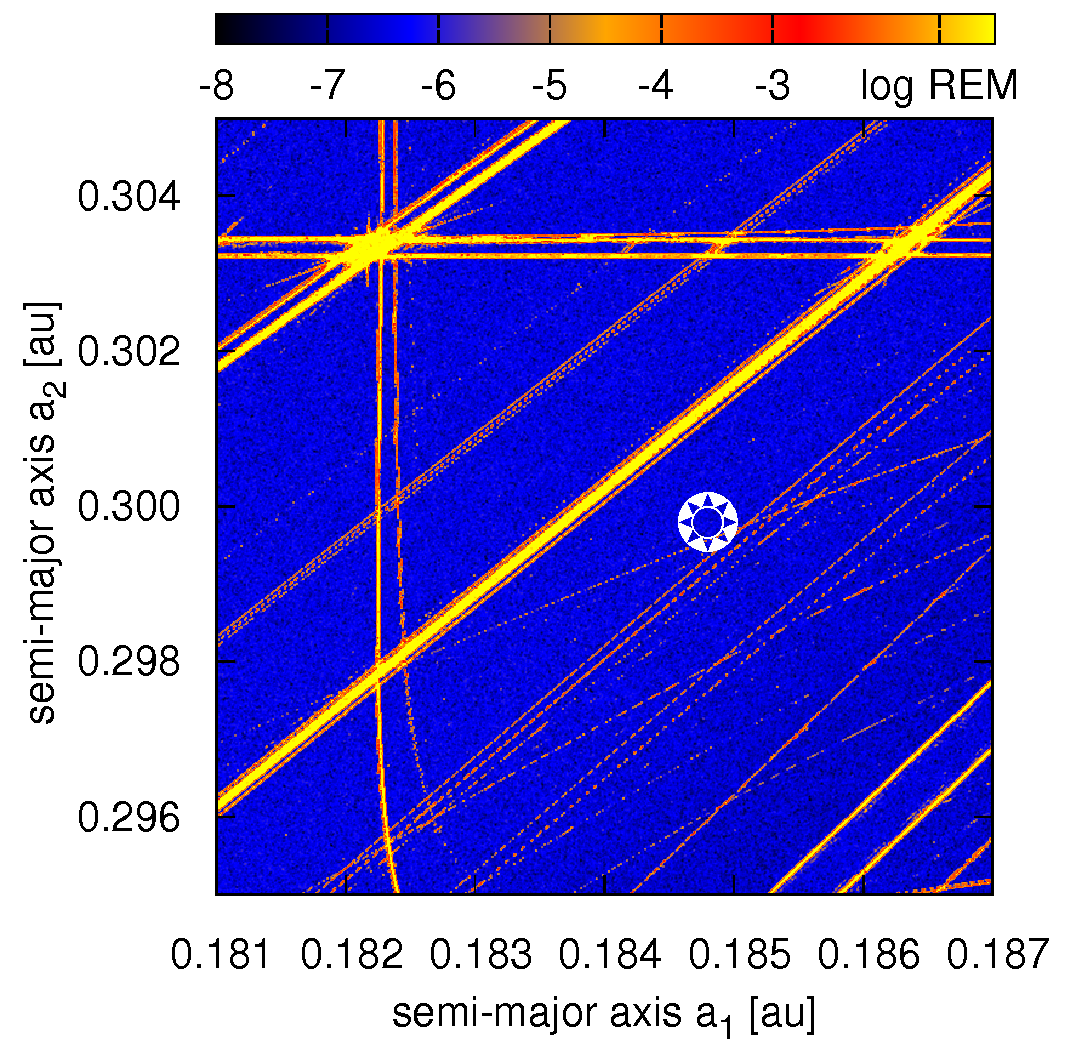}}
 }    
\caption{
 {\em Top panel:} the REM dynamical map in the $(a_{\rm 2},a_{\rm 3})$-plane. The osculating orbital parameters of the  best-fitting configuration marked with the black star-like symbol are listed in Tab.~\ref{tab:MCMCb}. 
 {\em Bottom panel:} the REM dynamical map  in the $(a_{\rm 1},a_{\rm 2})$-plane.  The grid resolution is $800\times800$ for the top panel with the total integration time of $92$~kyrs and the time-step of $1.0$~day. For the bottom panel the grid resolution is $640\times640$  with the integration time of $72$~kyrs and the time-step of $1.0$~day. The REM leapfrog indicator with the Keplerian drift \citep{Wisdom2016} and symplectic correctors \citep{Wisdom2006} was used here.
}
\label{fig:figure6}
\end{figure}

\begin{figure}
\centerline{
\vbox{
 \hbox{\includegraphics[width=0.47\textwidth]{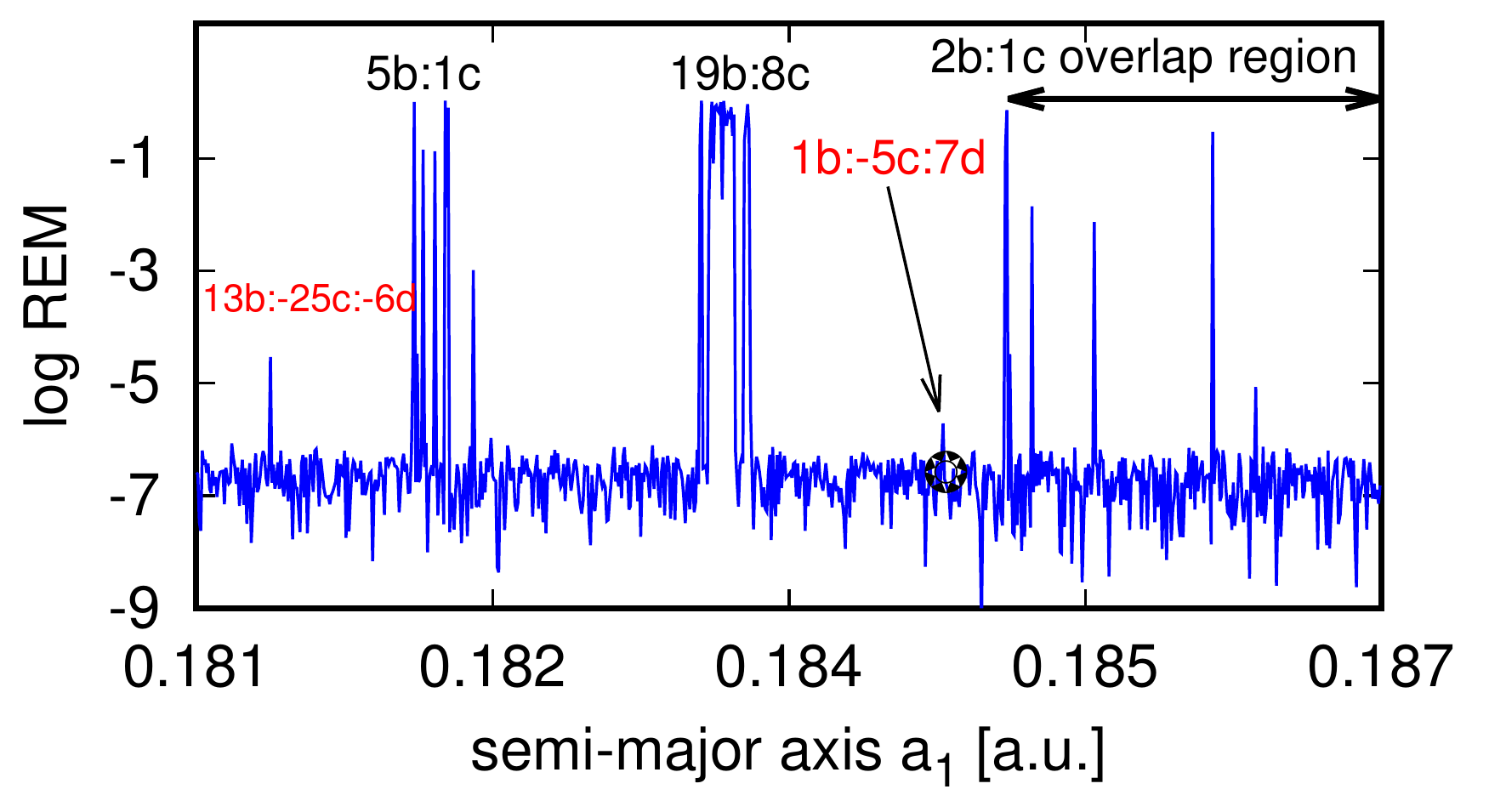}}
 \hbox{\includegraphics[width=0.47\textwidth]{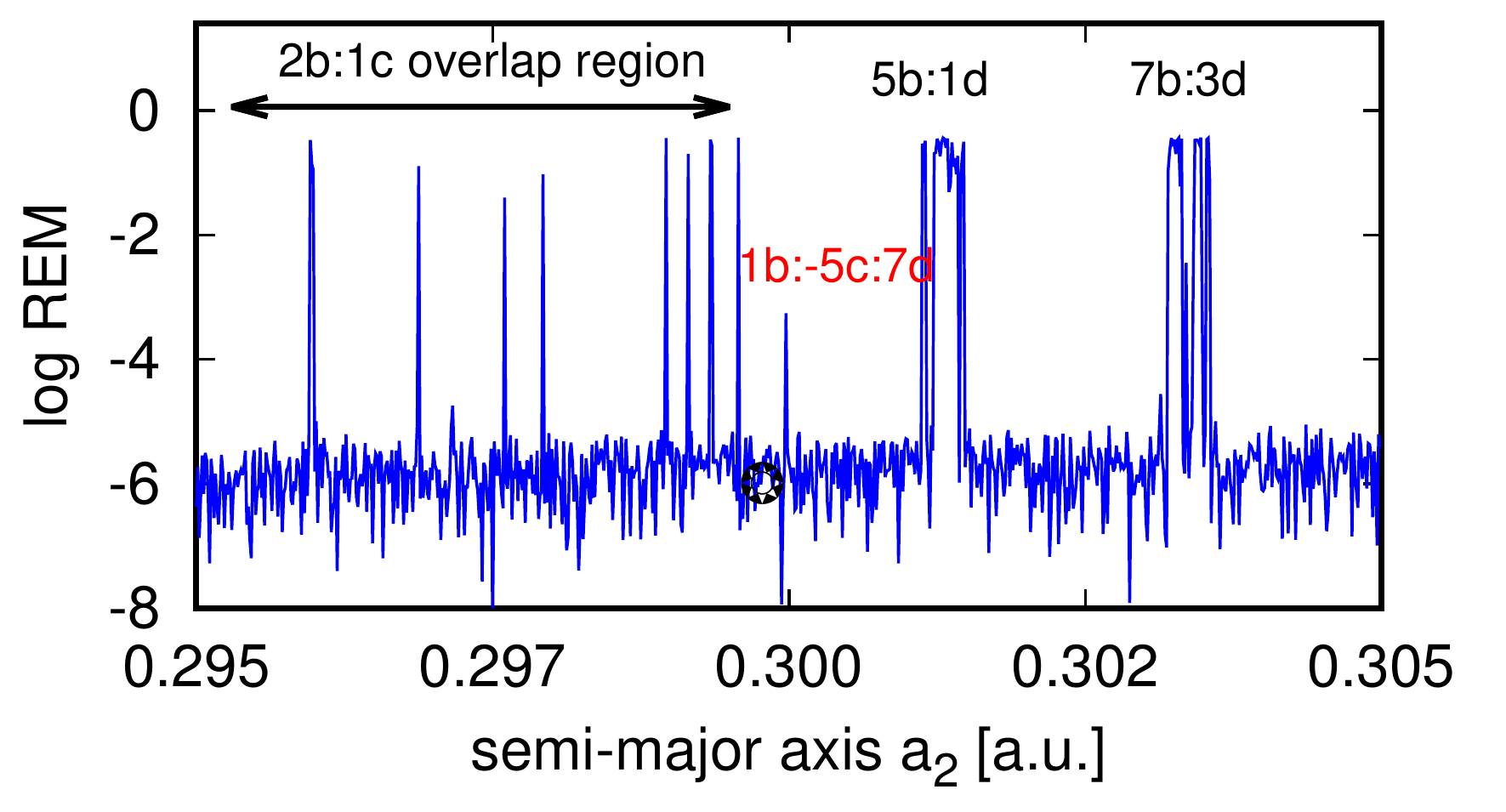}} 
 \hbox{\includegraphics[width=0.47\textwidth]{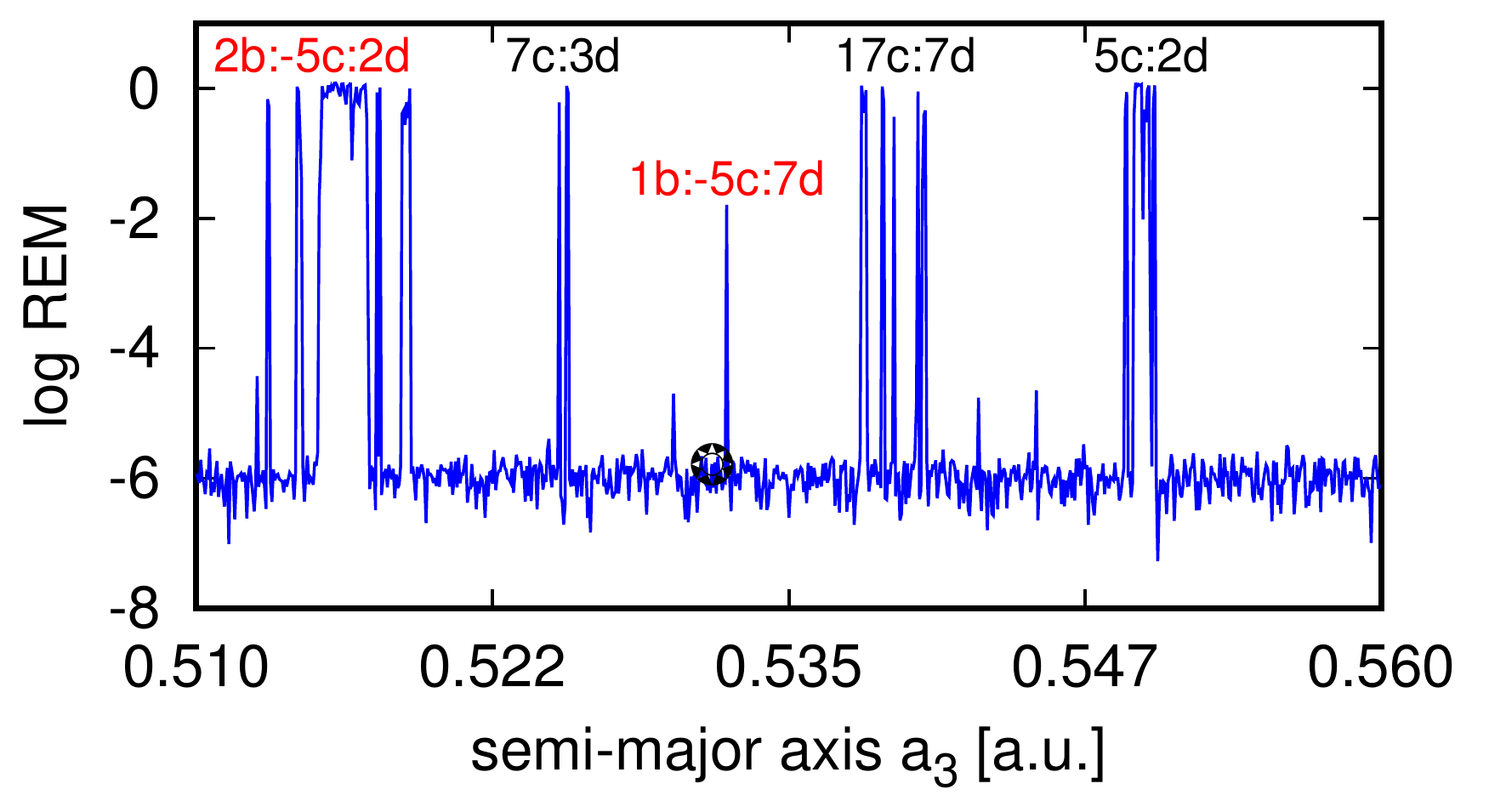}}
}
}
\caption{
One--dimensional scan along the semi-major axes.  {\em The top-panel:} a REM scan along the semi-major axis of the innermost planet (Kepler-30b), labeled are the two-body (black) and three-body (red) MMRs between the planets, the black star-like point marks the best-fitting model. {\em The middle-panel:} a REM scan along the semi-major axis of the middle planet (Kepler-30c). {\em The bottom-panel:} a REM scan along the semi-major axis of the outermost planet (Kepler-30d). The integration time for each scan is defined in the text, all the three scans are computed with the second-order leapfrog and the time-step of $1.0$~day.
}
\label{fig:figure7}
\end{figure}

The complex structure of the two-body and three-body MMRs may be better seen in two--dimensional REM maps in the semi-major axis -- semi-major axis planes. Such dynamical maps are shown in Fig.~\ref{fig:figure6} in a narrow  region selected nearby the best-fitting Model~I. The high resolution of these maps helps to reveal many fine MMRs structures. We note that the selected range of the semi-major axis $a_{\rm 2}$ in both panels of Fig.~\ref{fig:figure6} corresponds to a grey region marked in Fig.~\ref{fig:figure5}.  In the $(a_{\rm 1},a_{\rm 2})$-plane (bottom panel), the 7:3~MMR between the middle and the outermost planet are detected as unstable oblique strips across the map. The 19:8~MMR between the inner pair of planets is visible at the top of the figure while the border of the 2:1~MMR appears at the bottom right side. Other strips corresponding to the (1:-5:7)~MMR are visible too. Such a complex, fractal-like  network of unstable strips in the two--dimensional maps may be interpreted as the Arnold web, as described in \cite{Guzzo2005} for the Solar System, and in the binary $\nu$-Octantis \citep{Gozdziewski2013}.  
To understand even better the position of the system in the parameter space, we present  one--dimensional REM scans along the semi-major axis for all the three planets separately (Fig.~\ref{fig:figure7}).  In this way we may identify precisely the positions of the MMRs and detect their separatrices. We use a forward integration time of $36$~kyrs for the first scan and $48$~kyrs for the second and third scans. We use the same time-step and a modified leapfrog scheme for computing both the one--dimensional scans and dynamical maps in Fig.~\ref{fig:figure6}.
When dealing with higher order two-body or three-body MMRs, it is important to account for a long saturation time of the indicators as discussed in our recent paper \citep{Panichi2017}. Based on this experience, we integrated the equations of motion for $100$~kyrs ($2.5\times10^5$ outermost periods) to examine and  achieve possibly certain saturation of the indicator regarding unstable structures present in the dynamical maps. This experiment guarantees the proper representation and the true dynamical nature of tested orbits. We confirmed that patterns of unstable motions in Fig.~\ref{fig:figure7} are also present in the one--dimensional scans with longer integration time. 

\corre{Interpretation of unstable peaks indicating various MMRs in the one--dimensional scans is a crucial step to understand the dynamical structure of the system. To identify the resonances, we used the direct numerical integration of the equations of motion for the initial conditions selected  from 1- and 2-dim scans. The time-series of canonical orbital elements make it possible to determine temporal evolution of the critical angles and the fundamental frequencies\corre{, $f_i, i =1,2,3$} (the proper mean motions). We analyzed both the 3-planet as well as synthetic 2-planet systems in which we skipped one of the planets. 

\corre{The first two } panels in Fig.~\ref{fig:figure7} show that 1--dim scans along the inner semi-major axes are dominated by the inner 2:1~MMR. Fine structures in this region are generated by overlap of this resonance with other two- and three-body MMRs involving all planets.  
}
\corre{
To understand the emerging complex Arnold web structure, as detected in one-- and two--dimensional scans along the semi-major axes, as well as proximity of the system to the 2:1~MMR, we computed the amplitude of two resonant angles, the proper mean motions (fundamental frequencies) ratio and the MEGNO indicator along one--dimensional scan of the semi-major axis of the middle planet.} 

\begin{figure}
\centerline{
 \vbox{
 \hbox{\includegraphics[width=0.47\textwidth]{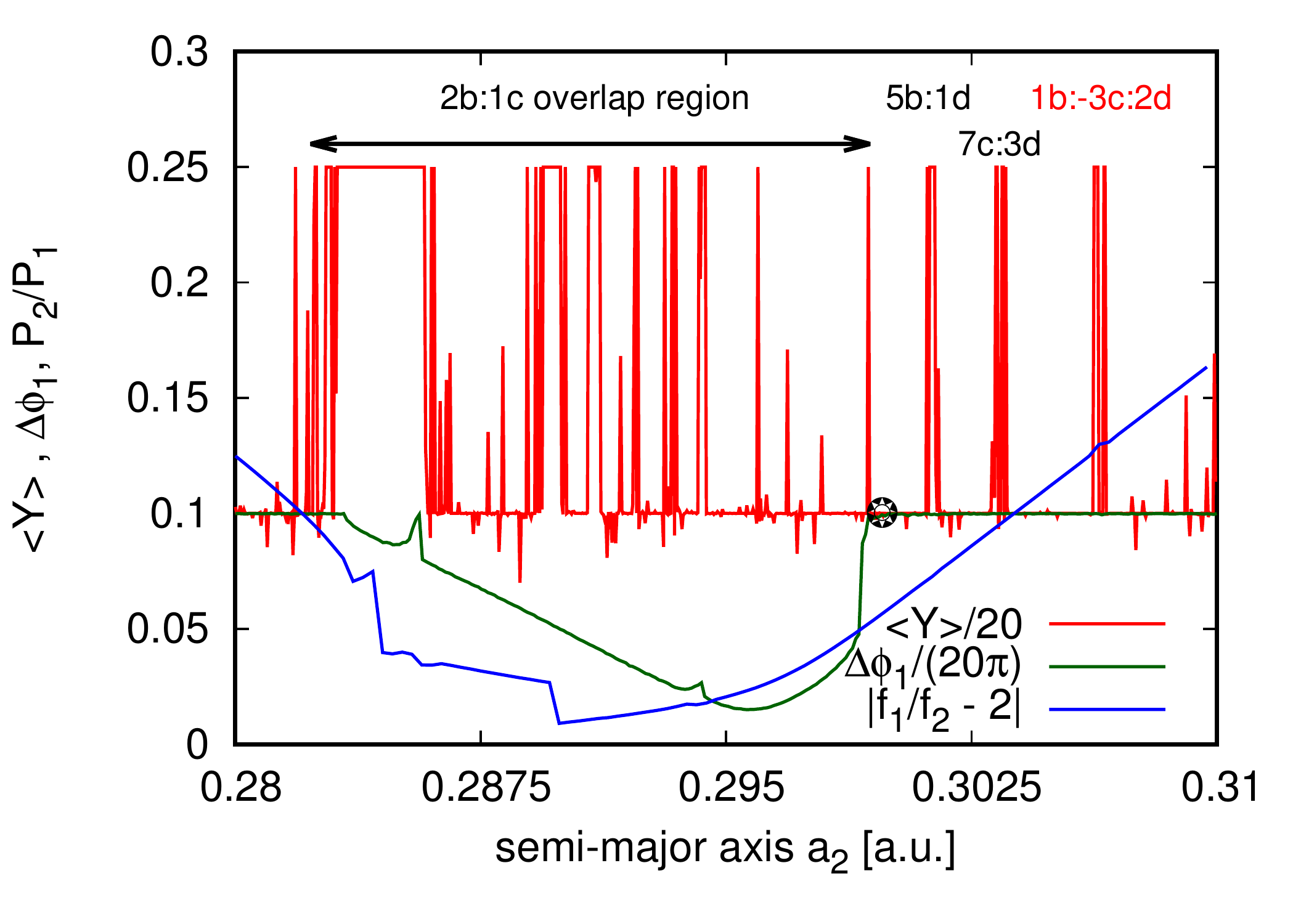}}
 }
}
\caption{
\corre{One--dimensional scan along the semi-major axis of Kepler-30c. The scan shows the amplitude of a selected critical angle of the inner 2:1~MMR (green curve), the ratio of the proper mean motions (blue curve) and the MEGNO value (red curve) for each of the $10^4$ initial conditions. Those values are appropriately scaled to fit the same $y$-axis range. The 2:1~MMR region overlaps with other two- and three-body MMRs emerging due to the presence of Kepler-30d. The best-fitting solution (black star-like point) lies at the border of this region. See the text for more details.
}
}
\label{fig:figure8}
\end{figure}

\corre{
In Fig.~\ref{fig:figure8} we report the amplitude scan $(\Delta\phi_1)$ for a selected critical angle $\phi_{2:1} =   \lambda_1 - 2 \lambda_2 +  \varpi_1$ for $10^4$~points to resolve a fine structure of the resonance.  The fundamental frequencies are determined via the Frequency Modified Fourier Transform \citep[FMFT,][]{Nesvorny1996} of the time series defined through $\left\{ a_i(t) \exp[ \mbox{i} \lambda_i(t)] \right\}$,
where $a_i(t)$ and $\lambda_i(t)$ are the osculating, canonical semi-major axis and the true longitude of the $i$-th planet, respectively. The total integration time is equal to $2^{18}$ time steps of $1.0$~days  ($\simeq 2 \times 10^3 P_3$). Both the MEGNO and amplitude scans are computed for the integration time of $10$~kyrs, with a time-step of $1.0$~day, and the fourth order SABA$_4$ scheme.}
\corre{
As Fig.~\ref{fig:figure8} suggests,  the best-fitting system is apparently very close to the dynamical border of the 2:1~MMR, which we identify as the end of libration range of the critical argument across $a_2$. The separation of the nominal system from this border $\simeq 6 \times 10^{-4}$~au is significant, as it is 300 times larger than the uncertainty of the semi-major axis equal to $2 \times 10^{-6}$~au (see Table~\ref{tab:MCMCb}).} 

\begin{figure}
\centerline{
 \vbox{
 \hbox{\includegraphics[width=0.47\textwidth]{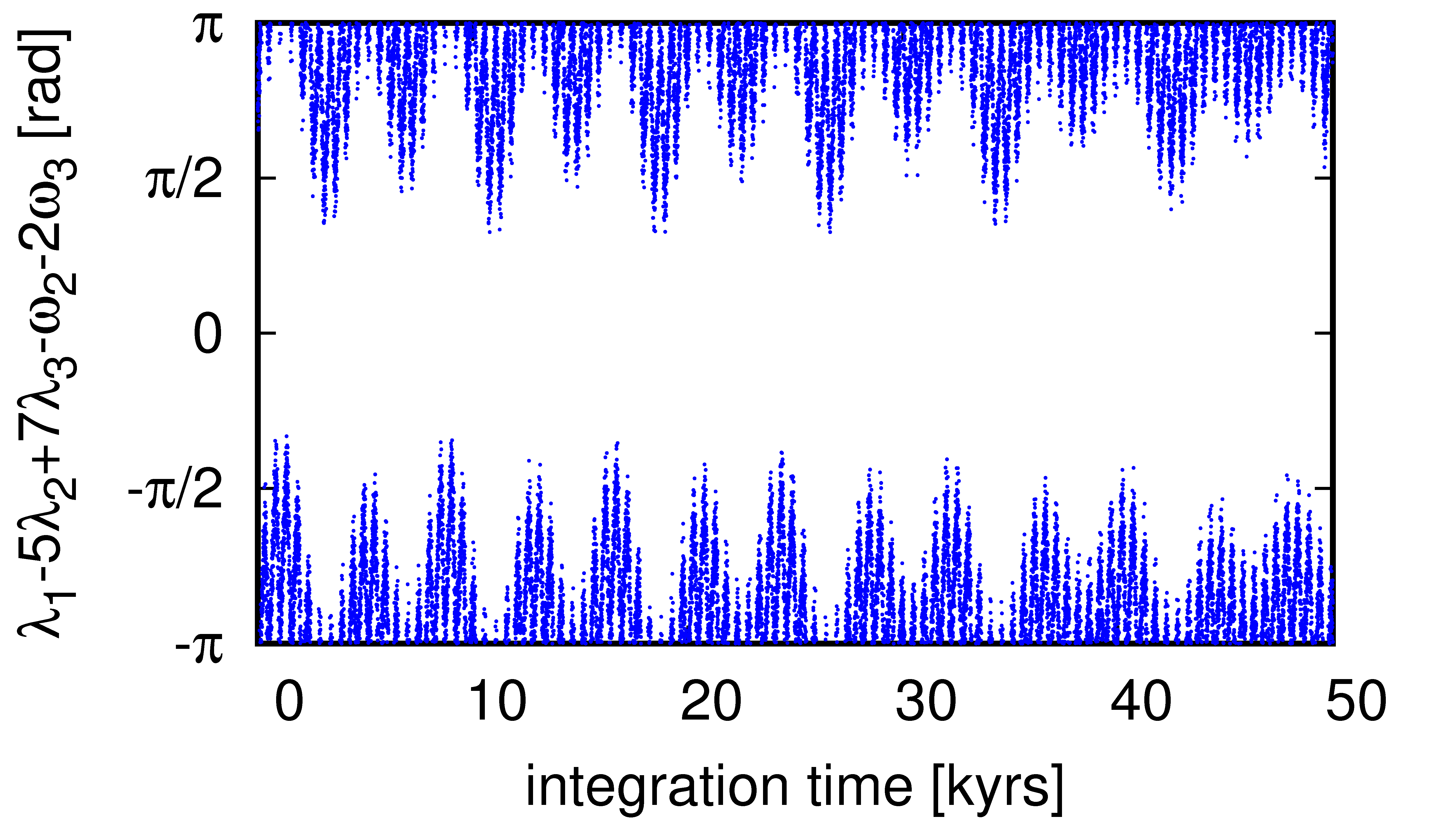}}
 \hbox{\includegraphics[width=0.47\textwidth]{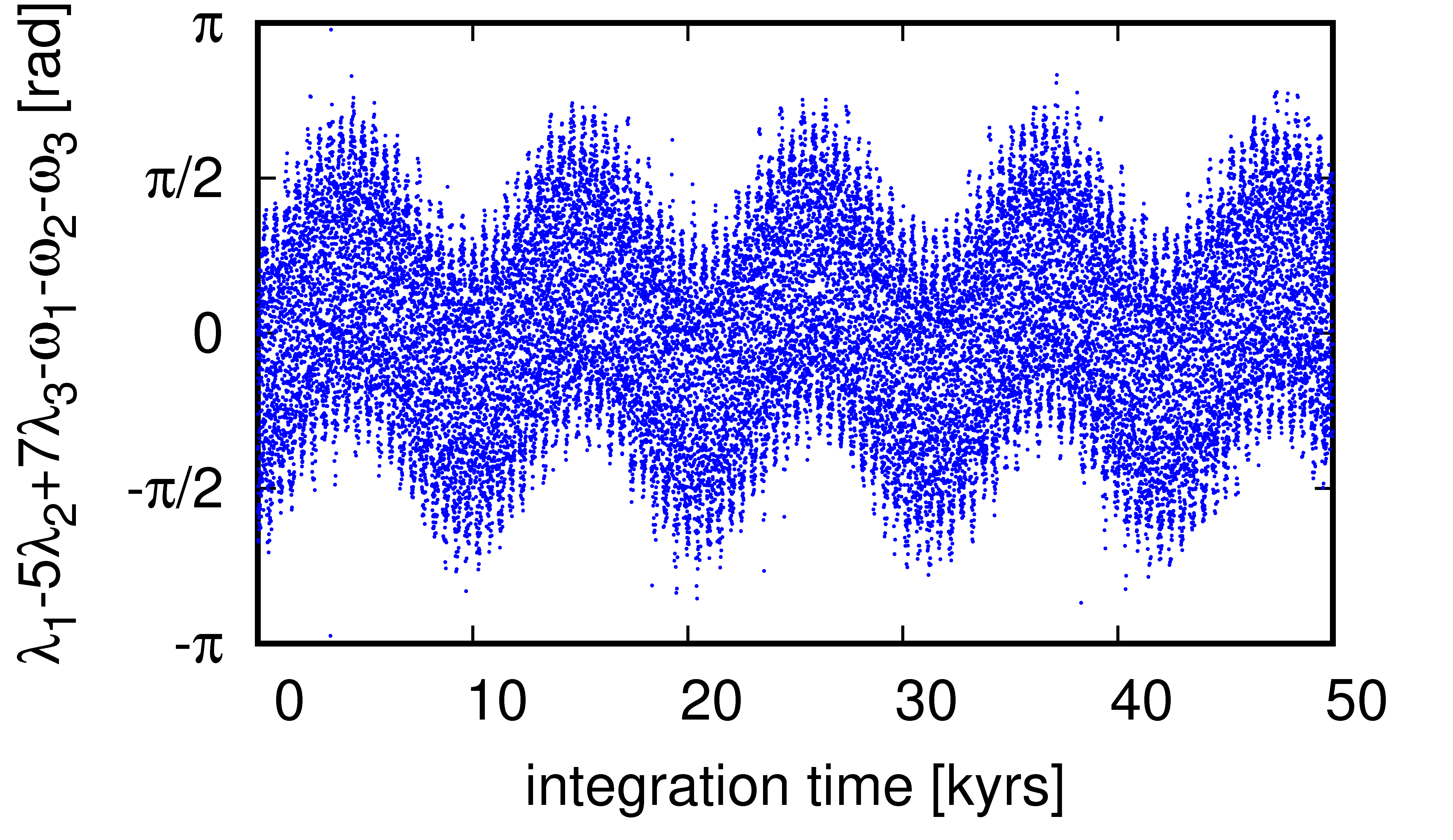}}
 }
}
\caption{
Time evolution of the (1:-5:7) three-body MMR critical  angles. {\em The top-panel:} only the semi-major axis of Kepler-30b is shifted of around $0.2\%$ its nominal value, the evolution of $\phi_{1:-5:7,2}$ is presented.  {\em The bottom-panel:} the semi-major axis of Kepler-30b is shifted of around $0.002\%$ its nominal value. A quasi-libration is shown for  $\phi_{1:-5:7,1}$. As before, all the other orbital parameters and masses are left unaltered.
}
\label{fig:figure9}
\end{figure}

\corre{
Moreover,  Fig.~\ref{fig:figure7} illustrates that the Kepler-30 system is also close to the (1:-5:7)~MMR.  We selected two critical angles
\begin{eqnarray*}
 \phi_{1:-5:7,1} &=&  \lambda_1 - 5 \lambda_2 + 7 \lambda_3 - \varpi_1 - \varpi_2 - \varpi_3,\\
 \phi_{1:-5:7,2} &=&  \lambda_1 - 5 \lambda_2 + 7 \lambda_3 - \varpi_2 - 2 \varpi_3,
\end{eqnarray*}
and their temporal evolution with large amplitude librations is  presented in Fig.~\ref{fig:figure9} for $50$~kyrs.  Actually, the (1:-5:7)~MMR is the closest low-order MMR with unstable zone to the nominal configuration. Yet, the separation
as small as $0.000014$~au is seven times larger than the uncertainty.
}

%
%
\section{Kepler-30 hosts heterogeneous planets}
\label{sec4:radiusMass}
\begin{figure}
\centering
\vbox{
   \hbox{
   \includegraphics[width=0.49\textwidth]{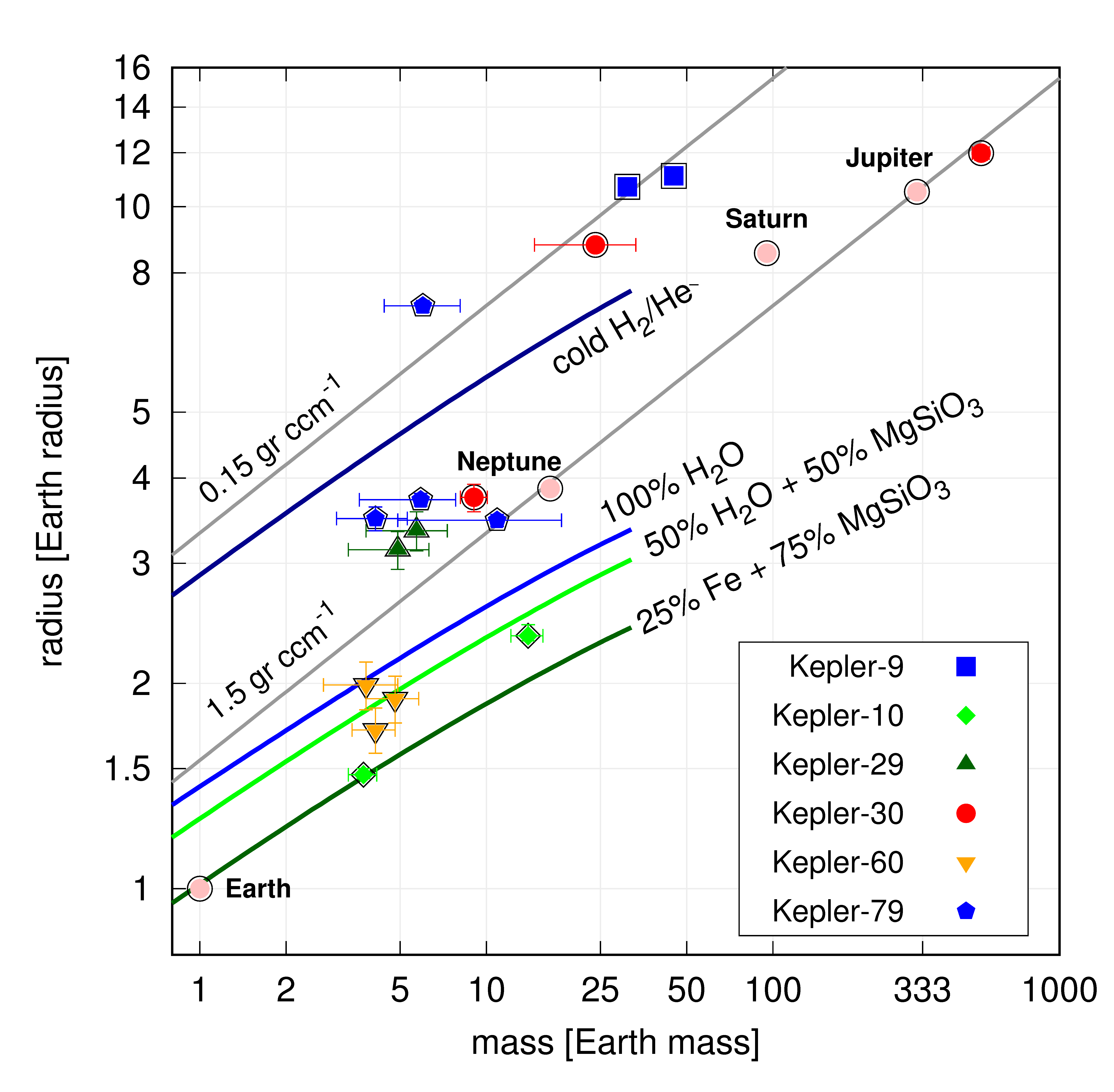} 
   }
}    
\caption{
The mass-radius diagram. The masses and radii of Kepler-30 planets (red circles together with the estimated errors) are selected from the nominal Model~I (Tab.~\ref{tab:MCMCb} and Tab.~\ref{tab:radii}). Some other systems of interest are illustrated with different colours and symbols. Data for Kepler-9 (blue squares) are from \citep{Borsato2014}, the masses for Kepler-10 (light-green diamonds) are taken from \citep{Weiss2016} while the radii are from \citep{Dumusque2014}. For Kepler-29 (dark-green triangles) and Kepler-60 (orange triangles), the masses are from \citep[]{Migaszewski2017} and from \citep{Gozdziewski2016}, respectively, while the planet radii from \citep{JontofHutter2016}. Computations are based on the internal composition models in \citep{Zeng2016}.}
\label{fig:figure10}
 \end{figure}

%
In this section we aim to discuss the physical properties of the three planets of Kepler-30, namely their densities and  internal compositions. In the mass-radius diagram (Fig.~\ref{fig:figure10}) we indicate the positions of the Kepler-30 planets and compare them with those of other well characterized  \kepler{} systems reported in the literature. For the Kepler-30 system, we use the masses determined from our dynamical TTV analysis, and their radii derived from the re-analysis of the \kepler{} light curves reported in Sec.~\ref{sec2:Newfit}. The internal compositions of the super-Earths and Neptunes are determined on the basis of the two layer models taken from \citep{Zeng2016}.   

We check whether the limb-darkening coefficients and the transit uncertainties may influence the radii estimates. We find no significant impact of these two parameters on the radii. In Tab.~\ref{tab:radii} we report the radii, inclinations and related uncertainties calculated with the MCMC sampling of the transit model for the folded, binned light curves. The densities and related uncertainties, inferred from best-fitting solution (Model~I in Tab.~\ref{tab:MCMCb}) are also presented.
\begin{table}
\centering
\caption{
The best-fitting parameters for the folded, binned light curves of Kepler-30. The parameters and related uncertainties are obtained using the MCMC sampling. The radius of Kepler-30 star is adopted as $0.867~\Rsun$ \citep[]{Rowe2015}. The densities are calculated using the masses and related errors from Model~I in Tab.~\ref{tab:MCMCb}.}
\begin{tabular}{l c c r}
\hline
\hline
System & $R\,[\RE]$ & $I$\,[deg] & $\rho$\,[\gccm{}] \\
\hline
Kepler-30~b  & 3.75$\pm$0.18 & 89.51$\pm$0.32 & 0.96$\pm$0.15 \\
 Kepler-30~c  & 11.98$\pm$0.28 & 89.74$\pm$0.02 & 1.71$\pm$0.13   \\
 Kepler-30~d  & 8.79$\pm$0.13 & 89.81$\pm$0.02 & 0.19$\pm$0.01 \\
\hline\hline
\end{tabular}
\label{tab:radii}
\end{table}
Our TTV analysis constrain the mass of Kepler-30b to $(9.2 \pm 0.1)~\mE$, i.e. roughly within 1\% and it is one of the most precise determinations of planet masses in the \kepler{} sample. Our re-analysis of the light curve results in a radius of $3.75\pm0.18~\RE$. With a mass of a Super-Earth planet and a radius similar to Neptune, Kepler-30b has a bulk density of $(0.96\pm0.15)$\,\gccm{}.  We stress that the radius quoted in \cite{Rowe2015} is almost two times smaller, yielding a density of $\simeq 7$\,\gccm{}, which is consistent with the density of a Super-Earth planet. Our reported density and radius better agree with data in the discovery paper \citep{Fabrycky2012}, as well as in \cite{SanchisOjeda2012}, who found a density of $1.02\pm0.13$\,\gccm{}, reporting a radius of $3.9\pm0.2~\RE$. Therefore  we classify Kepler-30b as a mini-Neptune planet rather than a rocky super-Earth. In the mass-radius diagram (Fig.~\ref{fig:figure10}), the Kepler-30b planet is localized near to the two planets of the Kepler-29 system with a bulk density smaller than that of Neptune but larger when compared to the density of Saturn. 

The Kepler-30c planet has a mass of $536 \pm 5~\mE$ and a radius of $11.98\pm0.28~\RE$.  The planet is $1.7$ times more massive than Jupiter while it has similar bulk density of $(1.71~\pm0.13)$\,\gccm{}. These data imply that this planet may be classified as a hot Jupiter.  The uncertainty of the mass of Kepler-30c is also within $\simeq 1\%$. This mass determination of a Jovian planet may be one of the most accurate ones derived for Jovian planets. For instance the uncertainties  are three times smaller when compared to the mass error of  hot Jupiters WASP-73b \citep{Delrez2014}, K2-31b \citep{Grziwa2016} and Kepler-412b \citep{Deleuil2014}. The uncertainty is smaller by one order of magnitude from mass uncertainties of Kepler-46b \citep{SaadOlivera2017}, Kepler-447b \citep{LilloBox2015} and HAT-P-49b \citep{Bieryla2014}. 

Regarding giant planets, the age of the parent star is critical to understand their internal structure \citep{Veras2016}. Since the age estimate of the Kepler-30 is uncertain, we cannot precisely characterize the internal structure of Kepler-30c. Considering a one layer structure model, this hot giant may have the internal structure similar to Jupiter.

The Kepler-30d planet with a mass of $23.7 \pm 1.3~\mE$ appears as one among a few Neptune-mass exoplanets \citep{Mazeh2016}. Moreover, with a radius of $8.79\pm 0.13 ~\RE$, it has a bulk density of  $(0.19 \pm 0.01)$\,\gccm{}, i.e., eight times smaller than Neptune. To put it more vividly, Kepler-30d has the radius of Saturn but the mass of a Neptune-like planet. This peculiar characteristics place this planet in a poorly populated region of the mass-radius diagram, where only a few similar \kepler{} planets could be found, like Kepler-9b and Kepler-9c. For an up-to-date list of low-density Neptune-like planets see \citep{Cubillos2017}. Their region lies between two  different density regimes: the non-degenerate regime, where the planet density decreases when the mass increase,  and the partially-degenerate regime, in which the density increases with the mass. The formal uncertainties of the radius and mass of Kepler-30d imply a very low bulk density with a high certainty. We note that it is similarly low as reported for Kepler-9b,c in \cite{Dreizler2014}. Such planets may be core-less \citep[]{AliDib2018}.

Following \citep[]{Mordasini2012b}, the Kepler-30d parameters and model are consistent with a planet with large primordial H$_2$/He envelope (see their Fig.~3 for a reference). Low density Neptune-like planets which may be similar to Kepler-30d are,  e.g., HATS-8b \citep{Bayliss2015}, and HAT-P-18b \citep{Hartman2011}. Their large radii might be explained by a strong irradiation and intense atmospheric evaporation expected for close-in planets \citep[]{Kurokawa2014}. This explanation, however, may be insufficient for Kepler-30d since it has relatively long orbital period of $\simeq 142$~days, which is 35 times longer than HATS-8b.

\corre{Recent statistical surveys \citep{Weiss2017,Millholland2017} indicate that planets in the same multiple \kepler{} system tend to have similar sizes. This is not the case for a few well characterised systems, including Kepler-30 studied in our paper.  As we found, the Kepler-30 system exhibits a large spread of planet masses -- the mass ratio between the most massive and the less massive planet is around 60.  So different masses and bulk densities together with the proximity to the double 2:1~MMR and bounding the system within $0.5~\au$, define the Kepler-30 system as one of the most heterogeneous and tightly packed extrasolar systems discovered to date. }

\corre{
The Kepler-87 system \citep{Ofir2014} may be another interesting multiple system which could be considered to be similar to Kepler-30, also because it exhibits equally large TTV amplitude for the outer planet of almost $2$~days. Two Kepler-87 planets are reported to have very different masses, more than 300 and 6 Earth masses, respectively, that implies low densities of 0.73 and 0.15\,\gccm{}; the latter is as small as in the Kepler-30 system\footnote{We note, however, that the TTVs signals for Kepler-87 are more noisy than in the Kepler-30 case and the low density planet Kepler-87c is marked as false positive in the NASA Exoplanet Archive Q1-Q17 DR25 release.}.  This Kepler-87 system is presumably located nearby the 5:3 MMR.}

Another member in the set of  well studied \kepler{} systems, which hosts planets with very different densities, is Kepler-36 \citep{Carter2012}. The masses of Kepler-36b and Kepler-36c are equal to 4.45 and 8.08$~\mE$, respectively. With the radii of 1.486 and 3.679$~\RE$ for planets~b and~c, their densities are 7.46\,\gccm{} and 0.89\,\gccm{}, respectively. 

\corre{Two systems hosting low density planets, which could be compared to Kepler-30b are WASP-47 \citep{Becker2015} and Kepler-79 \citep{JontofHutter2014}. All the four planets of Kepler-79 are similar and have low densities (see Fig.~\ref{fig:figure10}). The orbital configuration is near to the four-body MMR chain. The most massive planet in Kepler-79 is however more than 50 times less massive than Kepler-30c.
}

A dichotomy of planetary companions observed in the Kepler-87 or Kepler-36 systems is perhaps most striking in the case of Kepler-30 due to the presence of three different planets.  As described in \cite{Lopez2013}, the formation of such heterogeneous systems must be very complex. Strong constrains on their formation history, such as the initial core mass and the core accretion history \citep{Mordasini2009},  photo-evaporation flux and irradiation level from the star \citep{Rogers2011} as well as the protoplanetary disk structure \citep{Mordasini2012a} need to be taken into account.
%
%
\section{Formation of the Kepler-30 system}
\label{sec5:migration}
%
%
In this section we study a scenario of possible formation of the Kepler-30 system by means of migration. The 
period ratios of subsequent planets are relatively distant from small 
rational numbers in contrast with an expected outcome of the smooth convergent migration \citep[e.g.,][]{Snellgrove2001,Lee2002,Papaloizou2005}. Therefore, the observed architecture of Kepler-30 seems to be problematic in terms of the scenario discussed here. The problem reduces itself, though, when we notice that the period ratio of the two inner planets $\Pc/\Pb \approx 2.057$ is an equilibrium value of the smooth convergent migration for certain, rather realistic values of the parameters. This unexpected conclusion stems from the significant difference in the planets' masses, which will be explained later. 

The real discrepancy between the theoretical predictions and the observations is the period ratio of the two outer planets $\Pd/\Pc \approx 2.374$. There are three possible scenarios that leads to this value. The outermost planet could have started its migration with $\Pd/\Pc$ significantly greater than $2$ and so did not have enough time to reach the resonance. Another scenario is that the outermost planet could have been stopped with $\Pd/\Pc \sim 2.374$ without entering the resonance. The reason for the latter could be the planet-wake interaction \citep{PodlewskaGaca2012,Baruteau2013}. The third possibility is that the convergent migration led the system inside the Laplace resonance and then the migration of the outermost planet slowed down (because of the disc evolution or the planet-wake interaction).

The first scenario is difficult to be studied quantitatively and we will skip it. The second and the third scenarios require $\tau_3 > \tau_2$ (the migration time-scales of subsequent planets from the innermost to the outermost one are denoted by $\tau_1, \tau_2, \tau_3$) for certain amount of time or/and for certain $\Pd/\Pc$, possibly in a range of $(2, \sim 2.374)$. In order to study the range of $\Pd/\Pc$ in which the wake produced by the giant planet operates one would need to use hydrodynamical simulations. Similarly, the study of the disc evolution and possible change in the migration time-scale of the outermost planet is beyond the scope of this paper. 

The difference between scenarios 2 and~3 is that the system resides in the chain of resonances for the latter, before the outermost planet is forced to leave the chain. Slowing down the migration of this planet does not necessarily lead to an increase of $\Pd/\Pc$ without changing $\Pc/\Pb$. The divergent migration of three planets away from a given chain of MMRs can occur along certain paths at the period ratio--period ratio diagram \citep{Migaszewski2016}. It is then possible that $\Pc/\Pb$ would also increase, even though $\tau_1 > \tau_2$. Further in this section we try to find out if the system initially inside the chain of resonances can move towards the observed configuration of Kepler-30.

\corre{We use a parametric model of the migration in which the right-hand side of the $N$-body equations of motion
is completed with a term that mimics the planet-disc interaction \citep[e.g.,][]{Beauge2006,Moore2013}. The equations of motion reads
\begin{equation}
\ddot{\vec{r}}_i = \vec{f}_{\idm{N-body}, i} + \vec{f}_{\idm{migr}, i},
\end{equation}
where
\begin{equation}
\vec{f}_{\idm{N-body}, i} = -\frac{G (m_0 + m_i)}{r_i^3} \vec{r}_i - \sum_{j=1, j \neq i}^{N_{\idm{pl}}} G m_j \left( \frac{\vec{r}_i - \vec{r}_j}{\|\vec{r}_i - \vec{r_j}\|^3} + \frac{\vec{r}_j}{r_j^3}\right)
\end{equation}
and
\begin{equation}
\vec{f}_{\idm{migr}, i} = -\frac{\vec{v}_i}{2\,\tau_i} - \frac{\vec{v}_i - \vec{v}_{c,i}}{\kappa_i^{-1} \tau_i},
\end{equation}
In the $N$-body part of the equations $N_{\idm{pl}}=3$ and it stands for the number of planets in the Kepler-30 system. The position vectors of the planets are denoted by $\vec{r}_i$ and $\vec{r}_j$ ($i, j = 1,2,3$), $G$ is the gravitational constant.  The astrocentric velocity of planet~$i$ is denoted by $\vec{v}_i$, while $\vec{v}_{c,i}$ means the velocity of the planet at circular orbit at a distance $r_i \equiv \| \vec{r}_i \|$. }

We use indices of $1, 2, 3$ for the innermost, middle and outermost planets, respectively, when we refer to the simulated system instead of b, c, d that refer to the observed configuration. The
time-scale of migration for planet~$i$ is denoted with $\tau_i$, while the respective  circularization time-scale equals $\kappa_i^{-1} \tau_i$.
For the convergent migration of three planets $\tau_i$ should obey the inequality $\tau_1 > \tau_2 > \tau_3$. The scenario studied in this section requires that at some moment of time $\tau_3$ increases and becomes greater than $\tau_2$ or even $\tau_1$. We assume that $\tau_1$ and $\tau_2$ are constant, while $\tau_3$ changes according to the following formula:
\begin{equation}
\tau_3(t) = \tau_{3,\idm{conv}} \, e^{-\left( t/t_0 \right)^\gamma} + \tau_{3,\idm{div}} \bigg[ 1 - e^{-\left( t/t_0 \right)^\gamma} \bigg],
\label{eq:tau3}
\end{equation}
where $\tau_{3,\idm{conv}}$ and $\tau_{3,\idm{div}}$ denote the time-scales of migration corresponding to the convergent and divergent migration, respectively. The transition between the two regimes of migration occurs at time $t_0$, while the parameter $\gamma$ controls how smooth the passage is (we used $\gamma = 8$ in the simulations).
\begin{figure*}
\hbox{
\includegraphics[width=0.32\textwidth]{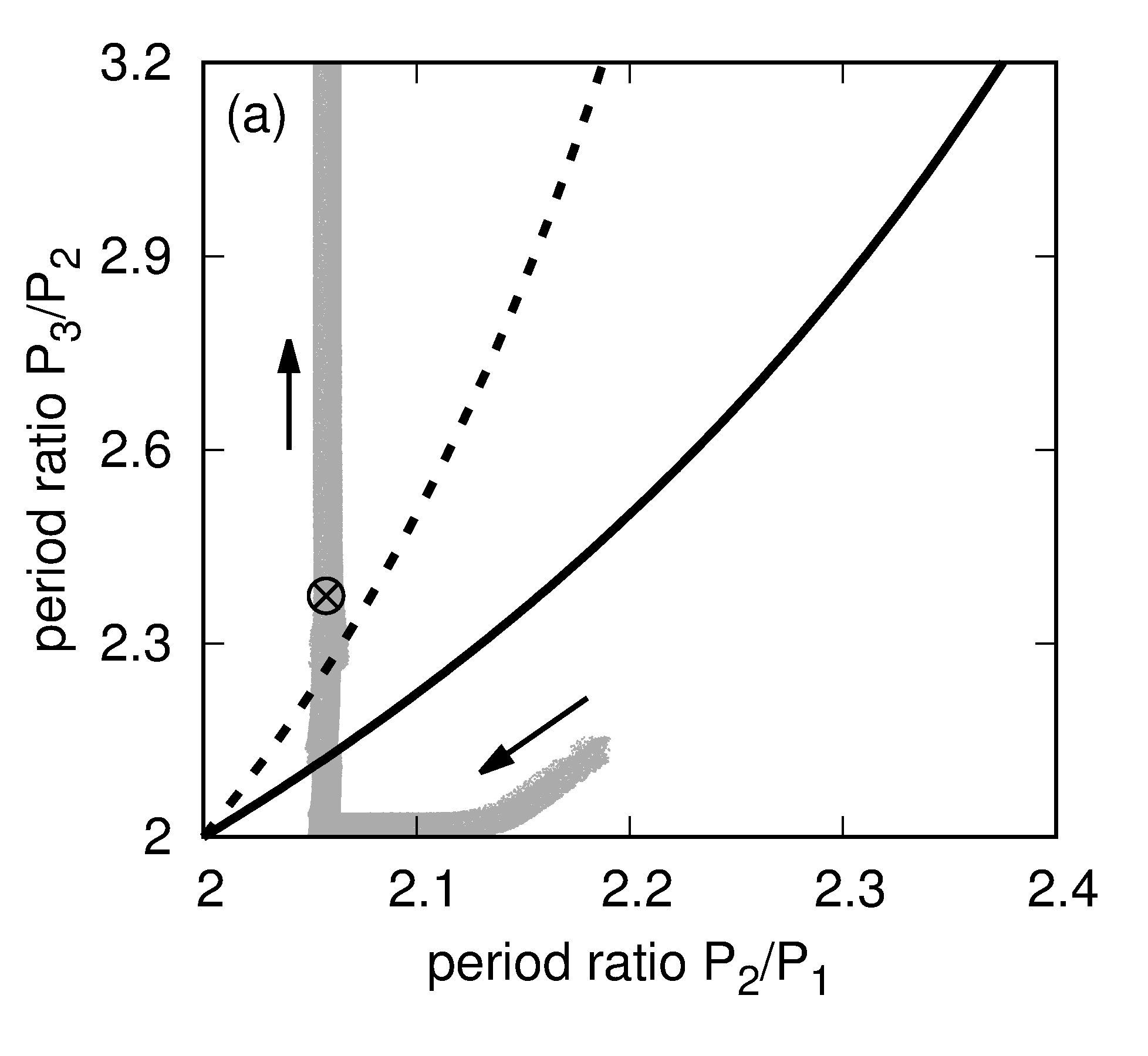}
\includegraphics[width=0.32\textwidth]{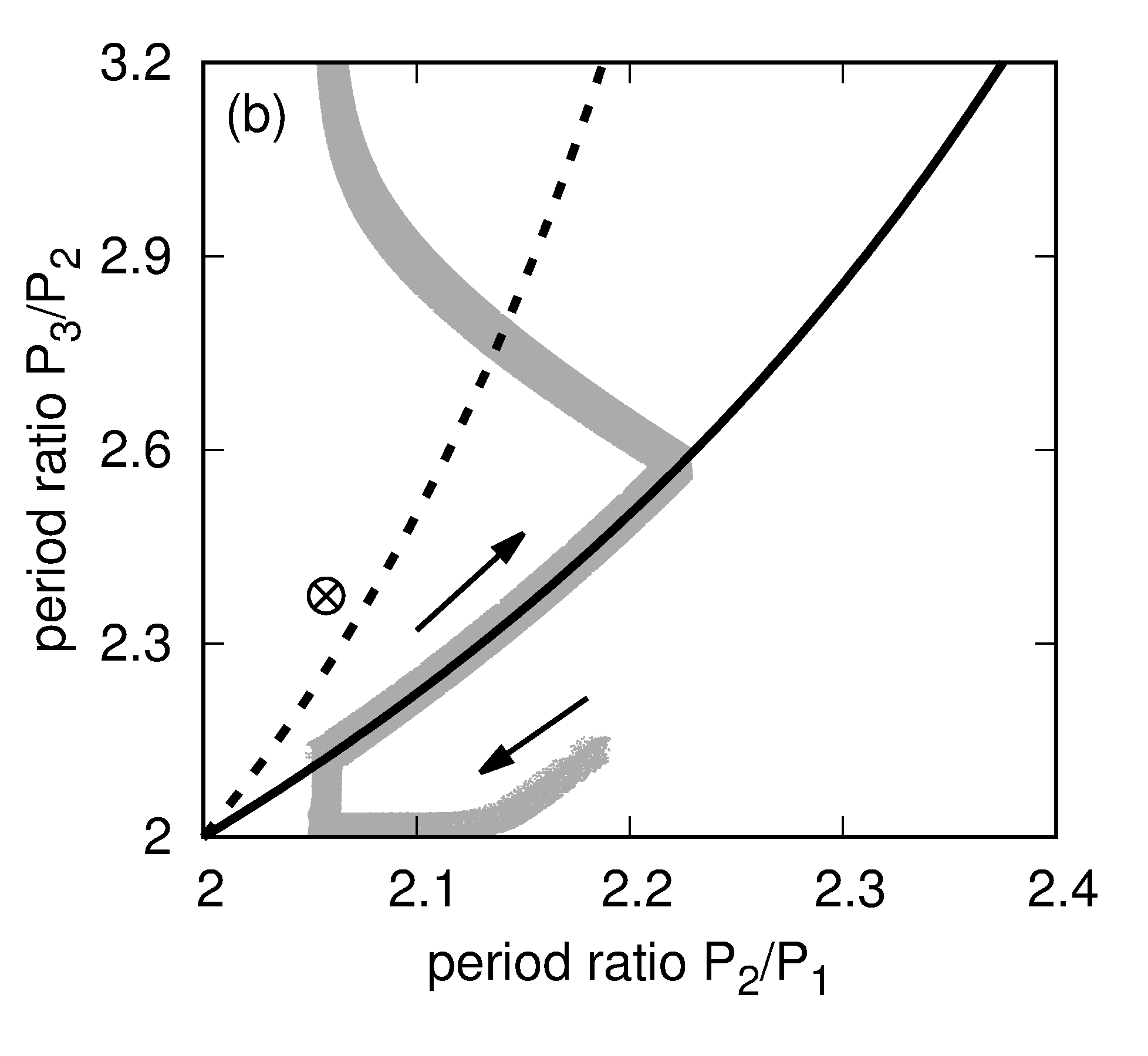}
\includegraphics[width=0.32\textwidth]{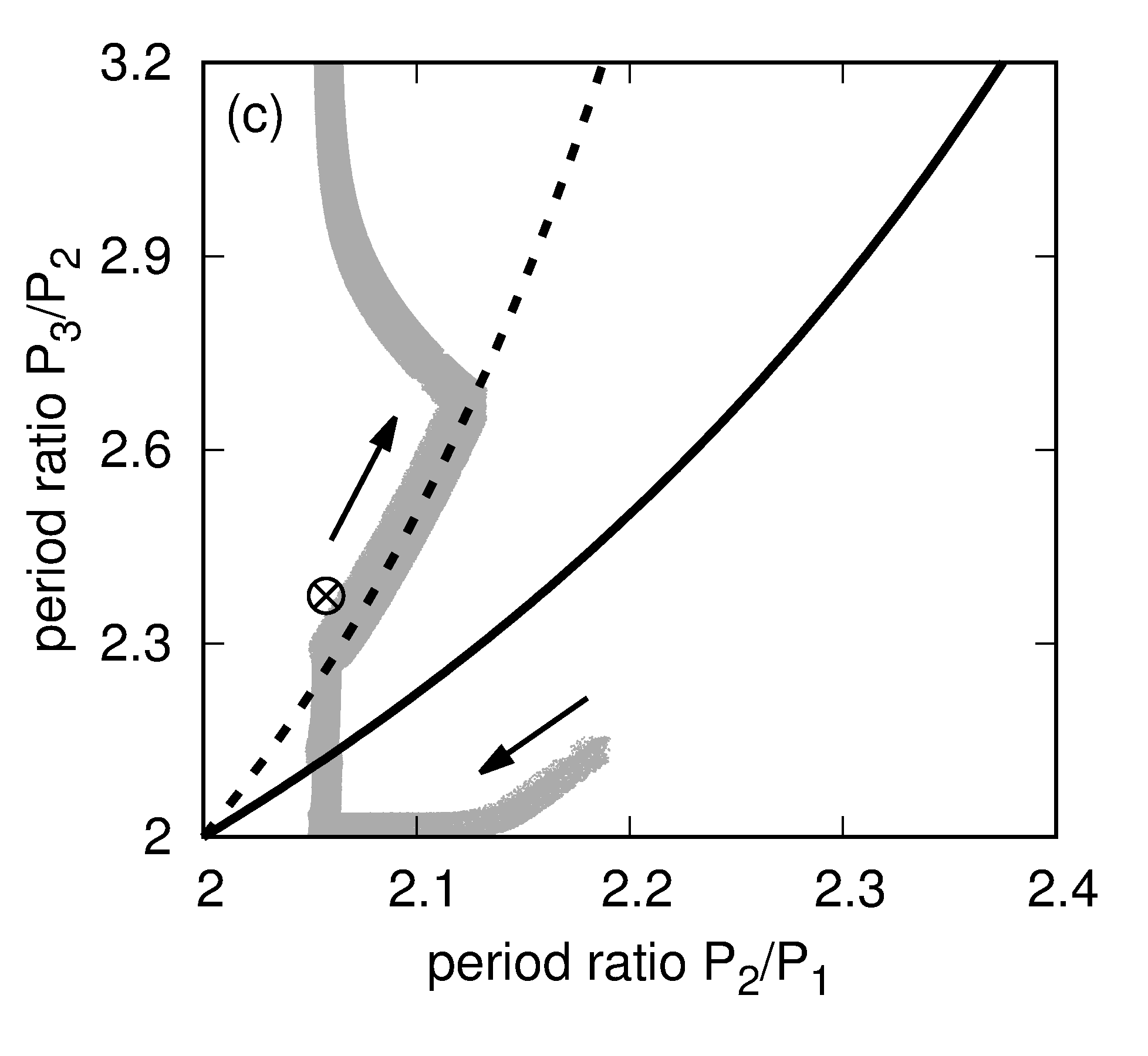}
}
\caption{
An illustration of the evolution of a system of masses of the best-fitting Kepler-30 system (listed in Tab.~\ref{tab:MCMCb}) resulting from the migration (first convergent, then divergent) of three different sets of the parameters. The migration is convergent for $t \lesssim t_0$ and divergent for $t \gtrsim t_0$ ($t_0 = 80\,$kyrs). The migration parameters for the convergent evolution regime are the same for all three simulations, i.e., $\kappa_1 = \kappa_2 = \kappa_3 = 100$, $\tau_1 = 100\,$kyrs, $\tau_2 = 77\,$kyrs, $\tau_3 = 59\,$kyrs. The only parameter that changes in time is $\tau_3$ (see Eq.~\ref{eq:tau3}). For subsequent simulations presented in panels~(a), (b) and~(c): $\tau_{3, \idm{div}} = 1100, 120, 130\,$kyrs, respectively ($\tau_{3, \idm{conv}} = 59\,$kyrs). The evolution is presented at the period ratio--period ratio diagram with a grey colour. The arrows point the evolution direction. Black solid and dashed curves show the evolution paths for a system in the Laplace resonance and the three-body resonance of the first order, respectively. See the text for details. The circle/cross symbol shows the position of Kepler-30 system at the diagram.
}
\label{fig:figure11} 
\end{figure*}
Figure~\ref{fig:figure11} presents the results of three simulations that differ one from another by the value of $\tau_{3,\idm{div}}$ only, therefore the convergent part of the evolution is the same (the values of the parameters are listed in the caption of Fig.~\ref{fig:figure11}). All three systems reach the chain of 2:1, 2:1~MMRs. The period ratio of the two inner planets stops at a value significantly different from $2$. As it is known, smooth migration of two planets leads to the evolution of the system along a branch of periodic orbits, that corresponds to the branch of equilibria of the averaged system \citep{Beauge2003,Beauge2006,Hadjidemetriou2006,Migaszewski2015}. Therefore the structure of the branch determines the relation between the eccentricities and the period ratio. Because $m_2/ m_1 \gg 1$, for a given value of $P_2/P_1$ the equilibrium values of the eccentricities $e_1 \gg e_2$. For $P_2/P_1 \sim 2.057$ $e_1$ is already relatively high $\sim 0.035$ and the disc-induced eccentricity damping balances the excitation caused by the resonance approach. The value of $\sim 2.057$ is an equilibrium value for the migration parameters used in the simulations. For all the planets $\kappa_i = 100$ which is a realistic value that stems from the hydrodynamical simulations as well as analytical models of planet--disc interactions \citep[e.g.,][]{Muto2011,Ketchum2011}.

After an increase of $\tau_3$ (the outermost planet starts to migrate slower than both inner planets) $P_3/P_2$ starts to increase, while $P_2/P_1$ keeps constant (see Fig.~\ref{fig:figure11}a). After some time $P_3/P_2$ gets close to the nominal value of Kepler-30 system. If, for some reason, the migration stopped there, the simulated system would resemble the observed configuration (marked with a cross/circle symbol). This result was obtained for $\tau_{3,\idm{div}} = 1100\,$kyrs, i.e., the migration of the outermost planet was almost stopped.

Nevertheless, the picture presented above is not the only one possible. The system can also enter a three-body resonance and then evolve along a curve of this resonance during the divergent migration \citep{Migaszewski2016}. Fig.~\ref{fig:figure11}b presents the results for $\tau_{3,\idm{div}} = 120\,$kyrs, that is only slightly more than $\tau_1$. Therefore the migration of the outermost planet is divergent with respect to the inner pair of planets, but the divergence is relatively slow. For $t > t_0$ the system moves at the beginning in a similar way than in the previous example, but then reaches a black solid curve and moves along it. The curve corresponds to the Laplace three-body resonance, for which the mean motions of the planets obey the relation $n_1 - 3\,n_2 + 2\,n_3 = 0$. A three-planet system that is initially involved in this chain of MMRs would evolve divergently along a curve $y^{-1} = 3/2 - x/2$, where $x \equiv P_2/P_1 = n_1/n_2$ and $y \equiv P_3/P_2 = n_2/n_3$ \citep{Migaszewski2016}. We can see that even though $\tau_1 > \tau_2$, $P_2/P_1$ increases and the observed configuration is omitted.

The third example presented in Fig.~\ref{fig:figure11}c has similar $\tau_{3,\idm{div}} = 130\,$kyrs, but the evolution differs significantly from the one illustrated in Fig.~\ref{fig:figure11}b. The system passes through the Laplace resonance, reaches a dashed curve and evolves further along it. The curve corresponds to the three-body resonance in which the mean motions vary according to the relation $2\,n_1 - 5\,n_2 + 2\,n_3 = 0$. The formula for this curve reads $y^{-1} = 5/2 - x$. A characteristic feature of this commensurability is that the relation is fulfilled for both $n_1/n_2 = 2$ and $n_2/n_3 = 2$, which is true also for the Laplace resonance. 

It can be shown that for a general three-body resonance $p_1 \, n_1 + p_2 \, n_2 + p_3 \, n_3 = 0$, where $p_1 + p_2 + p_3 = r$ ($p_1, p_2, p_3$ are non-zero integers and the order of the resonance $r$ is an integer $\geq 0$), there is only one combination of $p_1, p_2, p_3$ (apart from its multiplicities) for a given $r$ for which the relation is fulfilled for both $n_1/n_2 = 2$ and $n_2/n_3 = 2$. The resonance illustrated with a dashed curve is of the first order, while the Laplace resonance is of the zero order. The resonances of higher orders can be found as well, but none of them passes through the observed configuration of Kepler-30. In a series of simulations we did not observe that a system follows a resonance path different than the two illustrated in Fig.~\ref{fig:figure11}b,c. In conclusion of this part of the section we may say that Kepler-30 could have been formed within the scenario discussed here only for relatively fast divergence between the sub-system of the two inner orbits and the outer orbit. If the divergence was slower, the system would be trapped in one of the three-body resonances and would omit the observed configuration.
\begin{figure}
\vbox{
\hbox{\includegraphics[width=0.47\textwidth]{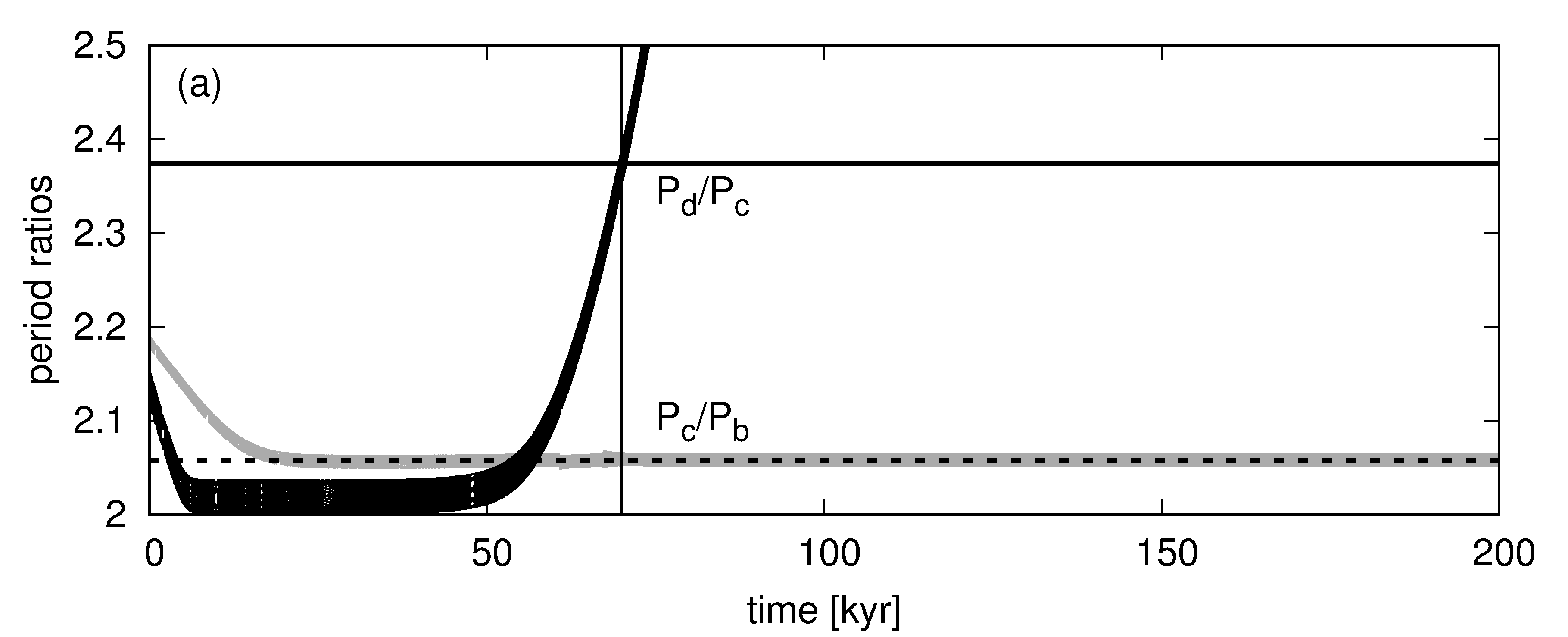}}
\hbox{\includegraphics[width=0.47\textwidth]{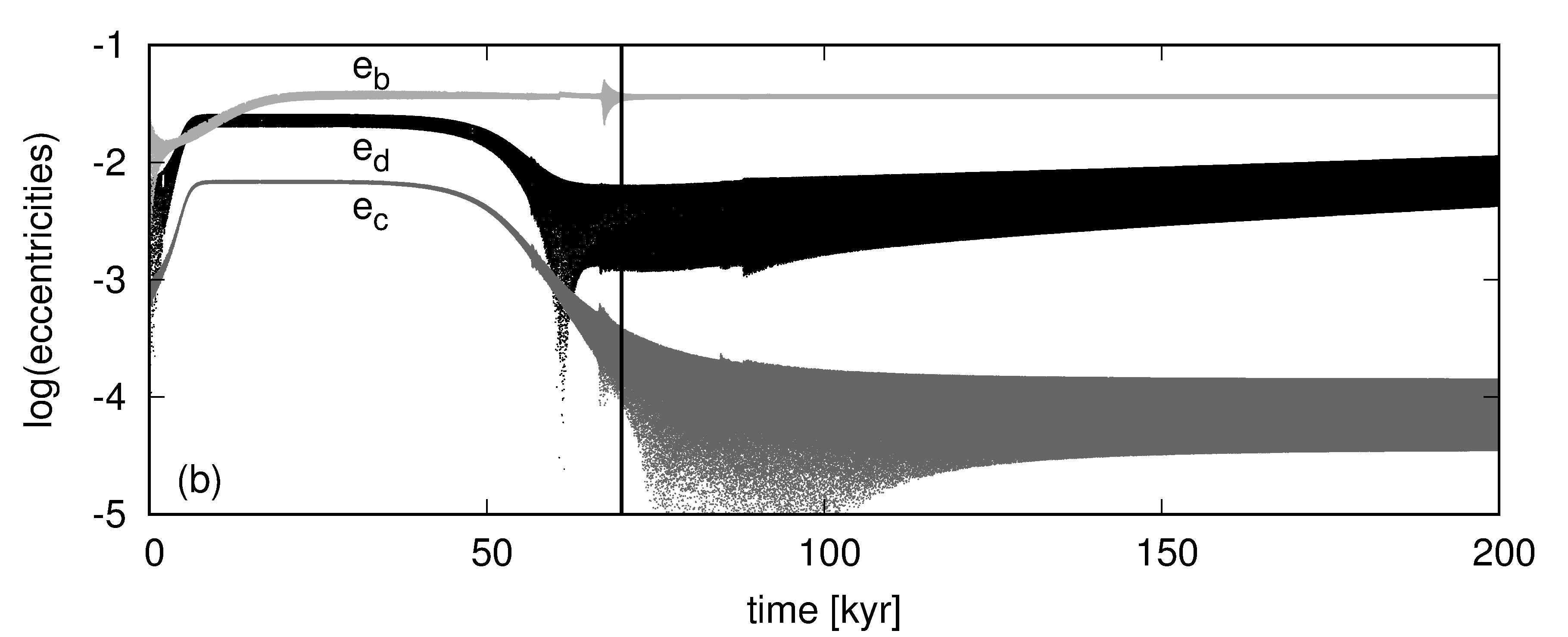}}
\hbox{\includegraphics[width=0.47\textwidth]{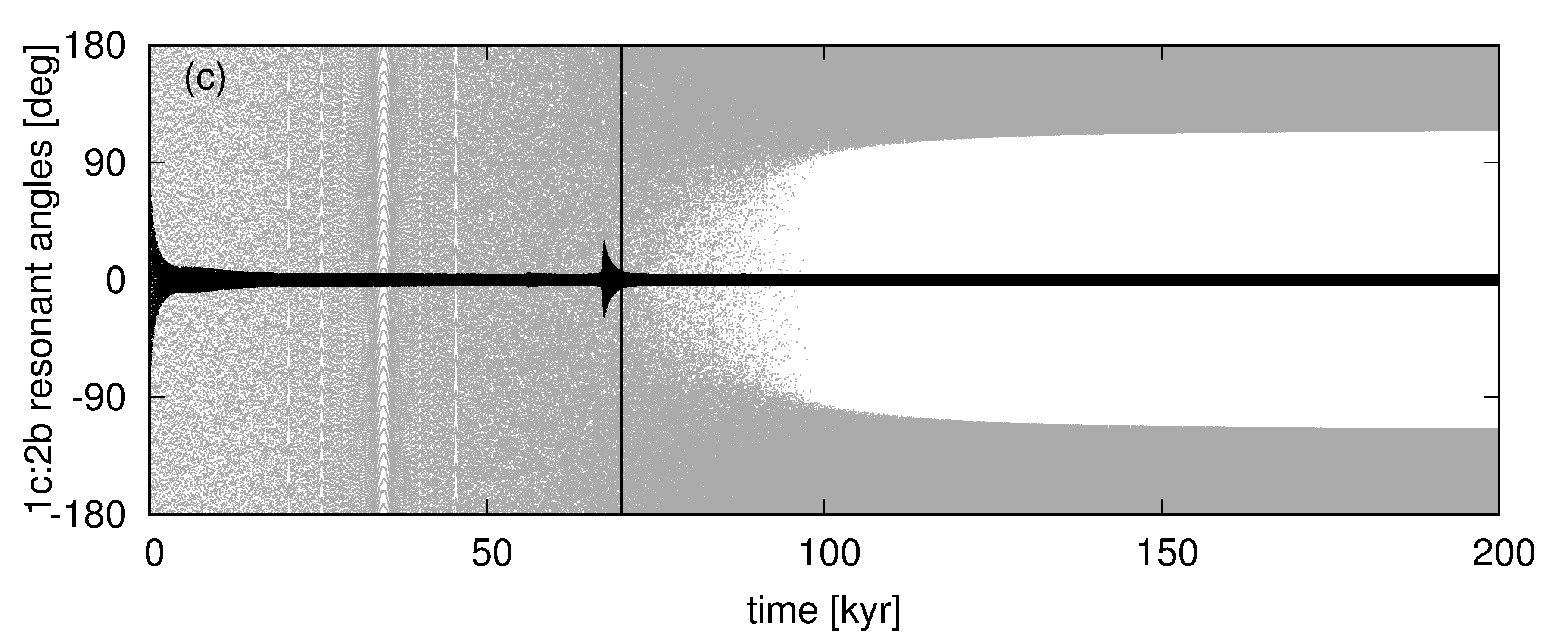}}
}
\caption{
The simulation illustrated in Fig.~\ref{fig:figure11}a presented as time evolution of the period ratios, eccentricities and the resonant angles of the 2:1~MMR between the first two planets. The evolution of particular parameters are shown with black or grey dots and labelled accordingly, apart from panel~(c), at which black dots represent the resonant angles $\phi_{2:1}^{(1)} = \lambda_1 - 2\,\lambda_2 + \varpi_1$, while grey dots denote $\phi_{2:1}^{(2)} = \lambda_1 - 2\,\lambda_2 + \varpi_2$. Horizontal lines at panel~(a) denote $P_2/P_1$ (dashed line) and $P_3/P_2$ (solid line) of Kepler-30. A vertical line points the time at which the simulated system has the same values of the period ratios as Kepler-30. 
}
\label{fig:figure12}
\end{figure}
The three simulations illustrated in Fig.~\ref{fig:figure11} are presented in Figs.~\ref{fig:figure12}, \ref{fig:figure13} and~\ref{fig:figure14} as time evolution of particular orbital elements. Subsequent figures correspond to panels~(a), (b) and~(c) of Fig.~\ref{fig:figure11}. 

The evolution of the period ratios for a simulation with $\tau_{3,\idm{div}} = 1100\,$kyrs is shown in Fig.~\ref{fig:figure12}a. After reaching the equilibrium values of $\sim 2.057$ for $P_2/P_1$ and $\sim 2$ for $P_3/P_2$ the period ratio of the outer pair starts to increase. Note that for $\tau_{3,\idm{div}}$ significantly greater than $\tau_1$, $\tau_3 > \tau_1$ for $t$ as low as $40\,$kyrs, thus the migration is divergent before $t = t_0$. At $t \sim 65\,$kyrs $P_3/P_2$ is close to the observed value. The period ratio $P_2/P_1$ does not change over the time of integration. Panel~(b) of Fig.~\ref{fig:figure12} shows the evolution of the eccentricities. They reach their equilibrium values after $20\,$kyrs and when the outermost planet starts to diverge from the inner pair, $e_2$ and $e_3$ decrease, while $e_1$ is constant. Interestingly, $e_3$ is kept at $\sim 0.01$ after initial decrease, even though $e_2$ decreases down to $0$ and $P_3/P_2$ increases up to very high values. This behaviour indicates that the structure of periodic orbits for the three-planet system is more complex than for the two-planet configuration. The last panel of Fig.~\ref{fig:figure12} illustrates the evolution of the resonant angles of 2:1~MMR between the inner pair of planets. The first resonant angle (shown with black colour) starts to librate around $0^{\circ}$ very quickly after the convergent migration begins. The second angle, however, starts librating around $180$ degrees when the outermost planet is far enough from the inner pair. For $t \sim 65\,$kyrs for which $P_3/P_2 \sim 2.374$ only the first angle librates, thus if the system was formed in the way described here, the inner pair of planets would be resonant but with only one critical angle librating.

\begin{figure}
\vbox{
\hbox{\includegraphics[width=0.47\textwidth]{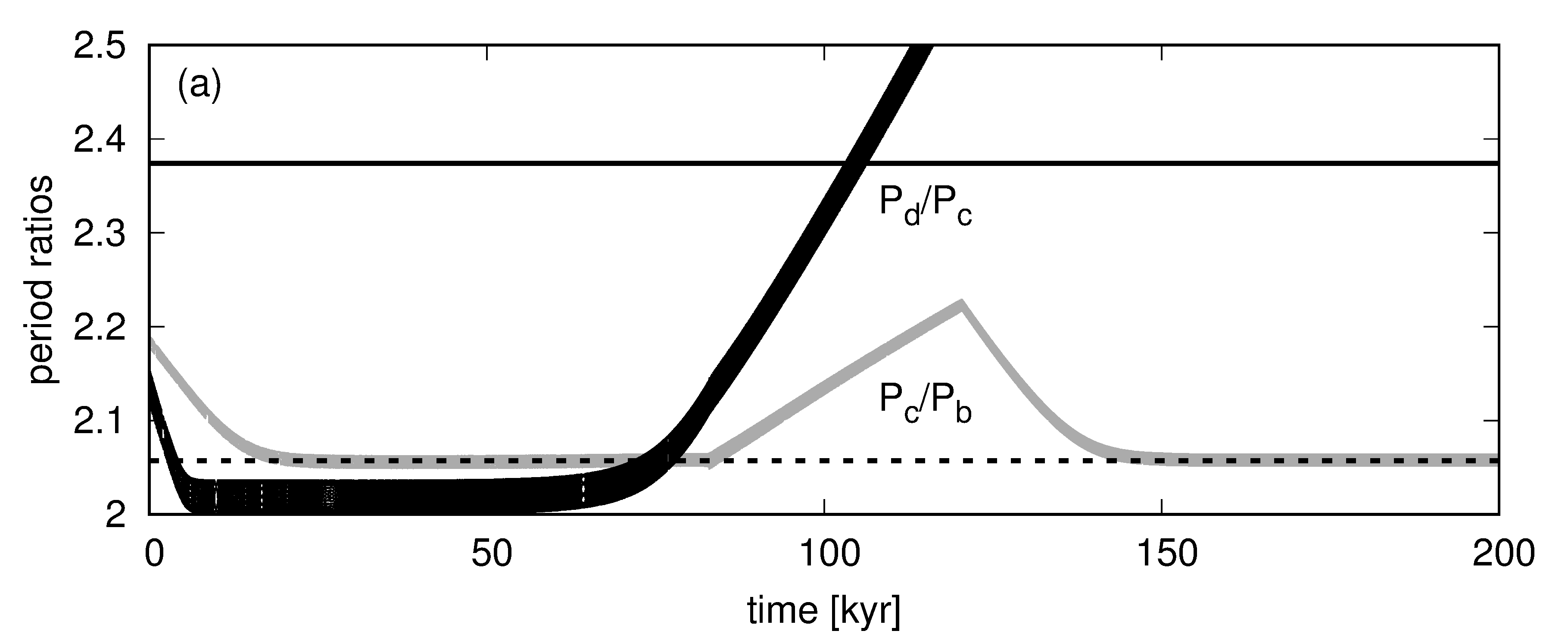}} 
\hbox{\includegraphics[width=0.47\textwidth]{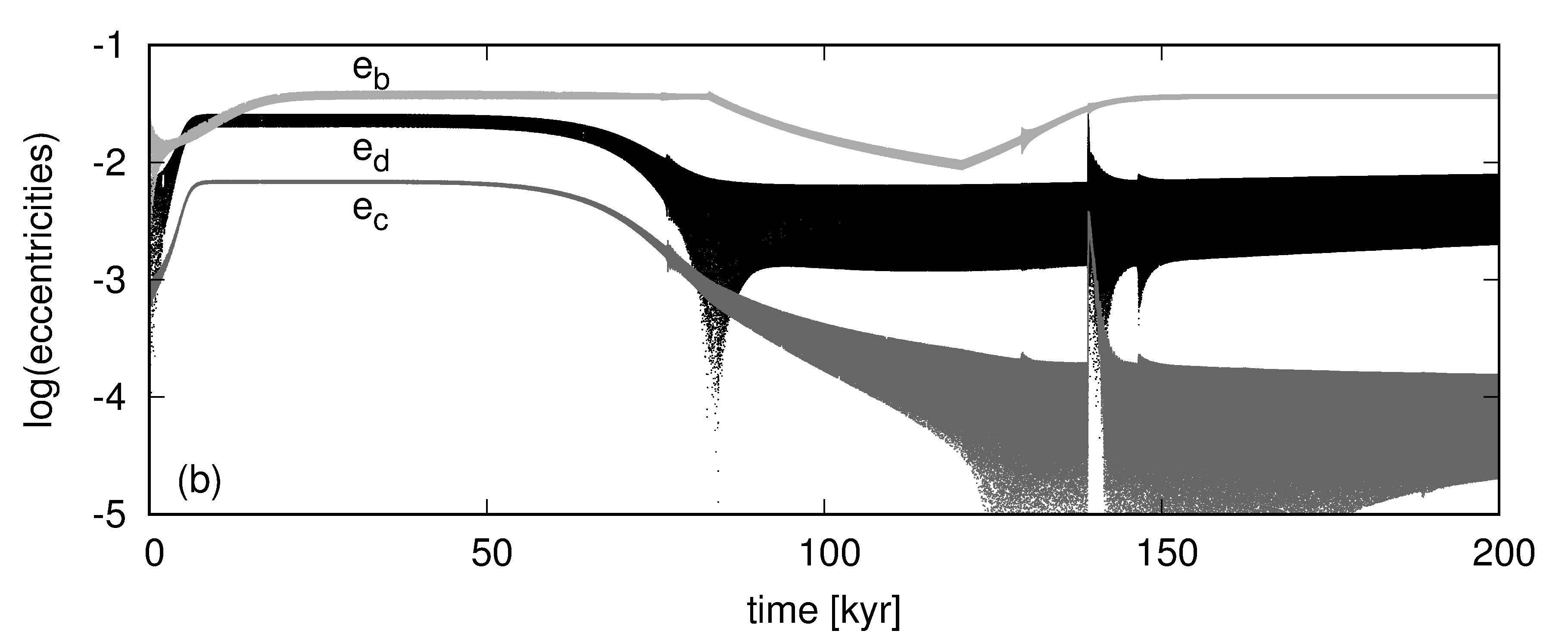}}
\hbox{\includegraphics[width=0.47\textwidth]{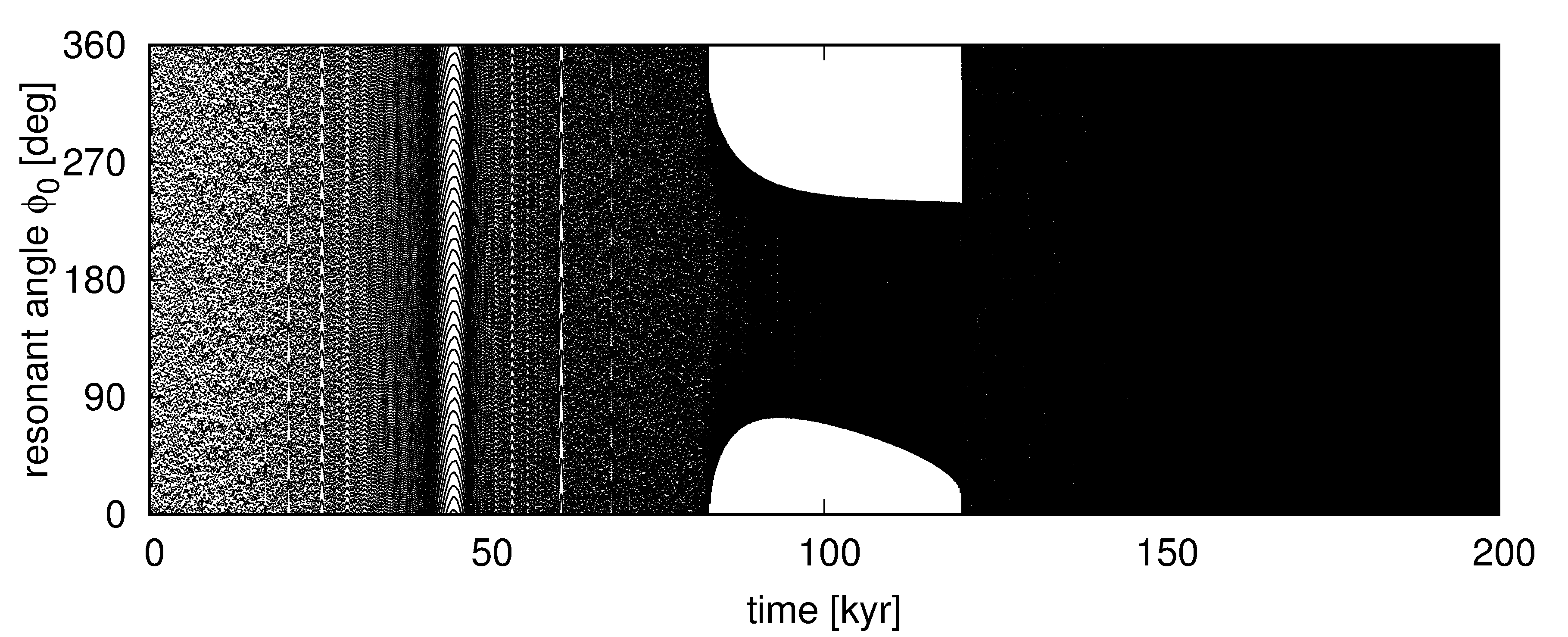}}
}
\caption{
The simulation illustrated in Fig.~\ref{fig:figure11}b presented in the same manner as in Fig.~\ref{fig:figure12}. Panel~(c) presents the evolution of the Laplace angle $\phi_0 = \lambda_1 - 3\,\lambda_2 + 2\,\lambda_3$.
}
\label{fig:figure13}
\end{figure}
Figure~\ref{fig:figure13} presents the results of simulation with $\tau_{3,\idm{div}} = 120\,$kyrs (see Fig.~\ref{fig:figure11}b for the evolution at the period ratio--period ratio diagram). Similarly to the previous example, after reaching the equilibrium values, $P_3/P_2$ starts to increase with $P_2/P_1$ being constant. At time $t \sim 90\,$kyrs $P_2/P_1$ starts to increase as well. It happens because the system enters the Laplace resonance, what can be observed in Fig.~\ref{fig:figure13}c. The Laplace resonant angle $\phi_0 = \lambda_1 - 3\,\lambda_2 + 2\,\lambda_3$ starts to librate around $180~$deg for $t \sim 90\,$kyrs. The libration amplitude increases and finally the system leaves the resonance, what can be observed at the period ratio--period ratio diagram (Fig.~\ref{fig:figure11}b). The evolution of the eccentricities (Fig.~\ref{fig:figure13}b) is similar to the example illustrated in Fig.~\ref{fig:figure12}, apart from the fact that after being locked in the Laplace resonance $e_1$ decreases, instead of being constant.

\begin{figure}
\vbox{
\hbox{\includegraphics[width=0.47\textwidth]{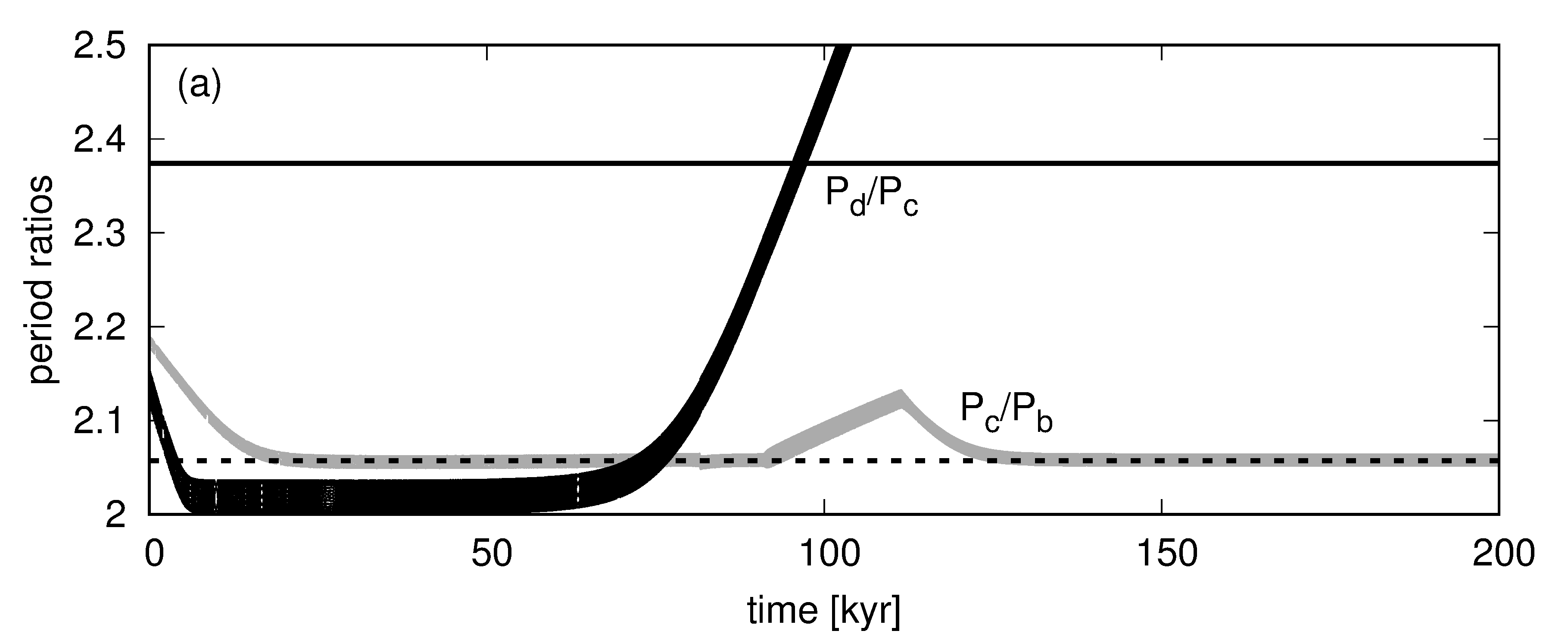}}
\hbox{\includegraphics[width=0.47\textwidth]{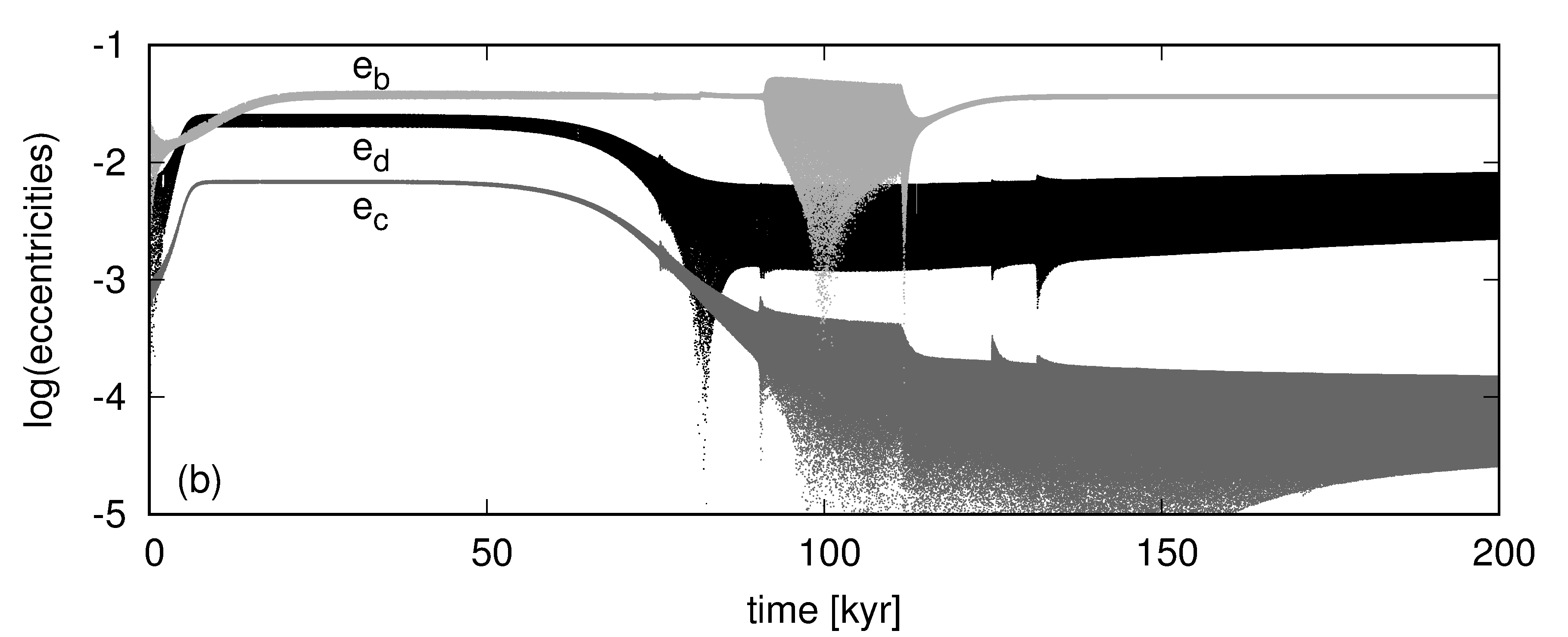}}
\hbox{\includegraphics[width=0.47\textwidth]{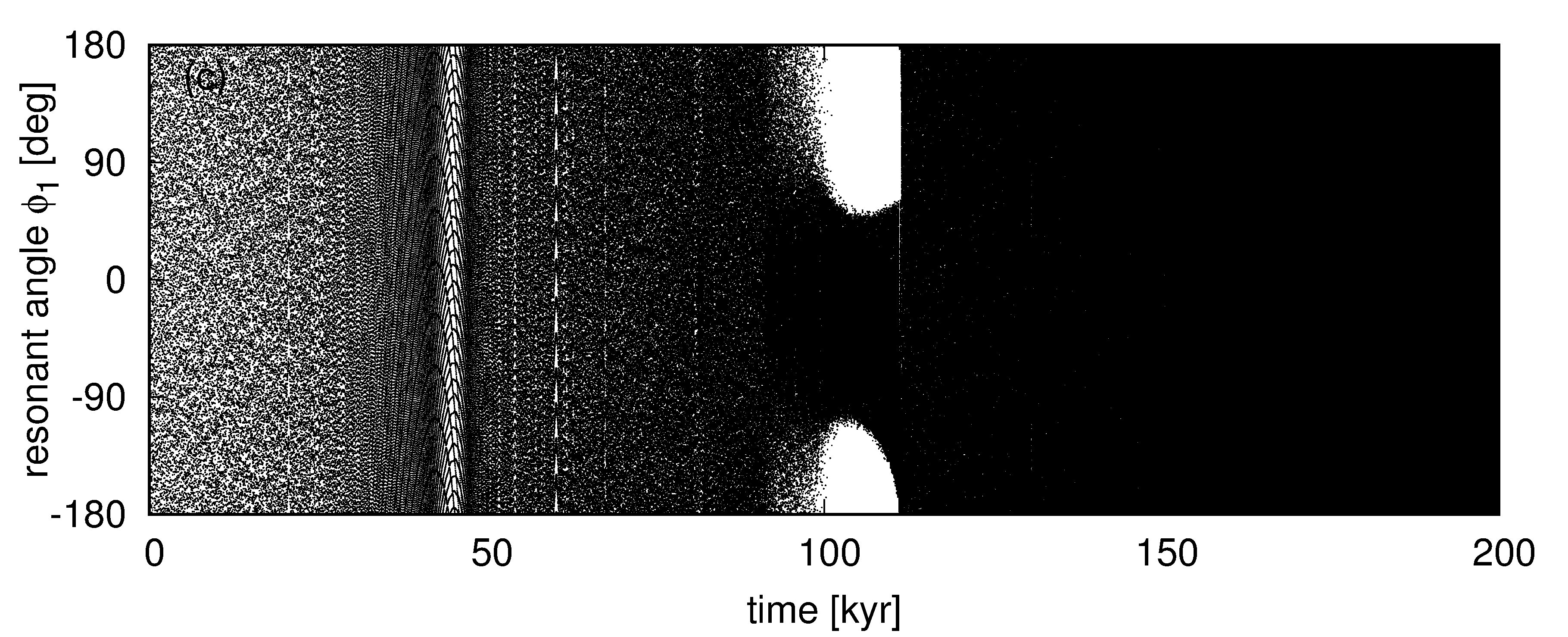}}
}
\caption{
The simulation illustrated in Fig.~\ref{fig:figure11}c presented in the same manner as in Fig.~\ref{fig:figure12}. Panel~(c) presents the evolution of a three-planet resonant angle of the first order $\phi_1 = 2\,\lambda_1 - 5\,\lambda_2 + 2\,\lambda_3 + \varpi_1$.
}
\label{fig:figure14}
\end{figure}
The last example of the evolution is illustrated in Fig.~\ref{fig:figure14}. The migration parameter $\tau_{3,\idm{div}} = 130\,$kyrs differs only slightly from the previous case, but the outcomes of the two simulations are significantly different. The system passes through the Laplace resonance without being locked in it. The period ratio $P_2/P_1$ starts to increase when the system enters a resonance whose critical angle $\phi_1 = 2\,\lambda_1 - 5\,\lambda_2 + 2\,\lambda_3 + \varpi_1$. The system stays in the first-order three-body MMR only for a time period of $\sim 20\,$kyrs (the resonant angle librates around $0$ for about $10\,$kyrs, see Fig.~\ref{fig:figure14}c, while the period ratios evolve along the resonant curve for about $20\,$kyrs) and then leaves it, with $P_2/P_1$ reaching again the equilibrium value and $P_3/P_2$ increasing further. After a series of simulations we find that the entrance in one of the two three-body resonances is a probabilistic process. Nevertheless, a deeper study is needed to find the probabilities of capturing the system in one of the two resonances during the divergent part of the evolution.

The analysis presented in this section shows that Kepler-30 system could have been formed within the scenario described here. The divergent migration of the outermost planet should be, however, fast enough, i.e., $P_3/P_2$ should increase at a rate high enough so the system avoids being trapped in one of the two three-body resonances. An interesting observation is that not only zero-order resonances, i.e., chains of two-body MMRs, can govern the divergent evolution of three-planet configurations, as shown in \cite{Migaszewski2016}. Other resonances, that are not chains of two-body MMRs, can work this way as well. It is a matter of further studies to investigate how important those resonances are in a process of formation of the period ratios distribution of the observed configurations, that is known to lack of significant picks at the resonant values \citep{Fabrycky2014}.

\section{Conclusions}
\label{conclusion}
%
The Kepler-30 planetary system is formed of three planets within the orbital radius of Mercury in the Solar System. 
Our work complements and extends the results reported in early papers regarding this system \citep{Fabrycky2012,SanchisOjeda2012}. It is important to stress that these works are based on the analysis of the Q1-Q6 quarters of \kepler{} photometry and  they also did  not focus on a dynamical analysis of this very compact planetary configuration. 
\corre{Moreover, we find that the TTVs datasets, spanning the Q1-Q17 \kepler{} quarters, in two recent TTV catalogues \citep{Rowe2015,Holczer2016} significantly differ from each other for the innermost planet Kepler-30b.}

Therefore, after a preliminary analysis of the TTV measurements in \cite{Holczer2016}, we recomputed transit parameters and TTVs from pre-reduced PDCSAP \kepler{} light curves of Kepler-30 available in the MAST archive. \corre{We could not identify one transit of Kepler-30b present in the earlier paper, since this particular transit seems to be missing in the PDCSAP light curves}. We determined new mid-transit times for all other events. Yet we obtained roughly two times larger uncertainties of these moments, as compared to the \cite{Holczer2016} catalogue. Regarding the transit parameters, we derived new estimates of the star-relative radii of the planets which are crucial to determine bulk densities  and to characterize their internal structure.

The large TTV signal of the innermost planet and the high-quality \kepler{} photometry make it possible to comprehensively characterize the Kepler-30 system, regarding  precise estimates of  masses of the planets. The masses of the two inner planets are determined with the formal uncertainty of $\simeq 1\%$ and the mass of the outermost planet is uncertain to $6\%$ only. These masses are smaller by $25\%$ and $20\%$, respectively, than reported in \cite{SanchisOjeda2012}. 

Having our new determination of the masses and radii, we find that the Kepler-30 system is composed of three planets \corre{with different bulk density}, i.e., an innermost mini-Neptune, a large Jupiter in the middle, and a very low density outermost Neptune with a radius twice as large as Saturn's. Among \kepler{}-detected planetary systems, Kepler-30 exhibits peculiar proprieties such as a wide range of bulk densities and, very likely, different internal compositions of the planets.  Our updated transit parameters may change the classification of the Kepler-30b. In \cite{Rowe2015}, the radius of this planet $1.9~\RE$ is quite different from  $3.9\pm0.2~\RE$ reported in \cite{SanchisOjeda2012}. Almost two times smaller radius implied that Kepler-30b is a Super-Earth rocky planet. Our results  favor a radius of 3.75$\pm 0.18~\RE$, much closer to the estimate in \cite{SanchisOjeda2012}.  With a mass of $9.2 \pm 0.1~\mE$  and a bulk density $(0.96\pm0.15)$\,\gccm{}, slightly smaller than the water, we may classify Kepler-30b as a mini-Neptune planet. 

The robust constrain on masses and orbital parameters derived with the Bayesian statistics make it possible to investigate the dynamical state of this system.  We considered four variants of the TTV coplanar model (Eq.~\ref{eq:Lfun}), with and without Gaussian priors set to Poincar\'e variables (eccentricities), as well as with and without statistical correction (error floor) of the TTVs uncertainties. A robust and self-consistent best-fitting solution to these models yields meaningfully non-zero eccentricities of $e_1\simeq 0.077$ for Kepler-30b, $e_2 \simeq 0.012$ for Kepler-30c, and  $e_3 \simeq 0.025$ for Kepler-30d, respectively. Moreover, our estimate of the Kepler-30b eccentricity is almost twice as large as reported in \cite{SanchisOjeda2012}.  We find that the Kepler-30 system is a very compact configuration of three relatively massive planets. 

\corre{We illustrate a complex dynamical structure of the phase space in the vicinity of the best-fitting TTVs models with the help of the fast indicators technique.} Strong mutual interactions between a Jovian planet almost two times more massive than Jupiter and two smaller Neptune-like planets create a complex net of two-body and three-body MMRs. Curiously, although the planets are pairwise close  to the 2:1~MMR (or to three-body Laplace resonance), \corre{the system seems to be} separated from these MMRs in the dynamical sense. 

\corre{The overlap of various 2-body and 3-body MMRs involving all planets
generates particularly complex and fine structure of the inner 2:1~MMR.   The fast indicators technique combined with the direct numerical integration, frequency analysis and scans of the critical arguments helped us to determine the width and dynamical properties of this resonance.  It is  characterized  by only one librating resonant argument.  A separation of the inner pair of planets in the best-fitting model from the 2:1 MMR is subtle, since the Kepler-30 system ``resides'' close to its border determined by the libration of one critical argument. Yet we found that the closest low-order resonance to the observed system is the three-body (1:-5:7)~MMR}.
The system may be deviated  from the exact resonance due to the density waves-planets interaction when a Jovian planet in a packed configuration excites density waves inside a protoplanetary disk \citep[]{PodlewskaGaca2012}.

We further study the formation of the Kepler-30 system using a simple prescription of the planet-disk interactions. We find that the Kepler-30 planets, once involved in a resonant chain, can evolve divergently and achieve an orbital configuration closely resembling the observed system. During the divergent evolution the system may be trapped in a three-body resonance of the zero or the first order, that would result in omitting the observed configuration. In order to avoid that, $P_3/P_2$ should increase relatively fast, i.e., the outermost planet should migrate an order of magnitude slower than the inner pair.

Comparing the Kepler-30 system with other multiple systems in the \kepler{}-sample, \corre{we refer to the recent papers by \cite{Weiss2017} and \cite{Millholland2017}} who predict that statistically these systems are composed of similar planets. This is however not the case for systems such as Kepler-36 or Kepler-10, and Kepler-30. The Kepler-30 system is unusual, since it is formed of three physically different planets with masses spanning almost two orders of magnitude and bulk densities spanning one order of magnitude. These characteristics could make the Kepler-30 system a challenging test-bed for studying the planet formation theory. 

\corre{In this paper we present a  comprehensive characterization of a peculiar multi-planet, extremely compact \kepler{} system.} 
Our work includes the \kepler{} light curves analysis, determination of dynamical masses and orbital parameters optimization, as well as the long-term stability analysis. We also interpret the initial conditions and a likely orbital setup on the grounds of the planetary migration. 
%
%
\section{Acknowledgements}
%
We thank the anonymous reviewer for comments that improved this paper. F.~P. is grateful to N\'e{}stor Espinoza for enlightening discussions on the light curve analysis. K.~G. thanks the staff of the Pozna\'n Supercomputer and Network Centre (PCSS, Poland) for generous and continuous support, and for providing computing resources (grant No.~313). This work has been supported by Polish National Science Centre MAESTRO grant DEC-2012/06/A/ST9/00276.
%
%
\bibliographystyle{mn2e}
\bibliography{ms}

\begin{thebibliography}{}

\bibitem[\protect\citeauthoryear{{Agol}, {Steffen}, {Sari} \&
  {Clarkson}}{{Agol} et~al.}{2005}]{Agol2005}
{Agol} E.,  {Steffen} J.,  {Sari} R.,    {Clarkson} W.,  2005, \mnras, 359, 567

\bibitem[\protect\citeauthoryear{{Ali-Dib} \& {Lakhlani}}{{Ali-Dib} \&
  {Lakhlani}}{2018}]{AliDib2018}
{Ali-Dib} M.,  {Lakhlani} G.,  2018, \mnras, 473, 1325

\bibitem[\protect\citeauthoryear{{Baluev}}{{Baluev}}{2008}]{Baluev2008}
{Baluev} R.~V.,  2008, Celestial Mechanics and Dynamical Astronomy, 102, 297

\bibitem[\protect\citeauthoryear{{Barros}, {D{\'{\i}}az}, {Santerne}, {Bruno},
  {Deleuil}, {Almenara}, {Bonomo}, {Bouchy}, {Damiani}, {H{\'e}brard},
  {Montagnier} \& {Moutou}}{{Barros} et~al.}{2014}]{Barros2014}
{Barros} S.~C.~C.,  {D{\'{\i}}az} R.~F.,  {Santerne} A.,  {Bruno} G.,
  {Deleuil} M.,  {Almenara} J.-M.,  {Bonomo} A.~S.,  {Bouchy} F.,  {Damiani}
  C.,  {H{\'e}brard} G.,  {Montagnier} G.,    {Moutou} C.,  2014, \aap, 561, L1

\bibitem[\protect\citeauthoryear{{Baruteau} \& {Papaloizou}}{{Baruteau} \&
  {Papaloizou}}{2013}]{Baruteau2013}
{Baruteau} C.,  {Papaloizou} J.~C.~B.,  2013, \apj, 778, 7

\bibitem[\protect\citeauthoryear{{Bayliss}, {Hartman}, {Bakos}, {Penev},
  {Zhou}, {Brahm}, {Rabus}, {Jord{\'a}n}, {Mancini}, {de Val-Borro}, {Bhatti},
  {Espinoza} \& et al.}{{Bayliss} et~al.}{2015}]{Bayliss2015}
{Bayliss} D.,  {Hartman} J.~D.,  {Bakos} G.~{\'A}.,  {Penev} K.,  {Zhou} G.,
  {Brahm} R.,  {Rabus} M.,  {Jord{\'a}n} A.,  {Mancini} L.,  {de Val-Borro} M.,
   {Bhatti} W.,  {Espinoza} N.,    et al. 2015, \aj, 150, 49

\bibitem[\protect\citeauthoryear{{Beaug{\'e}}, {Ferraz-Mello} \&
  {Michtchenko}}{{Beaug{\'e}} et~al.}{2003}]{Beauge2003}
{Beaug{\'e}} C.,  {Ferraz-Mello} S.,    {Michtchenko} T.~A.,  2003, \apj, 593,
  1124

\bibitem[\protect\citeauthoryear{{Beaug{\'e}}, {Michtchenko} \&
  {Ferraz-Mello}}{{Beaug{\'e}} et~al.}{2006}]{Beauge2006}
{Beaug{\'e}} C.,  {Michtchenko} T.~A.,    {Ferraz-Mello} S.,  2006, \mnras,
  365, 1160

\bibitem[\protect\citeauthoryear{{Becker}, {Vanderburg}, {Adams}, {Rappaport}
  \& {Schwengeler}}{{Becker} et~al.}{2015}]{Becker2015}
{Becker} J.~C.,  {Vanderburg} A.,  {Adams} F.~C.,  {Rappaport} S.~A.,
  {Schwengeler} H.~M.,  2015, \apjl, 812, L18

\bibitem[\protect\citeauthoryear{{Benettin}, {Galgani}, {Giorgilli} \&
  {Strelcyn}}{{Benettin} et~al.}{1980}]{Benettin1980}
{Benettin} G.,  {Galgani} L.,  {Giorgilli} A.,    {Strelcyn} J.-M.,  1980,
  Meccanica, 15, 9

\bibitem[\protect\citeauthoryear{{Bieryla}, {Hartman}, {Bakos}, {Bhatti},
  {Kov{\'a}cs} \& et al.}{{Bieryla} et~al.}{2014}]{Bieryla2014}
{Bieryla} A.,  {Hartman} J.~D.,  {Bakos} G.~{\'A}.,  {Bhatti} W.,  {Kov{\'a}cs}
  G.,    et al. 2014, \aj, 147, 84

\bibitem[\protect\citeauthoryear{{Borsato}, {Marzari}, {Nascimbeni}, {Piotto},
  {Granata}, {Bedin} \& {Malavolta}}{{Borsato} et~al.}{2014}]{Borsato2014}
{Borsato} L.,  {Marzari} F.,  {Nascimbeni} V.,  {Piotto} G.,  {Granata} V.,
  {Bedin} L.~R.,    {Malavolta} L.,  2014, \aap, 571, A38

\bibitem[\protect\citeauthoryear{{Carter}, {Agol}, {Chaplin}, {Basu},
  {Bedding}, {Buchhave}, {Christensen-Dalsgaard}, {Deck}, {Elsworth},
  {Fabrycky} \& {Ford}}{{Carter} et~al.}{2012}]{Carter2012}
{Carter} J.~A.,  {Agol} E.,  {Chaplin} W.~J.,  {Basu} S.,  {Bedding} T.~R.,
  {Buchhave} L.~A.,  {Christensen-Dalsgaard} J.,  {Deck} K.~M.,  {Elsworth} Y.,
   {Fabrycky} D.~C.,    {Ford} e.~a.,  2012, Science, 337, 556

\bibitem[\protect\citeauthoryear{{Cincotta} \& {Sim{\'o}}}{{Cincotta} \&
  {Sim{\'o}}}{2000}]{Cincotta2000}
{Cincotta} P.~M.,  {Sim{\'o}} C.,  2000, \aaps, 147, 205

\bibitem[\protect\citeauthoryear{{Conti}, {Rogers}, {Donaldson}, {Shiao},
  {Christian}, {Levay}, {White} \& {MAST Team}}{{Conti}
  et~al.}{2011}]{Conti2011}
{Conti} A.,  {Rogers} A.,  {Donaldson} T.,  {Shiao} B.,  {Christian} C.,
  {Levay} K.,  {White} R.,    {MAST Team} 2011, in American Astronomical
  Society Meeting Abstracts \#217 Vol.~43 of Bulletin of the American
  Astronomical Society, {The New MAST Portal}.
p. 344.07

\bibitem[\protect\citeauthoryear{{Cubillos}, {Erkaev}, {Juvan}, {Fossati},
  {Johnstone}, {Lammer}, {Lendl}, {Odert} \& {Kislyakova}}{{Cubillos}
  et~al.}{2017}]{Cubillos2017}
{Cubillos} P.,  {Erkaev} N.~V.,  {Juvan} I.,  {Fossati} L.,  {Johnstone} C.~P.,
   {Lammer} H.,  {Lendl} M.,  {Odert} P.,    {Kislyakova} K.~G.,  2017, \mnras,
  466, 1868

\bibitem[\protect\citeauthoryear{{Deck} \& {Agol}}{{Deck} \&
  {Agol}}{2015}]{Deck2015}
{Deck} K.~M.,  {Agol} E.,  2015, \apj, 802, 116

\bibitem[\protect\citeauthoryear{{Deleuil}, {Almenara}, {Santerne}, {Barros},
  {Havel}, {H{\'e}brard}, {Bonomo}, {Bouchy} \& et al.}{{Deleuil}
  et~al.}{2014}]{Deleuil2014}
{Deleuil} M.,  {Almenara} J.-M.,  {Santerne} A.,  {Barros} S.~C.~C.,  {Havel}
  M.,  {H{\'e}brard} G.,  {Bonomo} A.~S.,  {Bouchy} F.,    et al. 2014, \aap,
  564, A56

\bibitem[\protect\citeauthoryear{{Delrez}, {Van Grootel}, {Anderson},
  {Collier-Cameron}, {Doyle}, {Fumel}, {Gillon}, {Hellier} \& et al.}{{Delrez}
  et~al.}{2014}]{Delrez2014}
{Delrez} L.,  {Van Grootel} V.,  {Anderson} D.~R.,  {Collier-Cameron} A.,
  {Doyle} A.~P.,  {Fumel} A.,  {Gillon} M.,  {Hellier} C.,    et al. 2014,
  \aap, 563, A143

\bibitem[\protect\citeauthoryear{{Dreizler} \& {Ofir}}{{Dreizler} \&
  {Ofir}}{2014}]{Dreizler2014}
{Dreizler} S.,  {Ofir} A.,  2014, preprint (arXiv:1403.1372)

\bibitem[\protect\citeauthoryear{{Dumusque}, {Bonomo}, {Haywood}, {Malavolta},
  {S{\'e}gransan}, {Buchhave}, {Collier Cameron}, {Latham} \&
  {Molinari}}{{Dumusque} et~al.}{2014}]{Dumusque2014}
{Dumusque} X.,  {Bonomo} A.~S.,  {Haywood} R.~D.,  {Malavolta} L.,
  {S{\'e}gransan} D.,  {Buchhave} L.~A.,  {Collier Cameron} A.,  {Latham}
  D.~W.,    {Molinari} e.~a.,  2014, \apj, 789, 154

\bibitem[\protect\citeauthoryear{{Espinoza}, {Brahm}, {Jord{\'a}n}, {Jenkins},
  {Rojas}, {Jofr{\'e}}, {M{\"a}dler}, {Rabus}, {Chanam{\'e}}, {Pantoja},
  {Soto}, {Morzinski}, {Males}, {Ward-Duong} \& {Close}}{{Espinoza}
  et~al.}{2016}]{Espinoza2016}
{Espinoza} N.,  {Brahm} R.,  {Jord{\'a}n} A.,  {Jenkins} J.~S.,  {Rojas} F.,
  {Jofr{\'e}} P.,  {M{\"a}dler} T.,  {Rabus} M.,  {Chanam{\'e}} J.,  {Pantoja}
  B.,  {Soto} M.~G.,  {Morzinski} K.~M.,  {Males} J.~R.,  {Ward-Duong} K.,
  {Close} L.~M.,  2016, \apj, 830, 43

\bibitem[\protect\citeauthoryear{{Espinoza} \& {Jord{\'a}n}}{{Espinoza} \&
  {Jord{\'a}n}}{2015}]{Espinoza2015}
{Espinoza} N.,  {Jord{\'a}n} A.,  2015, \mnras, 450, 1879

\bibitem[\protect\citeauthoryear{{Fabrycky}, {Lissauer}, {Ragozzine}, {Rowe},
  {Steffen}, {Agol}, {Barclay}, {Batalha} \& {Borucki} W.~J.}{{Fabrycky}
  et~al.}{2014}]{Fabrycky2014}
{Fabrycky} D.~C.,  {Lissauer} J.~J.,  {Ragozzine} D.,  {Rowe} J.~F.,  {Steffen}
  J.~H.,  {Agol} E.,  {Barclay} T.,  {Batalha} N.,    {Borucki} W.~J. e.~a.,
  2014, \apj, 790, 146

\bibitem[\protect\citeauthoryear{{Fabrycky et al.}}{{Fabrycky et
  al.}}{2012}]{Fabrycky2012}
{Fabrycky et al.} 2012, \apj, 750, 114

\bibitem[\protect\citeauthoryear{{Faranda}, {Mestre} \& {Turchetti}}{{Faranda}
  et~al.}{2012}]{Faranda2012}
{Faranda} D.,  {Mestre} M.~F.,    {Turchetti} G.,  2012, International Journal
  of Bifurcation and Chaos, 22, 1250215

\bibitem[\protect\citeauthoryear{Foreman-Mackey}{Foreman-Mackey}{2016}]{corner}
Foreman-Mackey D.,  2016, The Journal of Open Source Software, 24

\bibitem[\protect\citeauthoryear{{Foreman-Mackey}, {Hogg}, {Lang} \&
  {Goodman}}{{Foreman-Mackey} et~al.}{2013}]{ForemanMackey2013}
{Foreman-Mackey} D.,  {Hogg} D.~W.,  {Lang} D.,    {Goodman} J.,  2013, \pasp,
  125, 306

\bibitem[\protect\citeauthoryear{{Gajdo{\v s}}, {Parimucha}, {Hamb{\'a}lek} \&
  {Va{\v n}ko}}{{Gajdo{\v s}} et~al.}{2017}]{Gajdos2017}
{Gajdo{\v s}} P.,  {Parimucha} {\v S}.,  {Hamb{\'a}lek} {\L}.,    {Va{\v n}ko}
  M.,  2017, \mnras, 469, 2907

\bibitem[\protect\citeauthoryear{{Goodman} \& {Weare}}{{Goodman} \&
  {Weare}}{2010}]{Goodman2010}
{Goodman} J.,  {Weare} J.,  2010, Communications in Applied Mathematics and
  Computational Science, Vol.~5, No.~1, p.~65-80, 2010, 5, 65

\bibitem[\protect\citeauthoryear{{Go{\'z}dziewski}, {Bois}, {Maciejewski} \&
  {Kiseleva-Eggleton}}{{Go{\'z}dziewski} et~al.}{2001}]{Gozdziewski2001}
{Go{\'z}dziewski} K.,  {Bois} E.,  {Maciejewski} A.~J.,    {Kiseleva-Eggleton}
  L.,  2001, \aap, 378, 569

\bibitem[\protect\citeauthoryear{{Go{\'z}dziewski}, {Breiter} \&
  {Borczyk}}{{Go{\'z}dziewski} et~al.}{2008}]{Gozdziewski2008}
{Go{\'z}dziewski} K.,  {Breiter} S.,    {Borczyk} W.,  2008, \mnras, 383, 989

\bibitem[\protect\citeauthoryear{{Go{\'z}dziewski}, {Migaszewski}, {Panichi} \&
  {Szuszkiewicz}}{{Go{\'z}dziewski} et~al.}{2016}]{Gozdziewski2016}
{Go{\'z}dziewski} K.,  {Migaszewski} C.,  {Panichi} F.,    {Szuszkiewicz} E.,
  2016, \mnras, 455, L104

\bibitem[\protect\citeauthoryear{{Go{\'z}dziewski}, {S{\l}onina}, {Migaszewski}
  \& {Rozenkiewicz}}{{Go{\'z}dziewski} et~al.}{2013}]{Gozdziewski2013}
{Go{\'z}dziewski} K.,  {S{\l}onina} M.,  {Migaszewski} C.,    {Rozenkiewicz}
  A.,  2013, \mnras, 430, 533

\bibitem[\protect\citeauthoryear{{Grziwa}, {Gandolfi}, {Csizmadia}, {Fridlund},
  {Parviainen}, {Deeg}, {Cabrera} \& et al.}{{Grziwa}
  et~al.}{2016}]{Grziwa2016}
{Grziwa} S.,  {Gandolfi} D.,  {Csizmadia} S.,  {Fridlund} M.,  {Parviainen} H.,
   {Deeg} H.~J.,  {Cabrera} J.,    et al. 2016, \aj, 152, 132

\bibitem[\protect\citeauthoryear{{Guzzo}}{{Guzzo}}{2005}]{Guzzo2005}
{Guzzo} M.,  2005, \icarus, 174, 273

\bibitem[\protect\citeauthoryear{{Hadden} \& {Lithwick}}{{Hadden} \&
  {Lithwick}}{2017}]{Hadden2017}
{Hadden} S.,  {Lithwick} Y.,  2017, \aj, 154, 5

\bibitem[\protect\citeauthoryear{{Hadjidemetriou}}{{Hadjidemetriou}}{2006}]{Hadjidemetriou2006}
{Hadjidemetriou} J.~D.,  2006, Celestial Mechanics and Dynamical Astronomy, 95,
  225

\bibitem[\protect\citeauthoryear{{Hartman}, {Bakos}, {Sato}, {Torres}, {Noyes},
  {Latham}, {Kov{\'a}cs}, {Fischer}, {Howard}, {Johnson} \& et al.}{{Hartman}
  et~al.}{2011}]{Hartman2011}
{Hartman} J.~D.,  {Bakos} G.~{\'A}.,  {Sato} B.,  {Torres} G.,  {Noyes} R.~W.,
  {Latham} D.~W.,  {Kov{\'a}cs} G.,  {Fischer} D.~A.,  {Howard} A.~W.,
  {Johnson} J.~A.,    et al. 2011, \apj, 726, 52

\bibitem[\protect\citeauthoryear{{Holczer}, {Mazeh}, {Nachmani},
  {Jontof-Hutter}, {Ford}, D. \& et al.}{{Holczer} et~al.}{2016}]{Holczer2016}
{Holczer} T.,  {Mazeh} T.,  {Nachmani} G.,  {Jontof-Hutter} D.,  {Ford} E.~B.,
  D. F.,    et al. 2016, The Astrophysical Journal Supplement Series, 225, 9

\bibitem[\protect\citeauthoryear{{Holman}, {Fabrycky}, {Ragozzine}, {Ford},
  {Steffen}, {Welsh}, {Lissauer}, {Latham}, {Marcy}, {Walkowicz} \& et
  al.}{{Holman} et~al.}{2010}]{Holman2010}
{Holman} M.~J.,  {Fabrycky} D.~C.,  {Ragozzine} D.,  {Ford} E.~B.,  {Steffen}
  J.~H.,  {Welsh} W.~F.,  {Lissauer} J.~J.,  {Latham} D.~W.,  {Marcy} G.~W.,
  {Walkowicz} L.~M.,    et al. 2010, Science, 330, 51

\bibitem[\protect\citeauthoryear{{Holman} \& {Murray}}{{Holman} \&
  {Murray}}{2005}]{Holman2005}
{Holman} M.~J.,  {Murray} N.~W.,  2005, Science, 307, 1288

\bibitem[\protect\citeauthoryear{{Johnson}, {Petigura}, {Fulton}, {Marcy},
  {Howard}, {Isaacson}, {Hebb}, {Cargile}, {Morton}, {Weiss}, {Winn}, {Rogers},
  {Sinukoff} \& {Hirsch}}{{Johnson} et~al.}{2017}]{Johnson2017}
{Johnson} J.~A.,  {Petigura} E.~A.,  {Fulton} B.~J.,  {Marcy} G.~W.,  {Howard}
  A.~W.,  {Isaacson} H.,  {Hebb} L.,  {Cargile} P.~A.,  {Morton} T.~D.,
  {Weiss} L.~M.,  {Winn} J.~N.,  {Rogers} L.~A.,  {Sinukoff} E.,    {Hirsch}
  L.~A.,  2017, \aj, 154, 108

\bibitem[\protect\citeauthoryear{{Jontof-Hutter}, {Ford}, {Rowe}, {Lissauer},
  {Fabrycky}, {Van Laerhoven}, {Agol}, {Deck}, {Holczer} \&
  {Mazeh}}{{Jontof-Hutter} et~al.}{2016}]{JontofHutter2016}
{Jontof-Hutter} D.,  {Ford} E.~B.,  {Rowe} J.~F.,  {Lissauer} J.~J.,
  {Fabrycky} D.~C.,  {Van Laerhoven} C.,  {Agol} E.,  {Deck} K.~M.,  {Holczer}
  T.,    {Mazeh} T.,  2016, \apj, 820, 39

\bibitem[\protect\citeauthoryear{{Jontof-Hutter}, {Lissauer}, {Rowe} \&
  {Fabrycky}}{{Jontof-Hutter} et~al.}{2014}]{JontofHutter2014}
{Jontof-Hutter} D.,  {Lissauer} J.~J.,  {Rowe} J.~F.,    {Fabrycky} D.~C.,
  2014, \apj, 785, 15

\bibitem[\protect\citeauthoryear{{Ketchum}, {Adams} \& {Bloch}}{{Ketchum}
  et~al.}{2011}]{Ketchum2011}
{Ketchum} J.~A.,  {Adams} F.~C.,    {Bloch} A.~M.,  2011, \apj, 726, 53

\bibitem[\protect\citeauthoryear{{Kinemuchi}, {Barclay}, {Fanelli}, {Pepper},
  {Still} \& {Howell}}{{Kinemuchi} et~al.}{2012}]{Kinemuchi2012}
{Kinemuchi} K.,  {Barclay} T.,  {Fanelli} M.,  {Pepper} J.,  {Still} M.,
  {Howell} S.~B.,  2012, \pasp, 124, 963

\bibitem[\protect\citeauthoryear{{Kipping} \& {Bakos}}{{Kipping} \&
  {Bakos}}{2011}]{Kipping2011}
{Kipping} D.,  {Bakos} G.,  2011, \apj, 730, 50

\bibitem[\protect\citeauthoryear{{Kreidberg}}{{Kreidberg}}{2015}]{Kreidberg2015}
{Kreidberg} L.,  2015, \pasp, 127, 1161

\bibitem[\protect\citeauthoryear{{Kurokawa} \& {Nakamoto}}{{Kurokawa} \&
  {Nakamoto}}{2014}]{Kurokawa2014}
{Kurokawa} H.,  {Nakamoto} T.,  2014, \apj, 783, 54

\bibitem[\protect\citeauthoryear{{Lanza}, {Das Chagas} \& {De
  Medeiros}}{{Lanza} et~al.}{2014}]{Lanza2014}
{Lanza} A.~F.,  {Das Chagas} M.~L.,    {De Medeiros} J.~R.,  2014, \aap, 564,
  A50

\bibitem[\protect\citeauthoryear{{Laskar}, {Quinn} \& {Tremaine}}{{Laskar}
  et~al.}{1992}]{Laskar1992}
{Laskar} J.,  {Quinn} T.,    {Tremaine} S.,  1992, \icarus, 95, 148

\bibitem[\protect\citeauthoryear{{Laskar} \& {Robutel}}{{Laskar} \&
  {Robutel}}{2001}]{Laskar2001}
{Laskar} J.,  {Robutel} P.,  2001, Celestial Mechanics and Dynamical Astronomy,
  80, 39

\bibitem[\protect\citeauthoryear{{Lee} \& {Peale}}{{Lee} \&
  {Peale}}{2002}]{Lee2002}
{Lee} M.~H.,  {Peale} S.~J.,  2002, \apj, 567, 596

\bibitem[\protect\citeauthoryear{{Lillo-Box}, {Barrado}, {Santos}, {Mancini},
  {Figueira}, {Ciceri} \& {Henning}}{{Lillo-Box} et~al.}{2015}]{LilloBox2015}
{Lillo-Box} J.,  {Barrado} D.,  {Santos} N.~C.,  {Mancini} L.,  {Figueira} P.,
  {Ciceri} S.,    {Henning} T.,  2015, \aap, 577, A105

\bibitem[\protect\citeauthoryear{{Lopez} \& {Fortney}}{{Lopez} \&
  {Fortney}}{2013}]{Lopez2013}
{Lopez} E.~D.,  {Fortney} J.~J.,  2013, \apj, 776, 2

\bibitem[\protect\citeauthoryear{{MacDonald}, {Ragozzine}, {Fabrycky}, {Ford},
  {Holman}, {Isaacson}, {Lissauer}, {Lopez}, {Mazeh}, {Rogers}, {Rowe},
  {Steffen} \& {Torres}}{{MacDonald} et~al.}{2016}]{MacDonald2016}
{MacDonald} M.~G.,  {Ragozzine} D.,  {Fabrycky} D.~C.,  {Ford} E.~B.,  {Holman}
  M.~J.,  {Isaacson} H.~T.,  {Lissauer} J.~J.,  {Lopez} E.~D.,  {Mazeh} T.,
  {Rogers} L.,  {Rowe} J.~F.,  {Steffen} J.~H.,    {Torres} G.,  2016, \aj,
  152, 105

\bibitem[\protect\citeauthoryear{{Masuda}}{{Masuda}}{2017}]{Masuda2017}
{Masuda} K.,  2017, \aj, 154, 64

\bibitem[\protect\citeauthoryear{{Mazeh}, {Holczer} \& {Faigler}}{{Mazeh}
  et~al.}{2016}]{Mazeh2016}
{Mazeh} T.,  {Holczer} T.,    {Faigler} S.,  2016, \aap, 589, A75

\bibitem[\protect\citeauthoryear{{Migaszewski}}{{Migaszewski}}{2015}]{Migaszewski2015}
{Migaszewski} C.,  2015, \mnras, 453, 1632

\bibitem[\protect\citeauthoryear{{Migaszewski}}{{Migaszewski}}{2016}]{Migaszewski2016}
{Migaszewski} C.,  2016, \mnras, 458, 2051

\bibitem[\protect\citeauthoryear{{Migaszewski}, {Go{\'z}dziewski} \&
  {Panichi}}{{Migaszewski} et~al.}{2017}]{Migaszewski2017}
{Migaszewski} C.,  {Go{\'z}dziewski} K.,    {Panichi} F.,  2017, \mnras, 465,
  2366

\bibitem[\protect\citeauthoryear{{Mikkola} \& {Innanen}}{{Mikkola} \&
  {Innanen}}{1999}]{Mikkola1999}
{Mikkola} S.,  {Innanen} K.,  1999, Celestial Mechanics and Dynamical
  Astronomy, 74, 59

\bibitem[\protect\citeauthoryear{{Millholland}, {Wang} \&
  {Laughlin}}{{Millholland} et~al.}{2017}]{Millholland2017}
{Millholland} S.,  {Wang} S.,    {Laughlin} G.,  2017, \apjl, 849, L33

\bibitem[\protect\citeauthoryear{{Mills}, {Fabrycky}, {Migaszewski}, {Ford},
  {Petigura} \& {Isaacson}}{{Mills} et~al.}{2016}]{Mills2016}
{Mills} S.~M.,  {Fabrycky} D.~C.,  {Migaszewski} C.,  {Ford} E.~B.,  {Petigura}
  E.,    {Isaacson} H.,  2016, \nat, 533, 509

\bibitem[\protect\citeauthoryear{{Moore}, {Hasan} \& {Quillen}}{{Moore}
  et~al.}{2013}]{Moore2013}
{Moore} A.,  {Hasan} I.,    {Quillen} A.~C.,  2013, \mnras, 432, 1196

\bibitem[\protect\citeauthoryear{{Mordasini}, {Alibert} \& {Benz}}{{Mordasini}
  et~al.}{2009}]{Mordasini2009}
{Mordasini} C.,  {Alibert} Y.,    {Benz} W.,  2009, \aap, 501, 1139

\bibitem[\protect\citeauthoryear{{Mordasini}, {Alibert}, {Georgy}, {Dittkrist},
  {Klahr} \& {Henning}}{{Mordasini} et~al.}{2012}]{Mordasini2012b}
{Mordasini} C.,  {Alibert} Y.,  {Georgy} C.,  {Dittkrist} K.-M.,  {Klahr} H.,
   {Henning} T.,  2012, \aap, 547, A112

\bibitem[\protect\citeauthoryear{{Mordasini}, {Alibert}, {Klahr} \&
  {Henning}}{{Mordasini} et~al.}{2012}]{Mordasini2012a}
{Mordasini} C.,  {Alibert} Y.,  {Klahr} H.,    {Henning} T.,  2012, \aap, 547,
  A111

\bibitem[\protect\citeauthoryear{{Muto}, {Takeuchi} \& {Ida}}{{Muto}
  et~al.}{2011}]{Muto2011}
{Muto} T.,  {Takeuchi} T.,    {Ida} S.,  2011, \apj, 737, 37

\bibitem[\protect\citeauthoryear{{Nesvorn{\'y}}, {Kipping}, {Terrell},
  {Hartman}, {Bakos} \& {Buchhave}}{{Nesvorn{\'y}} et~al.}{2013}]{Nesvorny2013}
{Nesvorn{\'y}} D.,  {Kipping} D.,  {Terrell} D.,  {Hartman} J.,  {Bakos}
  G.~{\'A}.,    {Buchhave} L.~A.,  2013, \apj, 777, 3

\bibitem[\protect\citeauthoryear{{Nesvorn{\'y}} \& {Morbidelli}}{{Nesvorn{\'y}}
  \& {Morbidelli}}{1998}]{Nesvorny1998}
{Nesvorn{\'y}} D.,  {Morbidelli} A.,  1998, \aj, 116, 3029

\bibitem[\protect\citeauthoryear{{Ofir}, {Dreizler}, {Zechmeister} \&
  {Husser}}{{Ofir} et~al.}{2014}]{Ofir2014}
{Ofir} A.,  {Dreizler} S.,  {Zechmeister} M.,    {Husser} T.-O.,  2014, \aap,
  561, A103

\bibitem[\protect\citeauthoryear{{Padovani}}{{Padovani}}{1998}]{Padovani1998}
{Padovani} P.,  1998, Space Telesc.~Sci.~Inst., Newsl., Vol.~15, No.~4, p.~14 -
  15, 15, 14

\bibitem[\protect\citeauthoryear{Panichi, Ciotti \& Turchetti}{Panichi
  et~al.}{2016}]{Panichi2016}
Panichi F.,  Ciotti L.,    Turchetti G.,  2016, Communications in Nonlinear
  Science and Numerical Simulation, 35, 53

\bibitem[\protect\citeauthoryear{{Panichi}, {Go{\'z}dziewski} \&
  {Turchetti}}{{Panichi} et~al.}{2017}]{Panichi2017}
{Panichi} F.,  {Go{\'z}dziewski} K.,    {Turchetti} G.,  2017, \mnras, 468, 469

\bibitem[\protect\citeauthoryear{{Papaloizou} \& {Szuszkiewicz}}{{Papaloizou}
  \& {Szuszkiewicz}}{2005}]{Papaloizou2005}
{Papaloizou} J.~C.~B.,  {Szuszkiewicz} E.,  2005, \mnras, 363, 153

\bibitem[\protect\citeauthoryear{{Podlewska-Gaca}, {Papaloizou} \&
  {Szuszkiewicz}}{{Podlewska-Gaca} et~al.}{2012}]{PodlewskaGaca2012}
{Podlewska-Gaca} E.,  {Papaloizou} J.~C.~B.,    {Szuszkiewicz} E.,  2012,
  \mnras, 421, 1736

\bibitem[\protect\citeauthoryear{{Rogers}, {Bodenheimer}, {Lissauer} \&
  {Seager}}{{Rogers} et~al.}{2011}]{Rogers2011}
{Rogers} L.~A.,  {Bodenheimer} P.,  {Lissauer} J.~J.,    {Seager} S.,  2011,
  \apj, 738, 59

\bibitem[\protect\citeauthoryear{{Rousseeuw} \& {Croux}}{{Rousseeuw} \&
  {Croux}}{1993}]{Rousseeuw1993}
{Rousseeuw} P.~J.,  {Croux} C.,  1993, Journal of the American Statistical
  Association, 88, 1273

\bibitem[\protect\citeauthoryear{{Rowe}, {Coughlin}, {Antoci}, {Barclay},
  {Batalha}, {Borucki}, {Burke} \& et al.}{{Rowe} et~al.}{2015}]{Rowe2015}
{Rowe} J.~F.,  {Coughlin} J.~L.,  {Antoci} V.,  {Barclay} T.,  {Batalha} N.~M.,
   {Borucki} W.~J.,  {Burke} C.~J.,    et al. 2015, \apjs, 217, 16

\bibitem[\protect\citeauthoryear{Ruci{\'n}ski, Izzo \& Biscani}{Ruci{\'n}ski
  et~al.}{2010}]{Izzo2010}
Ruci{\'n}ski M.,  Izzo D.,    Biscani F.,  2010, Parallel Computing, 36, 555

\bibitem[\protect\citeauthoryear{{Saad-Olivera}, {Nesvorn{\'y}}, {Kipping} \&
  {Roig}}{{Saad-Olivera} et~al.}{2017}]{SaadOlivera2017}
{Saad-Olivera} X.,  {Nesvorn{\'y}} D.,  {Kipping} D.~M.,    {Roig} F.,  2017,
  \aj, 153, 198

\bibitem[\protect\citeauthoryear{{Sanchis-Ojeda}, {Fabrycky}, {Winn}, {Barclay}
  \& et al.}{{Sanchis-Ojeda} et~al.}{2012}]{SanchisOjeda2012}
{Sanchis-Ojeda} R.,  {Fabrycky} D.~C.,  {Winn} J.~N.,  {Barclay} T.,    et al.
  2012, \nat, 487, 449

\bibitem[\protect\citeauthoryear{{\v{S}idlichovsk{\'y}} \&
  {Nesvorn{\'y}}}{{\v{S}idlichovsk{\'y}} \&
  {Nesvorn{\'y}}}{1996}]{Nesvorny1996}
{\v{S}idlichovsk{\'y}} M.,  {Nesvorn{\'y}} D.,  1996, Celestial Mechanics and
  Dynamical Astronomy, 65, 137

\bibitem[\protect\citeauthoryear{Sivia \& Skilling}{Sivia \&
  Skilling}{2006}]{Sivia2006}
Sivia D.,  Skilling J.,  2006, Data analysis: a Bayesian tutorial.
Oxford science publications, Oxford University Press

\bibitem[\protect\citeauthoryear{{Smith}, {Stumpe}, {Van Cleve}, {Jenkins},
  {Barclay}, {Fanelli}, {Girouard}, {Kolodziejczak}, {McCauliff}, {Morris} \&
  {Twicken}}{{Smith} et~al.}{2012}]{Smith2012}
{Smith} J.~C.,  {Stumpe} M.~C.,  {Van Cleve} J.~E.,  {Jenkins} J.~M.,
  {Barclay} T.~S.,  {Fanelli} M.~N.,  {Girouard} F.~R.,  {Kolodziejczak} J.~J.,
   {McCauliff} S.~D.,  {Morris} R.~L.,    {Twicken} J.~D.,  2012, \pasp, 124,
  1000

\bibitem[\protect\citeauthoryear{{Snellgrove}, {Papaloizou} \&
  {Nelson}}{{Snellgrove} et~al.}{2001}]{Snellgrove2001}
{Snellgrove} M.~D.,  {Papaloizou} J.~C.~B.,    {Nelson} R.~P.,  2001, \aap,
  374, 1092

\bibitem[\protect\citeauthoryear{Storn \& Price}{Storn \&
  Price}{1997}]{Storn1977}
Storn R.,  Price K.,  1997, J. of Global Optimization, 11, 341

\bibitem[\protect\citeauthoryear{{Stumpe}, {Smith}, {Van Cleve}, {Twicken},
  {Barclay}, {Fanelli}, {Girouard}, {Jenkins}, {Kolodziejczak}, {McCauliff} \&
  {Morris}}{{Stumpe} et~al.}{2012}]{Stumpe2012}
{Stumpe} M.~C.,  {Smith} J.~C.,  {Van Cleve} J.~E.,  {Twicken} J.~D.,
  {Barclay} T.~S.,  {Fanelli} M.~N.,  {Girouard} F.~R.,  {Jenkins} J.~M.,
  {Kolodziejczak} J.~J.,  {McCauliff} S.~D.,    {Morris} R.~L.,  2012, \pasp,
  124, 985

\bibitem[\protect\citeauthoryear{{Tingley}, {Palle}, {Parviainen}, {Deeg},
  {Zapatero Osorio}, {Cabrera-Lavers}, {Belmonte}, {Rodriguez}, {Murgas} \&
  {Ribas}}{{Tingley} et~al.}{2011}]{Tingley2011}
{Tingley} B.,  {Palle} E.,  {Parviainen} H.,  {Deeg} H.~J.,  {Zapatero Osorio}
  M.~R.,  {Cabrera-Lavers} A.,  {Belmonte} J.~A.,  {Rodriguez} P.~M.,  {Murgas}
  F.,    {Ribas} I.,  2011, \aap, 536, L9

\bibitem[\protect\citeauthoryear{{Veras}, {Mustill}, {G{\"a}nsicke},
  {Redfield}, {Georgakarakos}, {Bowler} \& {Lloyd}}{{Veras}
  et~al.}{2016}]{Veras2016}
{Veras} D.,  {Mustill} A.~J.,  {G{\"a}nsicke} B.~T.,  {Redfield} S.,
  {Georgakarakos} N.,  {Bowler} A.~B.,    {Lloyd} M.~J.~S.,  2016, \mnras, 458,
  3942

\bibitem[\protect\citeauthoryear{{Weiss}, {Marcy}, {Petigura}, {Fulton},
  {Howard}, {Winn}, {Isaacson}, {Morton}, {Hirsch}, {Sinukoff}, {Cumming},
  {Hebb} \& {Cargile}}{{Weiss} et~al.}{2017}]{Weiss2017}
{Weiss} L.~M.,  {Marcy} G.~W.,  {Petigura} E.~A.,  {Fulton} B.~J.,  {Howard}
  A.~W.,  {Winn} J.~N.,  {Isaacson} H.~T.,  {Morton} T.~D.,  {Hirsch} L.~A.,
  {Sinukoff} E.~J.,  {Cumming} A.,  {Hebb} L.,    {Cargile} P.~A.,  2017,
  preprint (arXiv:1706.06204)

\bibitem[\protect\citeauthoryear{{Weiss}, {Rogers}, {Isaacson}, {Agol},
  {Marcy}, {Rowe}, {Kipping}, {Fulton}, {Lissauer}, {Howard} \&
  {Fabrycky}}{{Weiss} et~al.}{2016}]{Weiss2016}
{Weiss} L.~M.,  {Rogers} L.~A.,  {Isaacson} H.~T.,  {Agol} E.,  {Marcy} G.~W.,
  {Rowe} J.~F.,  {Kipping} D.,  {Fulton} B.~J.,  {Lissauer} J.~J.,  {Howard}
  A.~W.,    {Fabrycky} D.,  2016, \apj, 819, 83

\bibitem[\protect\citeauthoryear{{Wisdom}}{{Wisdom}}{2006}]{Wisdom2006}
{Wisdom} J.,  2006, \aj, 131, 2294

\bibitem[\protect\citeauthoryear{{Wisdom} \& {Hernandez}}{{Wisdom} \&
  {Hernandez}}{2015}]{Wisdom2016}
{Wisdom} J.,  {Hernandez} D.~M.,  2015, \mnras, 453, 3015

\bibitem[\protect\citeauthoryear{{Zeng}, {Sasselov} \& {Jacobsen}}{{Zeng}
  et~al.}{2016}]{Zeng2016}
{Zeng} L.,  {Sasselov} D.~D.,    {Jacobsen} S.~B.,  2016, \apj, 819, 127

\end{thebibliography}
\label{lastpage}
%
\section*{On-line material}
%
\appendix
\setcounter{figure}{0}
\setcounter{table}{0}
\renewcommand{\thefigure}{A\arabic{figure}}
\renewcommand{\thetable}{A\arabic{table}}

{\bf We aim to publish this supplementary material on-line, if the manuscript could be accepted.}

\begin{table}
\centering
\caption{Midpoint transit times, TTVs and relative errors obtained with light curves analysis in this work. See the text for details. Listed also the transits affected by one or more star spots.}
\label{tab:tab3}
\begin{tabular}{crrrrr}
\hline\hline
Planet & Transit  & Transit time & TTV & $+1\sigma$ & $-1\sigma$  \\
\smallskip
[KOI] & number &  [BJD-2454900] & [min] & [min] & [min] \\
\hline\hline
806.01 & 0  & 83.442920   &   385.409   &   18.77   &   22.12  \\   
& 1  & 112.763042  &   129.103   &   15.60  &   16.17   \\
& 2 &  142.083165   &   -101.323   &   15.60   &   21.26   \\  
& 3 &  171.403288   &   -349.916   &   20.96   &   18.44  \\  
& 4 &  200.723411   &   -567.660   &   27.36   &   27.37 \\    
& 5 &  230.043534   &   -795.065   &   29.48   &   28.94 \\    
& 6 &  259.363656   &   -977.170  &   14.55   &   15.22 \\    
& 7 &  288.683779   &   -1158.293   &   11.17   &   10.38 \\    
& 9 &  347.324025   &   -1305.114   &   14.32   &   15.05  \\   
& 11 &  405.964271   &   -1271.939  &   12.73   &   20.61  \\   
& 12 &  435.284393   &   -1218.247   &   17.24   &   12.17 \\    
& 13 &  464.604516   &   -981.335   &   13.63   &   14.57 \\    
& 14 &  493.924639   &   -873.463   &   10.87   &   8.37 \\    
& 15 &  523.244762   &   -538.665   &   9.10  &   8.13 \\    
& 16 &  552.564885$^{(*)}$   &   -361.599   &   8.34   &   8.15 \\    
& 17 &  581.885008   &   18.738   &   9.57   &   10.55 \\    
& 18 &  611.205130   &   218.784   &   8.73   &   9.70 \\    
& 19 &  640.525253   &   581.897   &   13.32   &   13.50 \\    
& 20 &  669.845376  &   782.156   &   15.50   &   10.43 \\    
& 21 &  699.165499   &   1087.721   &   5.89   &   7.08 \\    
& 22 &  728.485622   &   1274.911  &   11.08   &   13.37 \\    
& 23 &  757.805745   &   1459.760   &   8.90   &   7.30 \\    
& 24 &  787.125867   &   1590.244  &   15.97   &   14.94 \\    
& 25 &  816.445990   &   1648.913   &   8.91   &   6.16 \\    
& 26 &  845.766113   &   1712.007   &   11.94  &   15.47 \\    
& 27 &  875.086236   &   1681.572   &   12.65   &   15.06 \\    
& 28 &  904.406359  &   1606.888   &   6.24   &   7.43 \\    
& 29 &  933.726482   &   1500.750   &   7.95   &   7.67 \\    
& 30 &  963.046604   &   1367.302   &   8.94   &   10.57 \\    
& 31 &  992.366727   &   1184.997   &   12.50   &   12.37 \\    
& 32 & 1021.686850   &   971.408   &   12.20   &   11.66 \\    
& 34 & 1080.327096   &   493.080   &   8.24   &   12.13 \\    
& 35 & 1109.647219   &   253.675   &   8.51  &   24.20 \\    
& 36 & 1138.967341   &   -46.622   &   21.85   &   7.89 \\    
& 37 & 1168.287464  &   -266.866   &   12.75   &   9.96 \\    
& 38 & 1197.607587   &   -539.452   &   21.09   &   13.87 \\    
& 40 & 1256.247833   &   -902.252   &   24.97   &   16.70 \\    
& 41 & 1285.567955   &   -1037.387   &   16.78   &   18.65 \\    
& 42 & 1314.888078   &   -1131.385   &   7.65   &   11.55 \\    
& 43 & 1344.208201   &   -1169.786   &   16.12   &   13.23 \\    
& 44 & 1373.528324   &   -1147.384   &   12.11  &   11.40 \\    
& 45 & 1402.848447   &   -1080.586   &   11.95   &   12.78 \\    
& 46 & 1432.168570   &   -900.686   &   14.94   &   17.35 \\    
& 47 & 1461.488692   &   -765.359   &   10.05   &   12.46 \\    
& 48 & 1490.808815   &   -461.761   &   8.99   &   9.00 \\
\hline 
806.02 & 1 & 176.892192$^{(*)}$ & -1.339 & 1.55 & 1.57 \\
& 2 & 237.217248$^{(*)}$ & 14.290 & 1.57 & 1.55 \\
& 3 & 297.542303$^{(*)}$ & 21.610 & 1.72 & 1.88 \\
& 4 & 357.867359 & 19.622 & 1.52 & 1.49 \\
& 5 & 418.192415 & 20.075 & 1.49 & 1.50 \\
& 6 & 478.517470$^{(*)}$ &  18.070 & 1.56 & 1.54 \\
& 7 & 538.842526 & 13.710 & 1.71 & 1.68 \\
& 8 & 599.167582 & 2.623  & 0.85 & 0.84 \\
& 10 & 719.817693 & -31.745 &  0.83 &  0.83 \\
\hline
&  & & & & Continue \\
\hline\hline
\end{tabular}
\end{table}
\begin{table}
\centering
\contcaption{Transit midpoint times, TTVs and relative errors obtained from the MCMC light curve analysis.}
\label{tab:tab4}
\begin{tabular}{crrrrr}
\hline\hline
Planet  & Transit  & Transit time & TTV & $+1\sigma$ & $-1\sigma$  \\
\smallskip
[KOI] & number &  [BJD-2454900] & [min] & [min] & [min] \\
\hline\hline
806.02 & 11 & 780.142749$^{(*)}$ & -38.616 & 0.81  & 0.81 \\
& 12 & 840.467804$^{(*)}$ & -43.313 & 0.87 & 0.85 \\
& 13 & 900.792860$^{(*)}$ & -34.586 & 0.89 & 0.87 \\
& 14 & 961.117916$^{(*)}$ & -29.075 & 0.85 & 0.85 \\
& 15 & 1021.442971 & -21.375 & 0.84 & 0.85 \\
& 16 & 1081.768027 & -13.556 & 0.85 & 0.86 \\
& 17 & 1142.093083 & -8.849 & 0.85 & 0.83 \\
& 18 & 1202.418138$^{(*)}$ & 6.934 & 0.86 & 0.87 \\
& 19 & 1262.743194 & 17.702 & 0.89 & 0.89 \\
& 20 & 1323.068250$^{(*)}$ & 31.140 &  0.85 &  0.85\\
& 21 & 1383.393305 &27.804  & 0.82 &  0.83\\
& 22 & 1443.718361 & 21.649 & 0.85 &  0.85\\
& 23 & 1504.043417$^{(*)}$ & 7.227 & 0.90 & 0.91  \\ 
\hline
806.03 & 0  & 87.227403 &  48.825 & 2.29 & 2.21 \\   
& 1  & 230.434853 & -80.491 & 2.53 & 2.59    \\
& 2 & 373.642304$^{(*)}$ & -28.747 & 2.70 & 2.50 \\   
& 3 & 516.849756 &  54.162 & 2.34 & 2.31  \\  
& 5 & 803.264658 &  14.240 & 1.78 & 1.75  \\  
& 6 & 946.472110 &  -2.202 & 1.95 & 2.01 \\    
& 7 & 1089.679561 & -22.343 & 1.69 & 1.70 \\   
& 8 & 1232.887012 &  75.089 & 1.98 & 2.01 \\   
& 9 & 1376.094463 & -58.534 & 1.68 & 1.74 \\
\hline\hline 
\multicolumn{6}{l}{$^{(*)}$ The identification of the spots is based on the visual inspection} \\
\multicolumn{6}{l}{ of each fragment of the light curve.}
\end{tabular}
\end{table}

\renewcommand{\thetable}{B\arabic{table}}
\begin{table*}
\centering
\caption{Planetary and stellar parameters inferred from the light curve analysis.}
\label{tab:tab5}
\begin{tabular}{lllc}
\hline
\multicolumn{4}{l}{Star parameters} \\
\hline
Quadratic limb-darkening coefficient $u_1$ & & 0.45$\pm~0.3$  &  \\
Quadratic limb-darkening coefficient $u_2$ & & 0.35$\pm~0.2$  &  \\
Linear limb-darkening coefficient $u$ & & 0.58$\pm~0.05$ & \\
Stellar density [\gccm{}] & & 1.9$\pm~0.1$& \\
\hline
Planet  parameters & Kepler-30b  & Kepler-30c & Kepler-30d \\
\hline
Mean Period [days] & 29.3201$\pm~0.0080$ & 60.3251$\pm~0.0005$& 143.208$\pm~0.004$\\
Mid-transit time [BJD-2454900] & 83.443$\pm~0.2$ & 116.567$\pm~0.007$& 87.225$\pm~0.02$ \\
line-of-sight orbital inclination, i [deg] & 89.81$\pm~0.32$ & 89.74$\pm~0.02$ & 89.51$\pm~0.02$   \\
Scaled planet radius $[R_p/R_*]$ & 0.039$\pm~0.008$ & 0.132 $\pm~0.004$ & 0.092$\pm~0.001$\\ 
Transit duration [h] & 4.7$\pm 0.2$ & 6.7$\pm~0.1$ & 8.3 $\pm~0.1$ \\
\hline\hline
\end{tabular}
\end{table*}

\newpage

\begin{figure*}
 \centerline{\includegraphics[width=1.\textwidth]{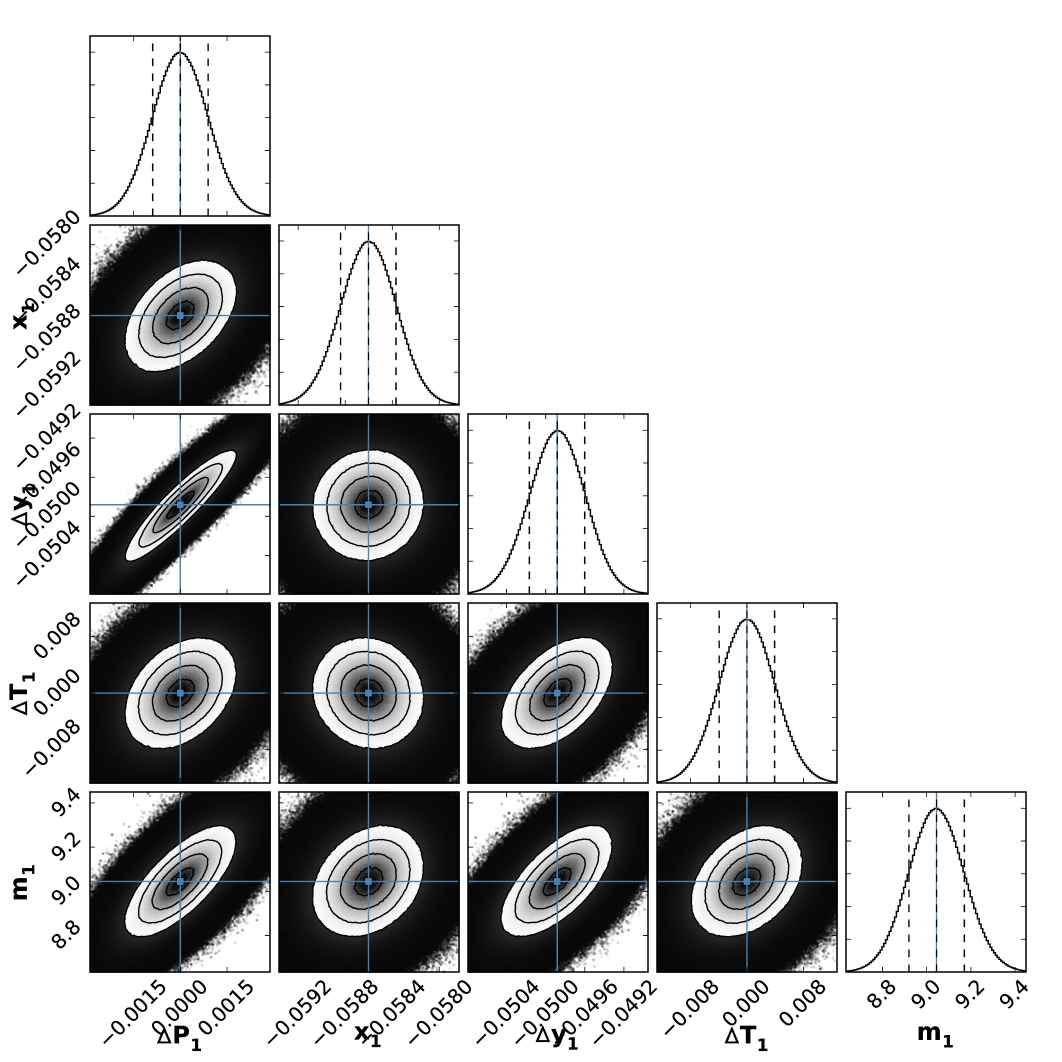}}
\caption{
One-- and two--dimensional projections of the posterior probability distribution for the TTV Model~I of Kepler-30 (see Tab.~\ref{tab:MCMCb} and the text for details). The error floor $\sigma_f$ is not included in the ${\cal L}$-function definition. The MCMC sampling was done for 560 initial conditions around the best fitting model derived by the GEA experiments, and continued for 192,000 iterations each.  If a parameter is labeled as $\Delta X$ then the zero-value for  this parameter $X$ in a posterior diagram is referred to as the nominal MCMC median in Table~\ref{tab:MCMCb}. Orbital periods $P_{1,2,3}$ and the first transit times $T_{1,2,3}$ are in days, and masses $m_{1,2,3}$ are expressed in the Earth masses. The blue lines mark the best-fitting parameters consistent with the maximum of the likelihood function ${\cal L}$ and found with the GEA search. Contours are for the uncertainties expressed through the 16th, 50th and 84th percentile of samples in the posterior distribution, respectively (10\% of initial, ``\corre{burn-in}'' samples are removed).  They are marked with contours in two--dimensional  diagrams. We used the \code{corner} package by \citep{corner} to plot the figures. 
} 
 \label{fig:figureA1}
\end{figure*}

\begin{figure*}
 \centerline{\includegraphics[width=1.\textwidth]{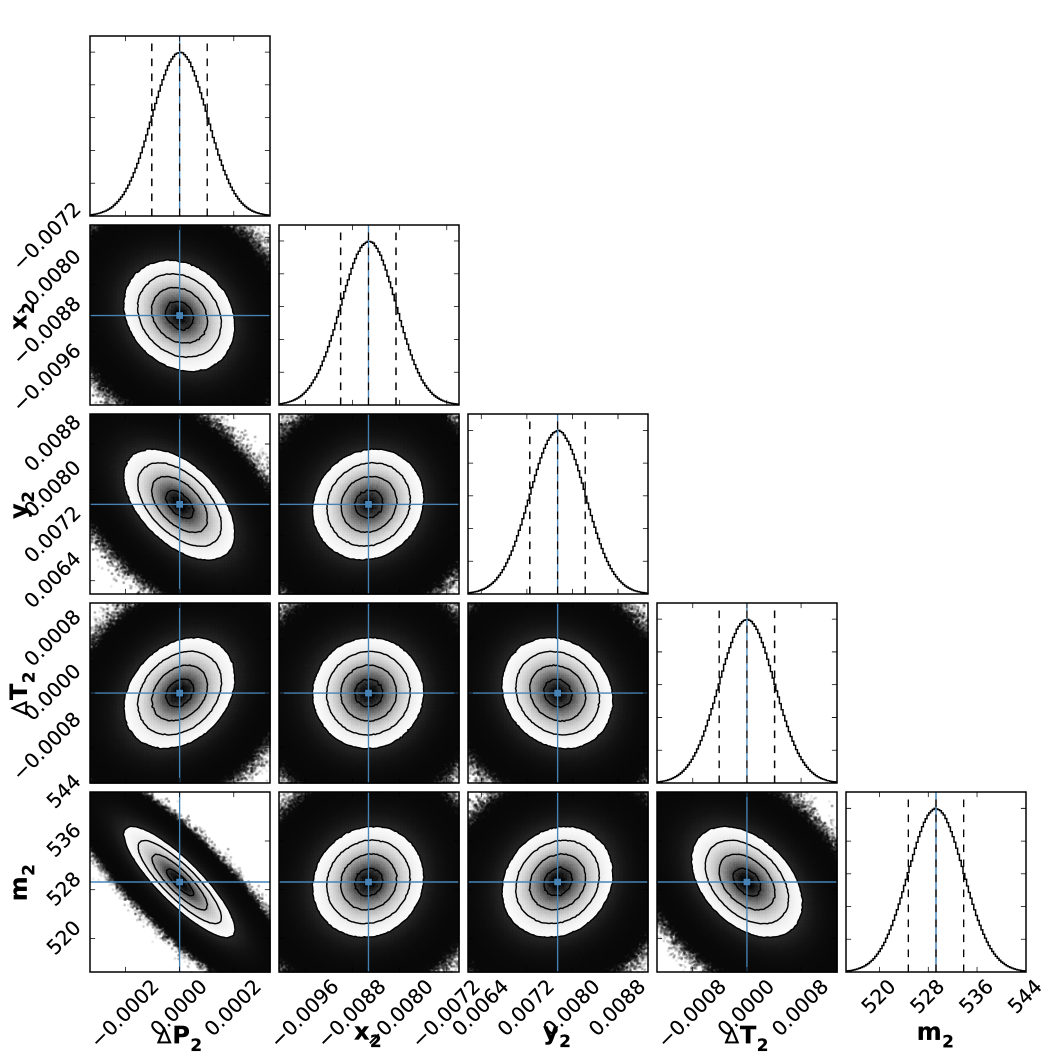}}
\caption{See caption to Fig.~\ref{fig:figureA1}.}  
 \label{fig:figureA2}
\end{figure*}

\begin{figure*}
 \centerline{\includegraphics[width=1.\textwidth]{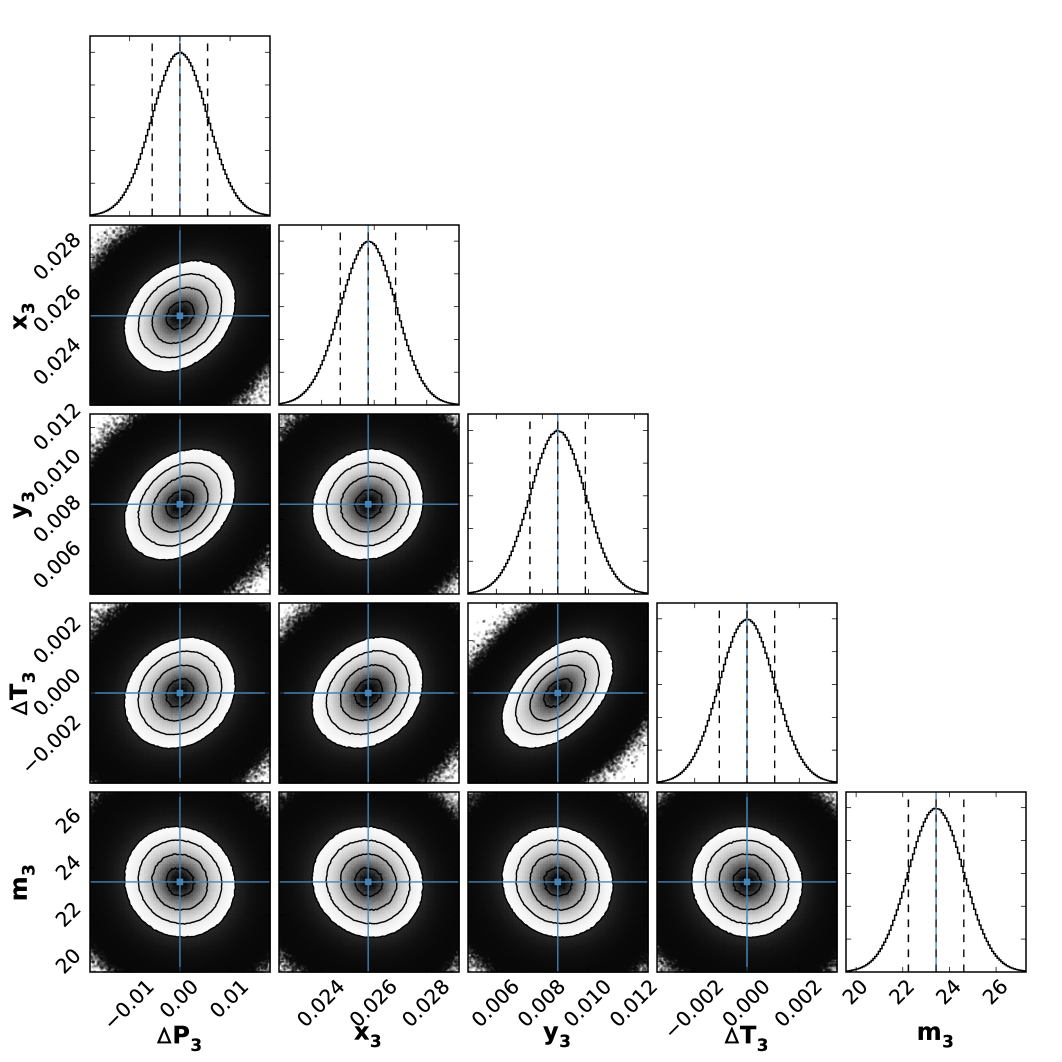}}
 \caption{See caption to Fig.~\ref{fig:figureA1}.} 
  \label{fig:figureA3}
\end{figure*} 
 
\begin{figure*}
 \centerline{\includegraphics[width=1.\textwidth]{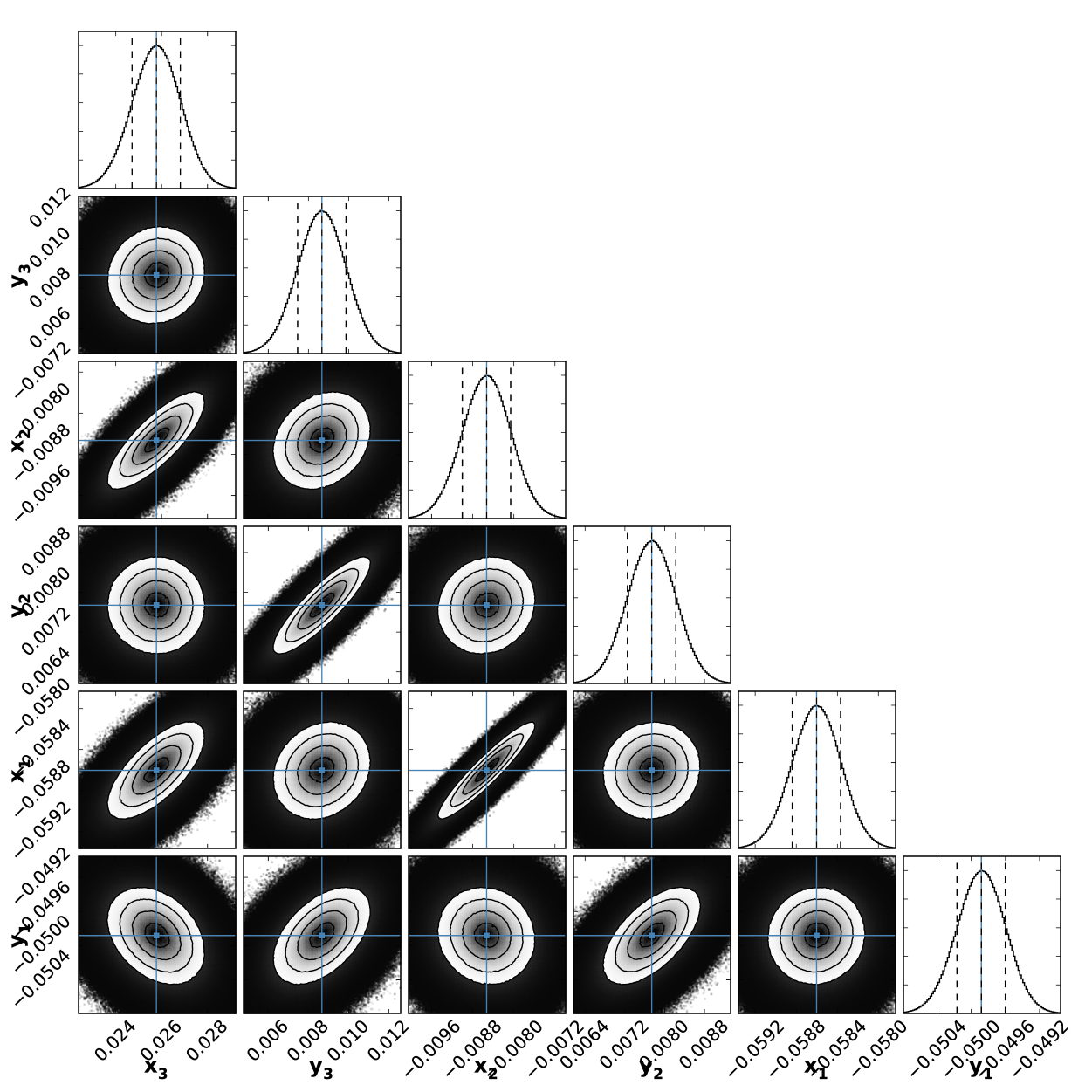}}
\caption{See caption to \ref{fig:figureA1}.}  
 \label{fig:figureA4}
\end{figure*}

\begin{figure*}
\centerline{\includegraphics[width=1.\textwidth]{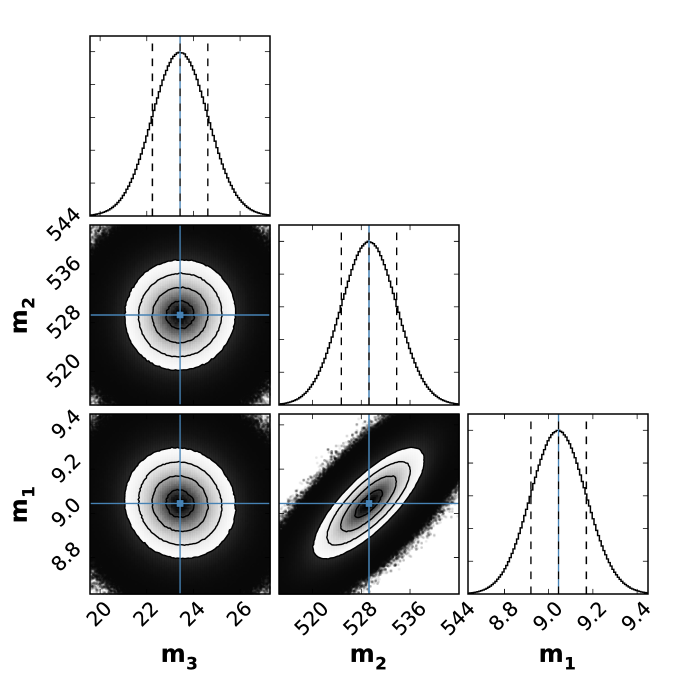}}
\caption{See caption to Fig.~\ref{fig:figureA1}.} 
\label{fig:figureA5}
\end{figure*}

\begin{figure*}
 \centerline{\includegraphics[width=1.\textwidth]{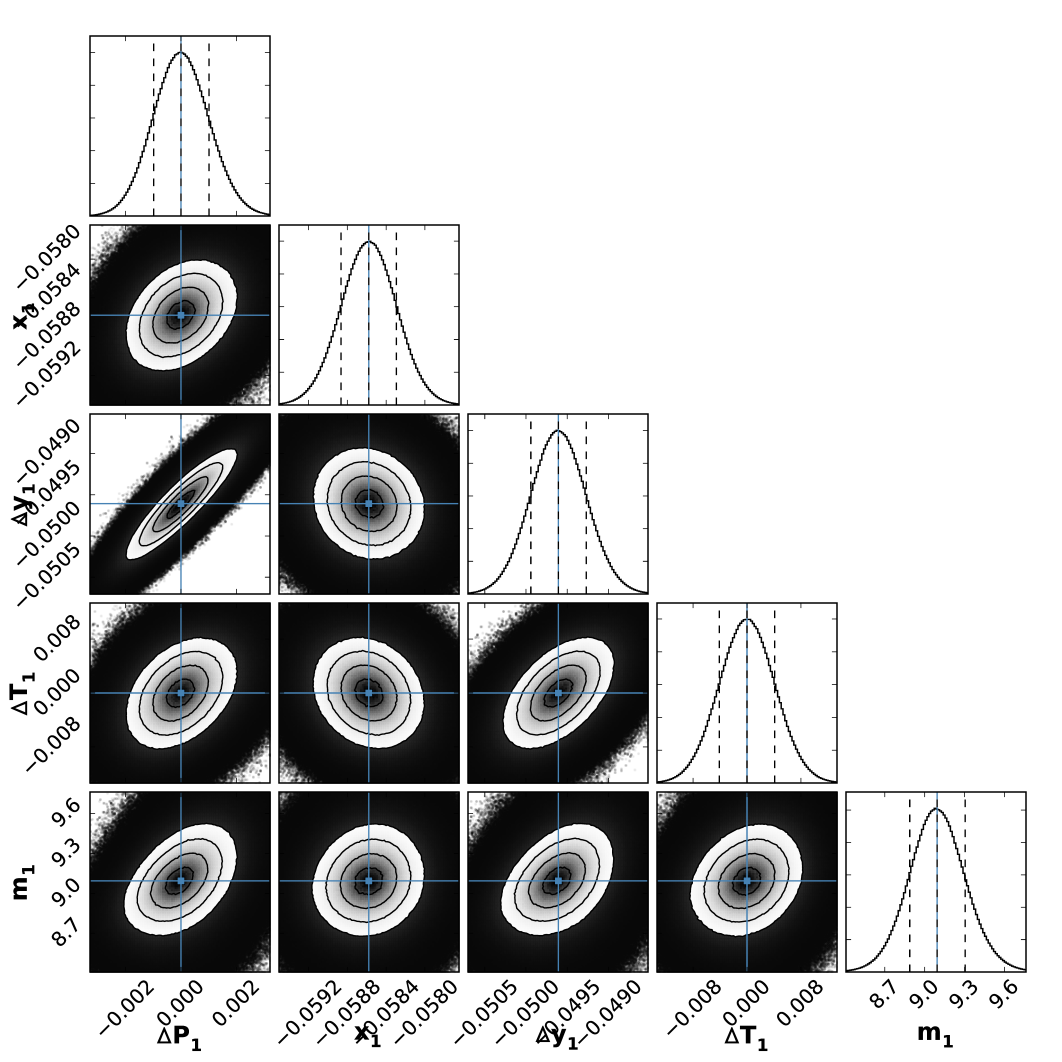}}
\caption{
One-- and two--dimensional projections of the posterior probability distribution for the TTV Model~II of Kepler-30 (see Tab.~\ref{tab:MCMCb} and the text for details). The error floor $\sigma_f$ is included in the ${\cal L}$-function definition. The MCMC sampling was preformed for 560 initial conditions around the best fitting model derived by the GEA experiments, and continued for 192,000 iterations each.  If a parameter is labeled as $\Delta X$ then the zero-value for  this parameter $X$ in a posterior diagram is referred to as the nominal MCMC median in Table~\ref{tab:MCMCb}. Orbital periods $P_{1,2,3}$ and the first transit times $T_{1,2,3}$ are in days, and masses $m_{1,2,3}$ are in the Earth masses. The blue lines mark the best-fitting parameters consistent with the maximum of the likelihood function ${\cal L}$ and found with the GEA search. Contours are for the uncertainties expressed through the 16th, 50th and 84th percentile of samples in the posterior distribution, respectively (10\% of initial, ``\corre{burn-in}'' samples are removed).  They are marked with contours in two--dimensional diagrams. We used the \code{ corner} package by \citep{corner} to plot the figures.
}
 \label{fig:figureB1}
\end{figure*}

\begin{figure*}
 \centerline{\includegraphics[width=1.\textwidth]{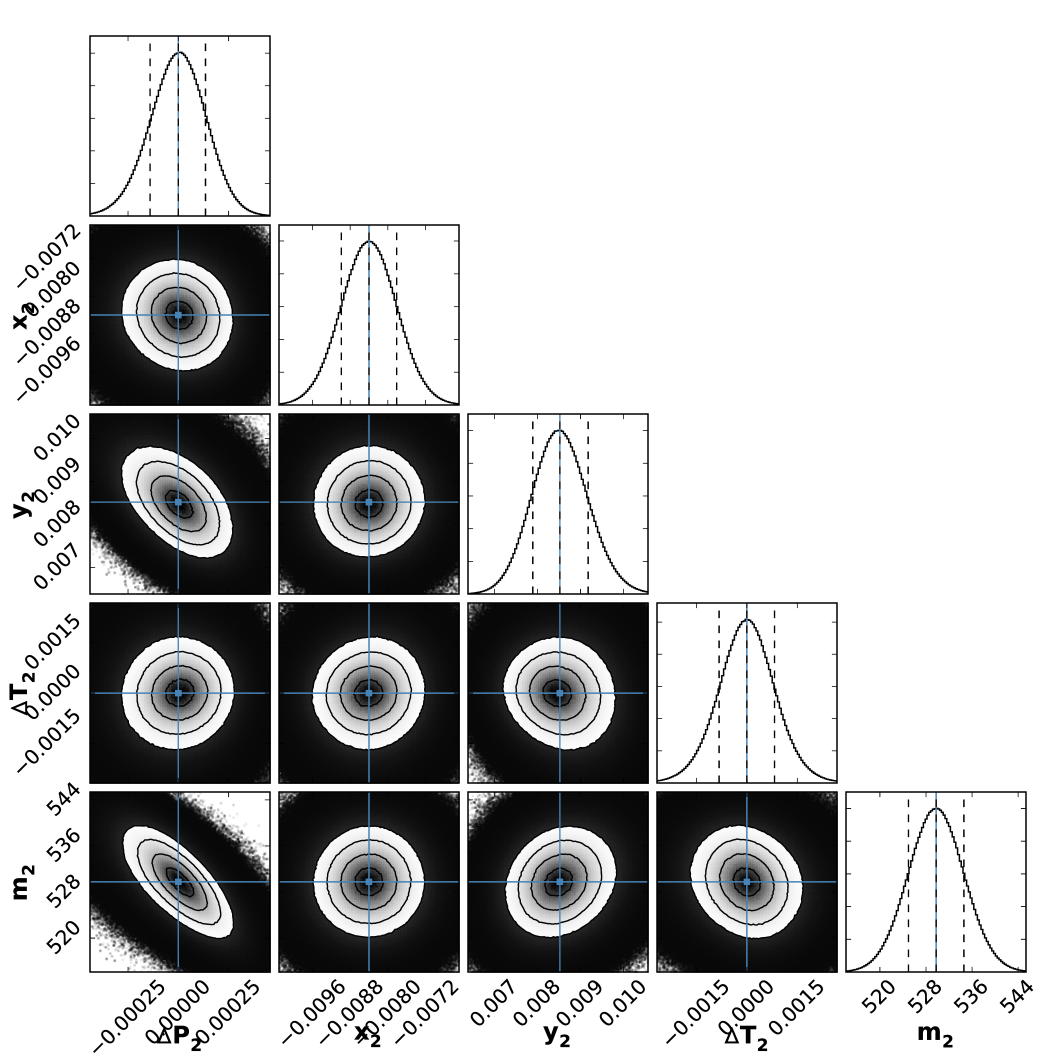}}
\caption{See caption to Fig.~\ref{fig:figureB1}.} 
 \label{fig:figureB2}
\end{figure*}

\begin{figure*}
 \centerline{\includegraphics[width=1.\textwidth]{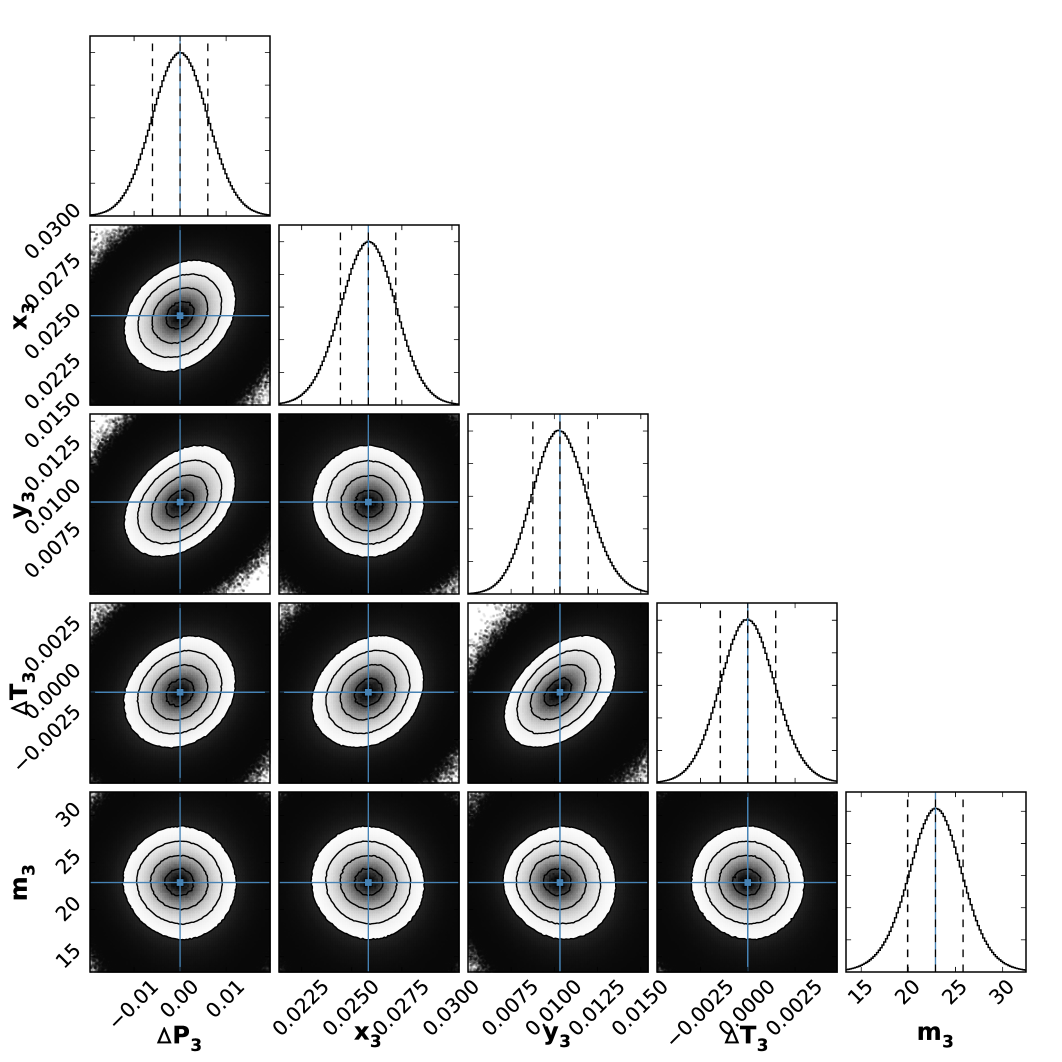}}
\caption{See caption to Fig.~\ref{fig:figureB1}.}  
 \label{fig:figureB3}
\end{figure*} 
 
\begin{figure*}
 \centerline{\includegraphics[width=1.\textwidth]{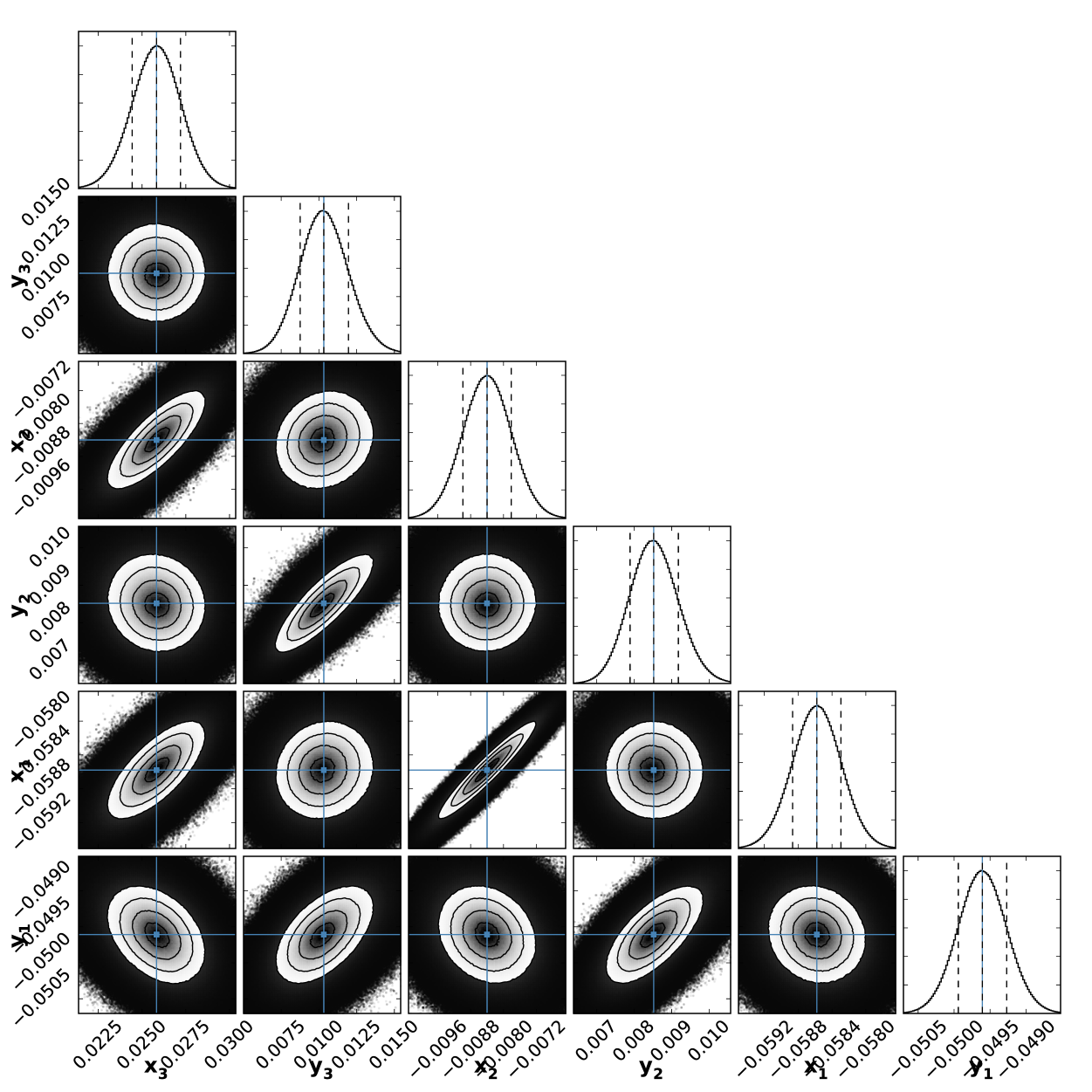}}
\caption{See caption to Fig.~\ref{fig:figureB1}.}  
 \label{fig:figureB4}
\end{figure*}

\begin{figure*}
\centerline{\includegraphics[width=1.\textwidth]{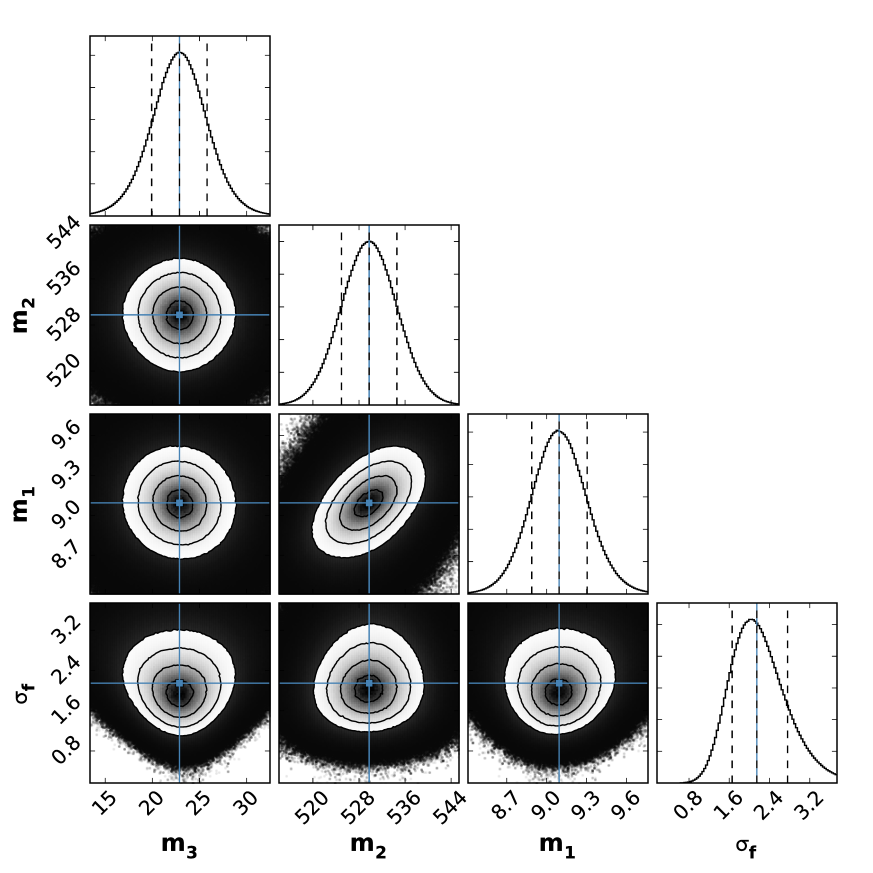}}
\caption{See caption to Fig.~\ref{fig:figureB1}. The error floor parameter $\sigma_f$ is included in the ${\cal L}$-function definition, and expressed in minutes.} 
\label{fig:figureB5}
\end{figure*}


\begin{figure*}
 \centerline{\includegraphics[width=1.\textwidth]{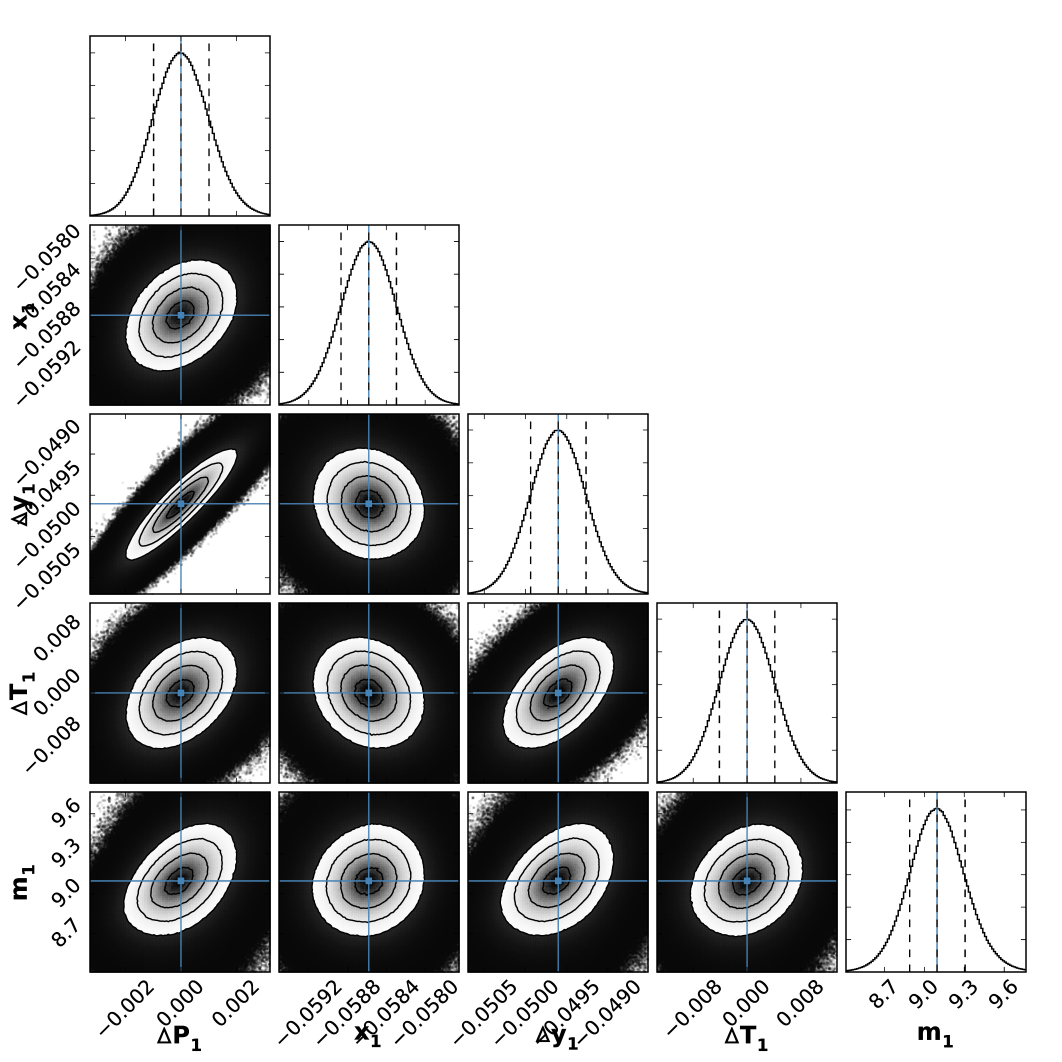}}
\caption{
One-- and two--dimensional projections of the posterior probability distribution for the TTV Model~III of Kepler-30 (see Tab.~\ref{tab:MCMCb} and the text for details). The error floor $\sigma_f$ is included in the ${\cal L}$-function definition. The MCMC sampling was preformed for 560 initial conditions around the best fitting model derived by the GEA experiments, and continued for 192,000 iterations each.  If a parameter is labeled as $\Delta X$ then the zero-value for  this parameter $X$ in a posterior diagram is referred to as the nominal MCMC median in Table~\ref{tab:MCMCb}. Orbital periods $P_{1,2,3}$ and the first transit times $T_{1,2,3}$ are in days, and masses $m_{1,2,3}$ are in the Earth masses.  Gaussian priors with the variance of 0.05 are imposed on the $(x_{1,2,3},y_{1,2,3})$, i.e., the Poincar\'e variables (eccentricities). The blue lines mark the best-fitting parameters consistent with the maximum of the likelihood function ${\cal L}$ and found with the GEA search. Contours are for the uncertainties expressed through the 16th, 50th and 84th percentile of samples in the posterior distribution, respectively (10\% of initial, ``\corre{burn-in}'' samples are removed).  They are marked with contours in two--dimensional  diagrams. We used the \code{corner} package by \citep{corner} to plot the figures.
} 
 \label{fig:figureC1}
\end{figure*}

\begin{figure*}
 \centerline{\includegraphics[width=1.\textwidth]{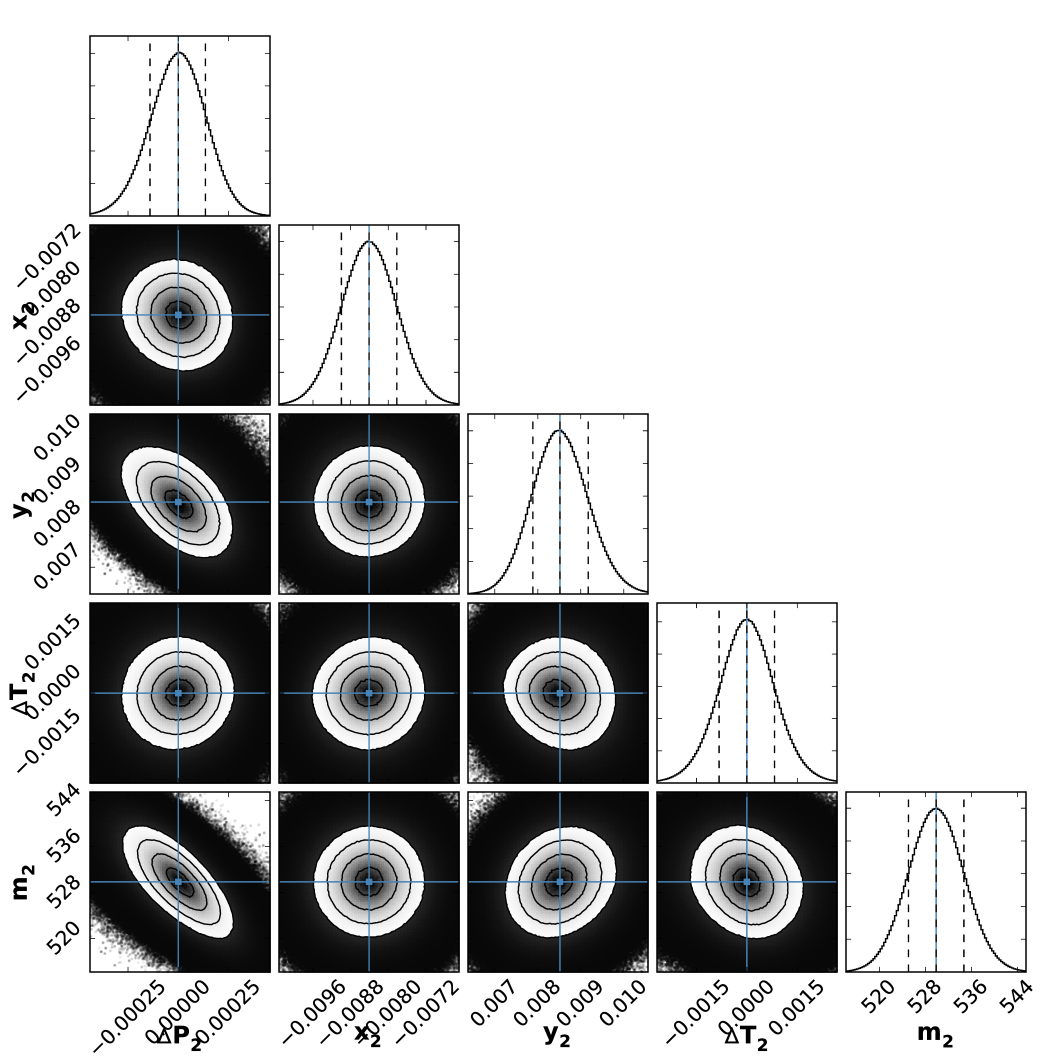}}
\caption{See caption to Fig.~\ref{fig:figureC1}.}
 \label{fig:figureC2}
\end{figure*}

\begin{figure*}
 \centerline{\includegraphics[width=1.\textwidth]{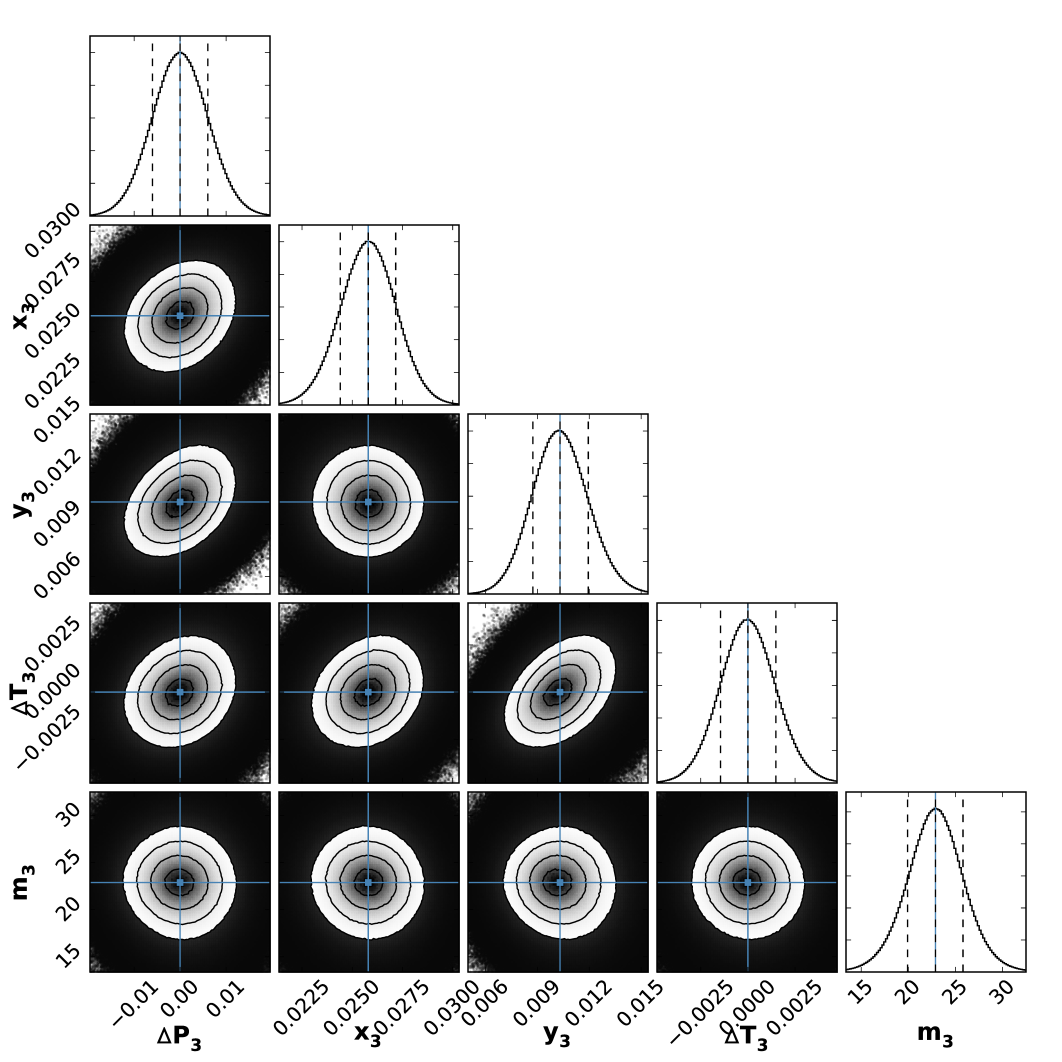}}
\caption{See caption to Fig.~\ref{fig:figureC1}.} 
 \label{fig:figureC3}
\end{figure*} 
 
\begin{figure*}
 \centerline{\includegraphics[width=1.\textwidth]{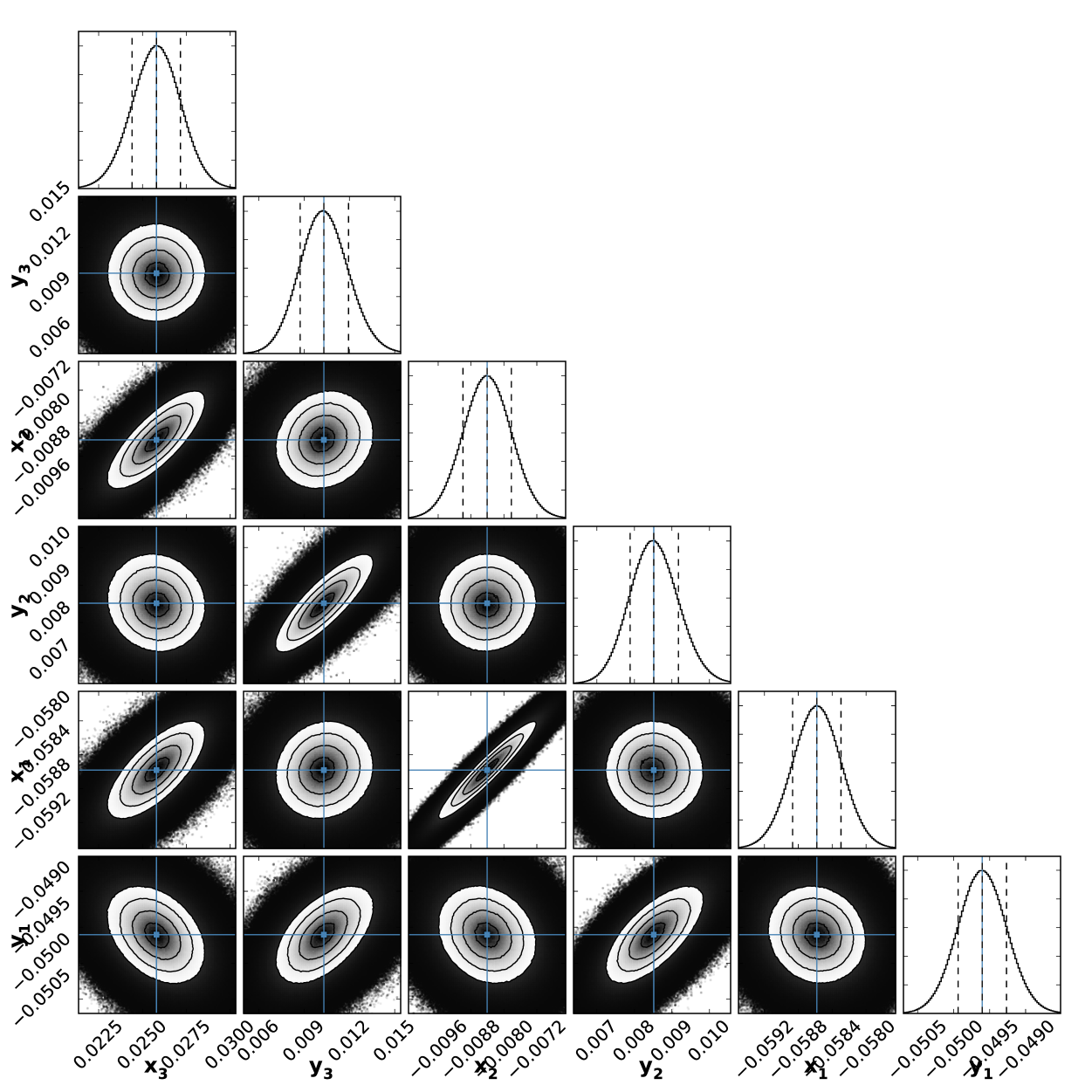}}
\caption{See caption to Fig.~\ref{fig:figureC1}.} 
 \label{fig:figureC4}
\end{figure*}

\begin{figure*}
\centerline{\includegraphics[width=1.\textwidth]{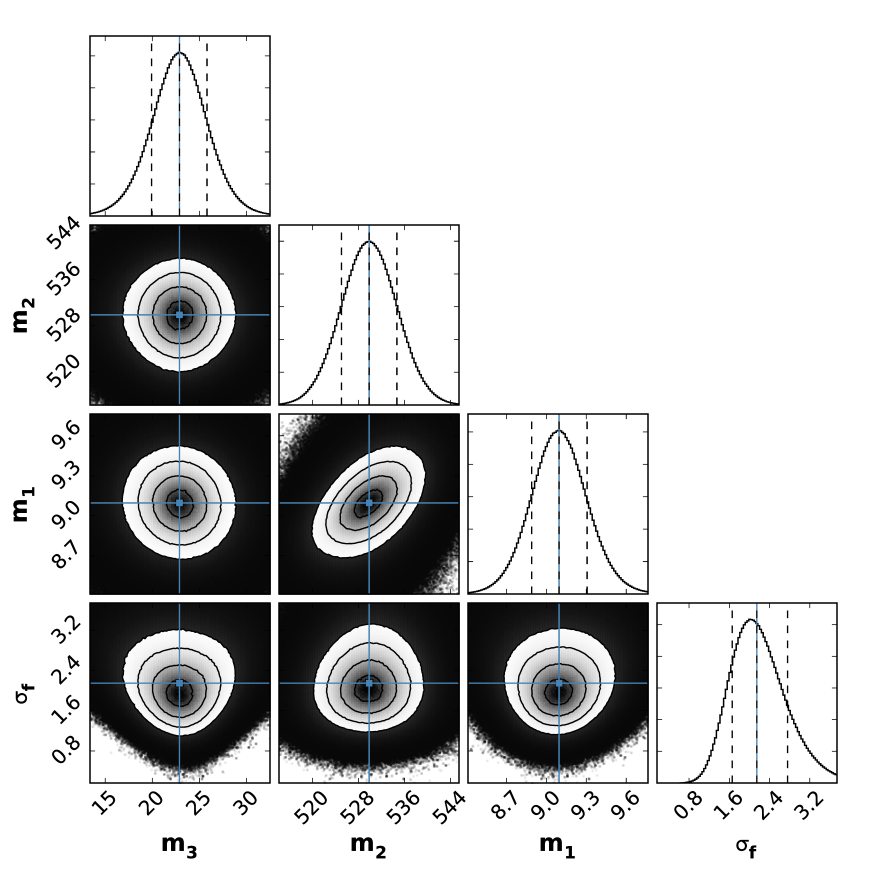}}
\caption{See caption to Fig.~\ref{fig:figureC1}. The error floor parameter $\sigma_f$ is included in the ${\cal L}$-function definition, and expressed in minutes.}
\label{fig:figureC5}
\end{figure*}

\begin{figure*}
 \centerline{\includegraphics[width=1.\textwidth]{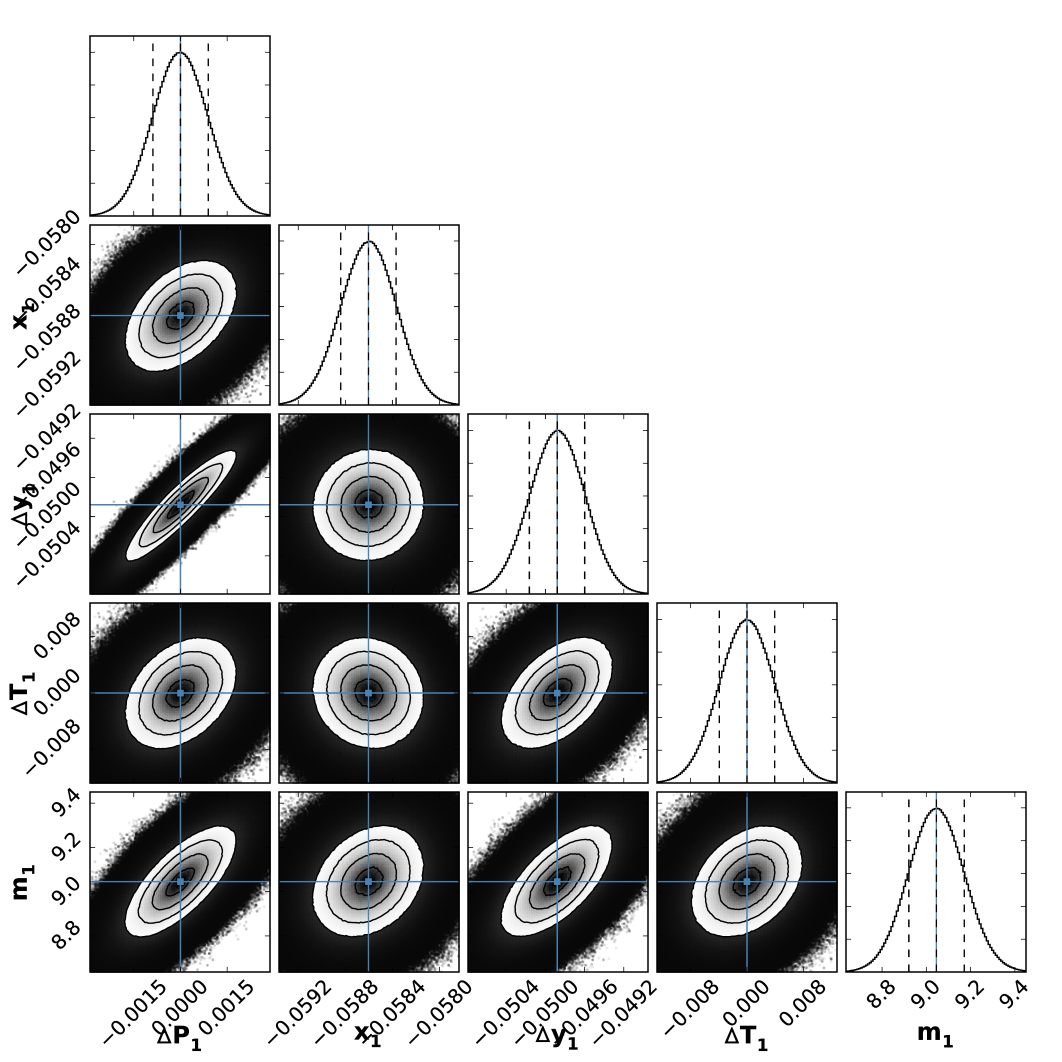}}
\caption{
One-- and two--dimensional projections of the posterior probability distribution for the TTV Model~IV of Kepler-30 (see Tab.~\ref{tab:MCMCb} and the text for details). The error floor $\sigma_f$ is not included in the ${\cal L}$-function definition. The MCMC sampling was preformed for 560 initial conditions around the best fitting model derived by the GEA experiments, and continued for 192,000 iterations each.  If a parameter is labeled as $\Delta X$ then the zero-value for  this parameter $X$ in a posterior diagram is referred to as the nominal MCMC median in Table~\ref{tab:MCMCb}. Orbital periods $P_{1,2,3}$ and the first transit times $T_{1,2,3}$ are in days, and masses $m_{1,2,3}$ are in the Earth masses. Gaussian priors with the variance of 0.05 are imposed on the $(x_{1,2,3},y_{1,2,3})$, i.e., the Poincar\'e variables (eccentricities). The blue lines mark the best-fitting parameteres consistent with the maximum of the likelihood function ${\cal L}$ and found with the GEA search. Contours are for the uncertainties expressed through the 16th, 50th and 84th percentile of samples in the posterior distribution, respectively (10\% of initial, ``\corre{burn-in}'' samples are removed). They are marked with contours in two--dimensional  diagrams. We used the \code{corner} package by \citep{corner} to plot the figures.
} 
 \label{fig:figureD1}
\end{figure*}

\begin{figure*}
 \centerline{\includegraphics[width=1.\textwidth]{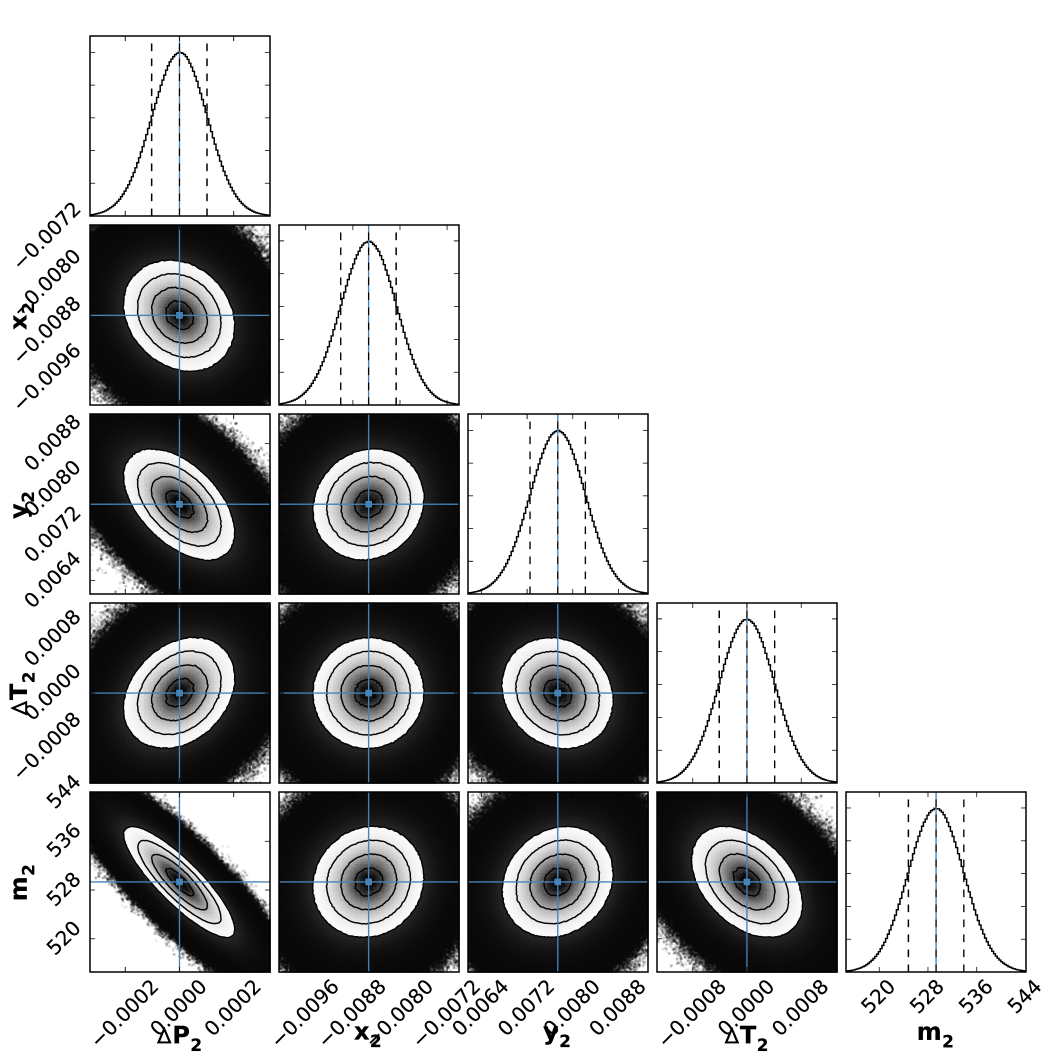}}
\caption{See caption to Fig.~\ref{fig:figureD1}.} 
 \label{fig:figureD2}
\end{figure*}

\begin{figure*}
 \centerline{\includegraphics[width=1.\textwidth]{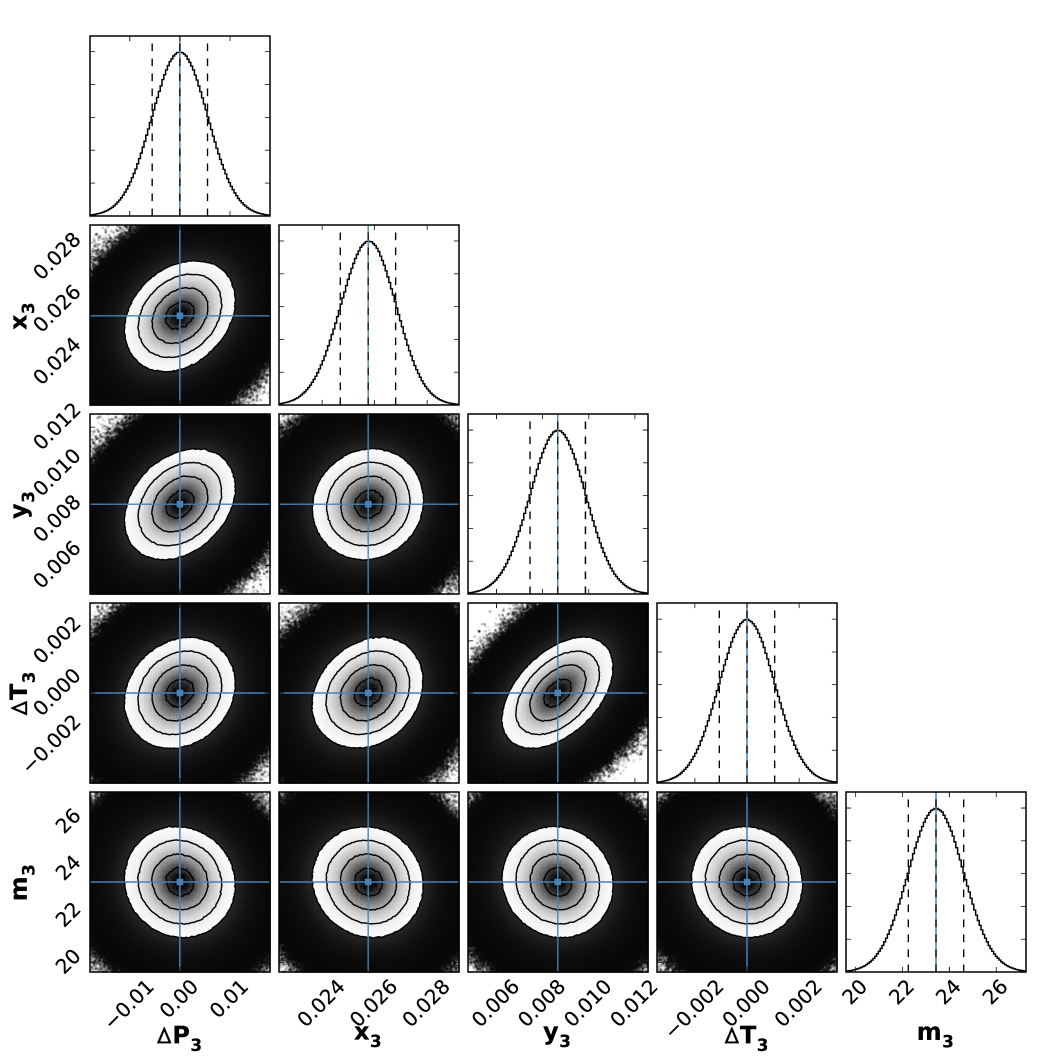}}
 \caption{See caption to Fig.~\ref{fig:figureD1}.}
  \label{fig:figureD3}
\end{figure*} 
 
\begin{figure*}
 \centerline{\includegraphics[width=1.\textwidth]{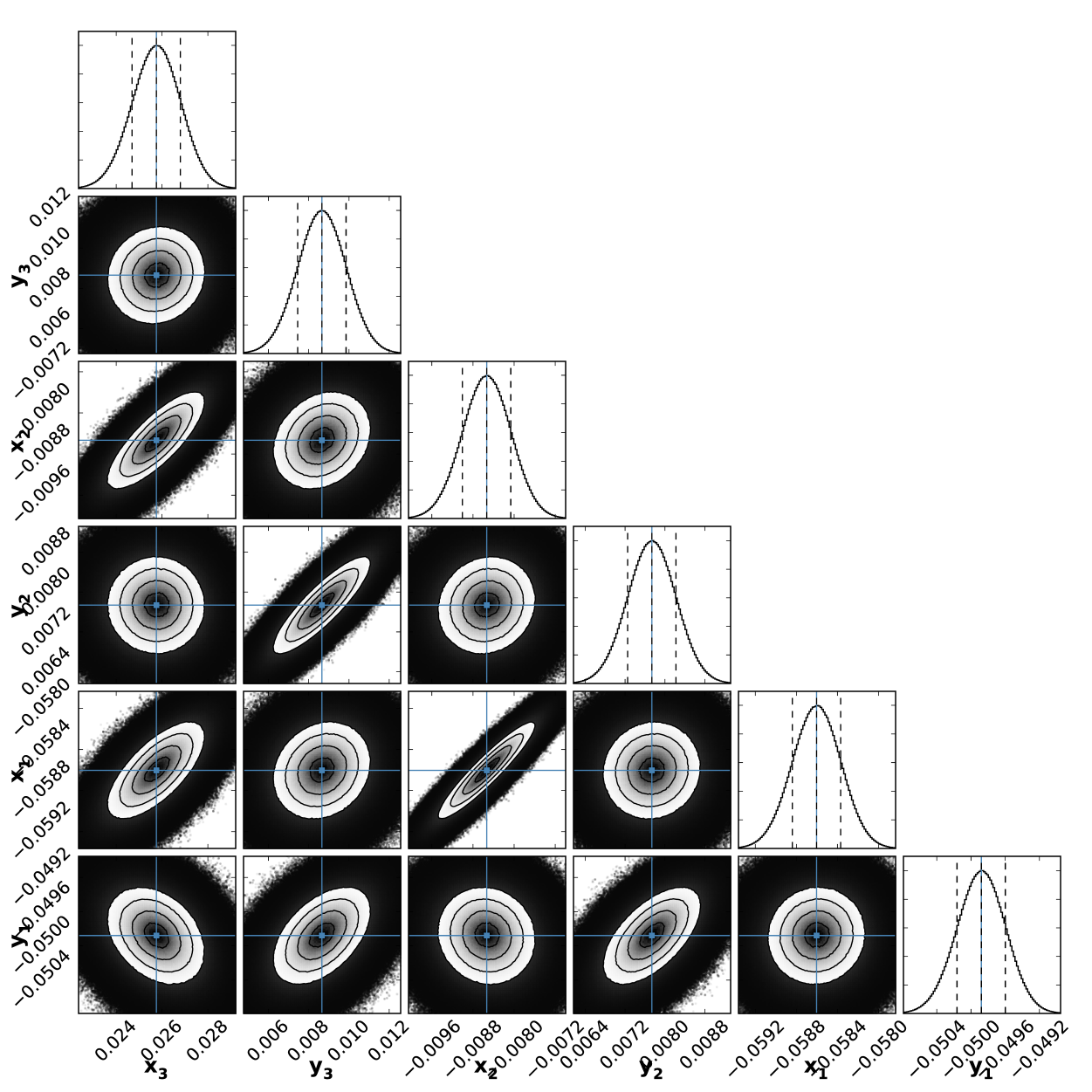}}
 \caption{See caption to Fig.~\ref{fig:figureD1}.}
  \label{fig:figureD4}
\end{figure*}

\begin{figure*}
\centerline{\includegraphics[width=1.\textwidth]{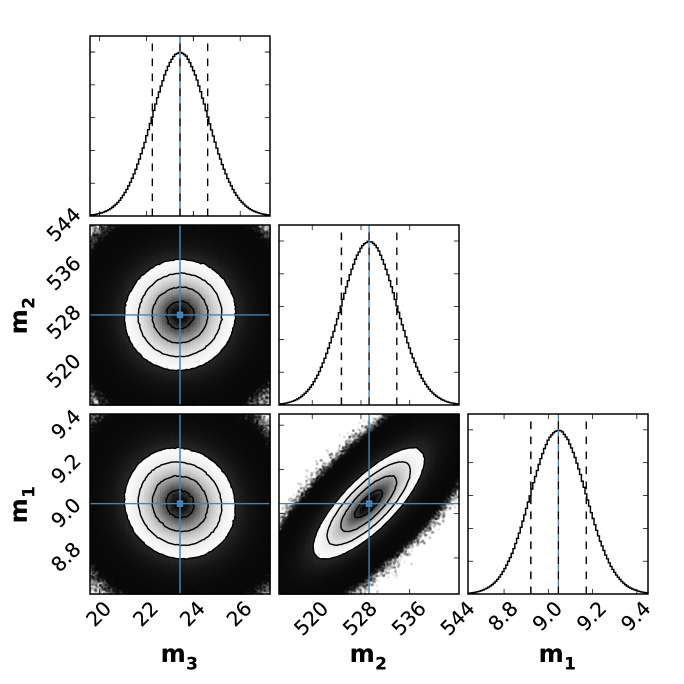}}
\caption{See caption to Fig.~\ref{fig:figureD1}.}
\label{fig:figureD5}
\end{figure*}

\end{document}